\documentclass[prd,twocolumn, nofootinbib, longbibliography,
]{revtex4-1}
\usepackage{hyperref}
\usepackage{tabularx}
\usepackage{amsmath}
\usepackage{float}
\usepackage{datetime}
\usepackage{textpos}
\usepackage{booktabs}
\usepackage{graphicx}
\usepackage{mathrsfs}
\newcommand\physrep{Phys. Rep.}
\newcommand\eq[1]{Eq.~(\ref{#1})}
\newcommand\eqs[2]{Eqs.~(\ref{#1}) and (\ref{#2})}

\newcommand\eqsss[4]{Eqs.~(\ref{#1}), (\ref{#2}), (\ref{#3})
and (\ref{#4})}

\newcommand{\ba}{\begin{eqnarray}}
\newcommand{\ea}{\end{eqnarray}}
\newcommand{\beq}{\begin{equation}}
\newcommand{\eeq}{\end{equation}}
\newcommand{\ee}{\end{equation}}
\newcommand{\be}{\begin{equation}}
\newcommand{\es}{\end{split}}
\newcommand{\bs}{\begin{split}}
\newcommand{\eg}{\end{gather}}
\newcommand{\bg}{\begin{gather}}
\newcommand{\eea}{\end{eqnarray}}
\newcommand{\bea}{\begin{eqnarray}}

\newcommand{\C}{C_{\rm syst}}

\begin{document}
\title{Dark Matter and Pulsar Model Constraints from Galactic Center Fermi-LAT Gamma Ray Observations }
\author{Chris Gordon}
\author{Oscar Macias}

\affiliation{Department of Physics and Astronomy, Rutherford Building, University of Canterbury, Private Bag 4800, Christchurch 8140, New Zealand}

\begin{abstract}
 
Employing  Fermi-LAT gamma ray observations, several  independent groups have found excess extended gamma ray emission at the Galactic center (GC).  Both, annihilating dark matter (DM) or a population of $\sim 10^3$ unresolved millisecond pulsars (MSPs) are regarded as well motivated possible explanations. However, there is significant uncertainties in the 
diffuse galactic background at the GC. 
We have performed a revaluation of these  two models for the extended gamma ray source at the GC by accounting for the systematic uncertainties of the Galactic diffuse emission model. We also
marginalize over point source and
diffuse background parameters
in the region of interest.
We show that the excess emission
is significantly more extended than a point source.
We find that the DM (or pulsars population) signal is larger than the systematic errors and therefore proceed to determine the sectors of parameter space that provide an acceptable fit to the data. We found that a population of order a 1000 MSPs with parameters consistent with the average spectral shape of Fermi-LAT measured MSPs was able to fit the GC excess emission. For DM, we found that a pure $\tau^+\tau^-$ annihilation channel is not a good fit to the data. But a mixture of $\tau^+\tau^-$  and $b\bar{b}$ with a $\left<\sigma v\right>$ of order the thermal relic value and a DM mass of around 20 to 60 GeV provides an adequate fit.
\end{abstract}

\maketitle

\section{Introduction}
\label{sec:introduction}
There is considerable  evidence 
that the majority of the matter in the Universe  consists of cold dark matter (DM) rather than 
Standard Model particles~\cite{bertonehoopersilk2005,cirelli2,clowebradacgonzalez2006,Bringmann}.
 Although, there are many dark matter candidates, one of the most strongly motivated are weakly interacting massive particles (WIMPs). Prompt production as well as decays, hadronization and radiative processes associated with the annihilation of WIMPs could result in a measurable signal of gamma ray photons which may be observable by the The Large Area Telescope (LAT) aboard the Fermi Gamma-Ray Space Telescope~\cite{Baltz:2008wd}. 
A promising location to search for WIMP annihilations is the central region of the Milky Way as it is relatively close by and has a high density of DM. However, the Galactic Center (GC) region also contains a large number of bright astrophysical sources. In particular, the interaction of energetic cosmic rays with the interstellar gas  constitutes the main source of Galactic diffuse emission. Unfortunately, there is significant uncertainty  about the propagation and origin of these cosmic rays, the distribution of the magnetic fields, radiation fields and the interstellar medium. In addition, due to the relatively low angular resolution of the LAT instrument ($\sim0.2^{\circ}$ at 10 GeV), several undetected point-like gamma ray sources could mimic diffuse gamma ray emission, consequently, the task of disentangling a tentative DM signal from the astrophysical background necessarily implies the implementation of detailed techniques to account for the uncertainties of the Galactic diffuse emission model.             

The GC hosts a supermassive black hole with a mass of $\sim 4\times10^6 \rm M_{\odot}$, called Sagittarius A* (Sgr A*). 
With the Fermi-LAT resolution, it can be modeled as point source 
with curved spectral shape~\cite{2FGL}. The interesting analysis performed in Ref.~\citep{Linden:2012iv} points out that the upcoming Cherenkov Telescope Array (CTA) will be key in the understanding of the physical mechanisms powering high energy photons from Sgr A*.

 Constraints on annihilating DM have been made using dwarf galaxies~\cite{GeringerSameth:2011iw,Ackermann:2011wa}
and  galaxy clusters 
(e.g., Refs.~\cite{Ando:2012vu,Han:2012uw,virgo}). Several studies have found an excess of gamma-rays in the GC 
that are consistent with roughly $10- 100\ \mathrm{GeV}$ DM mass annihilating into $\tau^+\tau^-$, $b\bar{b}$ final states or a combination of both~\cite{Goodenough:2009gk,Hooper:2010mq,Boyarsky:2010dr,hooperlinden2011,AK,AKerratum}. 
The Fermi-LAT Collaboration have not yet published a full analysis of  GC excess, but a preliminary study  by them using one year of data, reported an excess in observed counts around energies of $2-5\rm\ GeV$~\cite{Vitale:2009hr,2011NIMPA.630..147V} at the GC.

The signal was also shown to be consistent with a population of millisecond pulsars (MSPs) in the GC~\cite{Abazajian:2010zy,AK,AKerratum}.
Studies have also looked at the possibility of the signal arising from cosmic-ray interaction with gas in the GC \cite{hooperlinden2011,Linden:2012iv,yusef-zadehhewittwardle2013,AK,AKerratum}. 
In Ref.~\cite{AK,AKerratum}, they highlighted the need to marginalize over the point source (PS) parameters, due to their degeneracy with any 
proposed model for the excess GC emission.

In this article we extended the treatment of Ref.~\cite{AK,AKerratum} in a number of ways. In particular we estimate systematic errors for the galactic diffuse background. We also evaluate marginalized confidence intervals and determine the areas of parameter space that provide an acceptable fit to the data. In Sec.~\ref{sec:observations} we describe the data used and some initial goodness of fit tests. In Sec.~\ref{sec:maps}   we check the spatial fit of the models and evaluate the systematic errors in the diffuse Galactic background.  The results are given in Sec.~\ref{sec:Results} and the discussion and conclusions are given in Sec.~\ref{sec:discussion} and \ref{sec:conclusions}.

\section{Fermi-LAT observations and data reduction }
\label{sec:observations}
A detailed description of the characteristics and performance of the LAT instrument aboard Fermi is given in Ref.~\citep{FermiInstrument}. The LAT data used in this work were collected for about 45 months of continuous sky survey observations over the period August 4th 2008$-$June 6th 2012 (corresponding to mission elapsed time (MET) $239557417-360716517$). The \texttt{SOURCE} event class was chosen and photons beyond the earth zenith angle of $100^{\circ}$ were excluded to minimize Earth albedo gamma rays. Time periods during which the spacecraft rocking angle is larger than $52^{\circ}$ are also excluded as an additional guard against gamma-ray contamination. We further restrict the analysis to the photon energy range 200 MeV$-$100 GeV and make no distinction between \textit{Front} and \textit{Back} events.

We select all events within a squared region of interest (ROI) of size $7^{\circ} \times 7^{\circ}$ centred on $(\alpha, \delta)=(266\overcirc{.}417, -29\overcirc{.}008)$. This position coincides with the current best fit coordinates of the gamma-ray source \texttt{2FGLJ1745.6-2858} (Sgr A*). The analysis is performed using the LAT Science Tools package \texttt{v9r27p1} and the \texttt{P7$_{-}$V6} instrument response functions (IRFs).

 We model the Galactic background component using the LAT standard diffuse background model \texttt{gal$_{-}$2yearp7v6$_{-}$v0.fits}. The extragalactic and residual instrumental backgrounds, assumed as being isotropic, are fitted with the file \texttt{iso$_{-}$p7v6source.txt}. 

The analysis of the Fermi-LAT spectrum was performed using a binned likelihood technique~\cite{mattox} with the \textit{pyLikelihood} library in the Science Tools. The energy binning was set to 20 logarithmic evenly spaced bins.

We adopted the same fitting procedure followed in Ref.~\citep{AK,AKerratum}. This is a relaxation method which consists in freeing the spectral model parameters consecutively from their distance to Sgr A*. Normalizations are freed first, and then the full spectra within concentric regions: within $2^{\circ}$, then within the $7^{\circ} \times 7^{\circ}$ square region and finally in the full ROI and for all sources whose $TS>25$.     Where, the test statistic ($TS$) is defined as in Ref.~\cite{2FGL} 
\begin{eqnarray}\label{tsdef}
TS=2\left[\log \mathcal{L} (\mbox{new source})-\log \mathcal{L} (\mbox{NO-new source})\right]\mbox{,} \qquad 
\end{eqnarray}
where $\mathcal{L}$ stands for the maximum of the likelihood of the data given the model with or without the new source at a certain location of the ROI.
In the large sample limit, under the no source hypothesis, $TS$ has a $\chi^2/2$ distribution with the number of degrees of freedom equal to the number of 
parameters associated with the proposed positive amplitude new source  \cite{wilks,mattox} which in this case is two for position, one for amplitude, and one for spectral slope, so four in total. As the amplitude is restricted to be be non-negative, a $\chi^2/2$ distribution rather than the $\chi^2$ distribution is needed.

Using the \texttt{make2FGLxml.py} tool we generated all the relevant 2FGL sources that could contribute to the ROI and applied to it the aforementioned relaxation method, this is called the ``baseline'' model~\citep{AK,AKerratum}. 

\subsection{Detection of an Extended Source at the Galactic Center }
\label{subsec:ExtSrcdetection} 

\begin{figure}[!t]
\centering
\includegraphics[width=1.0\linewidth]{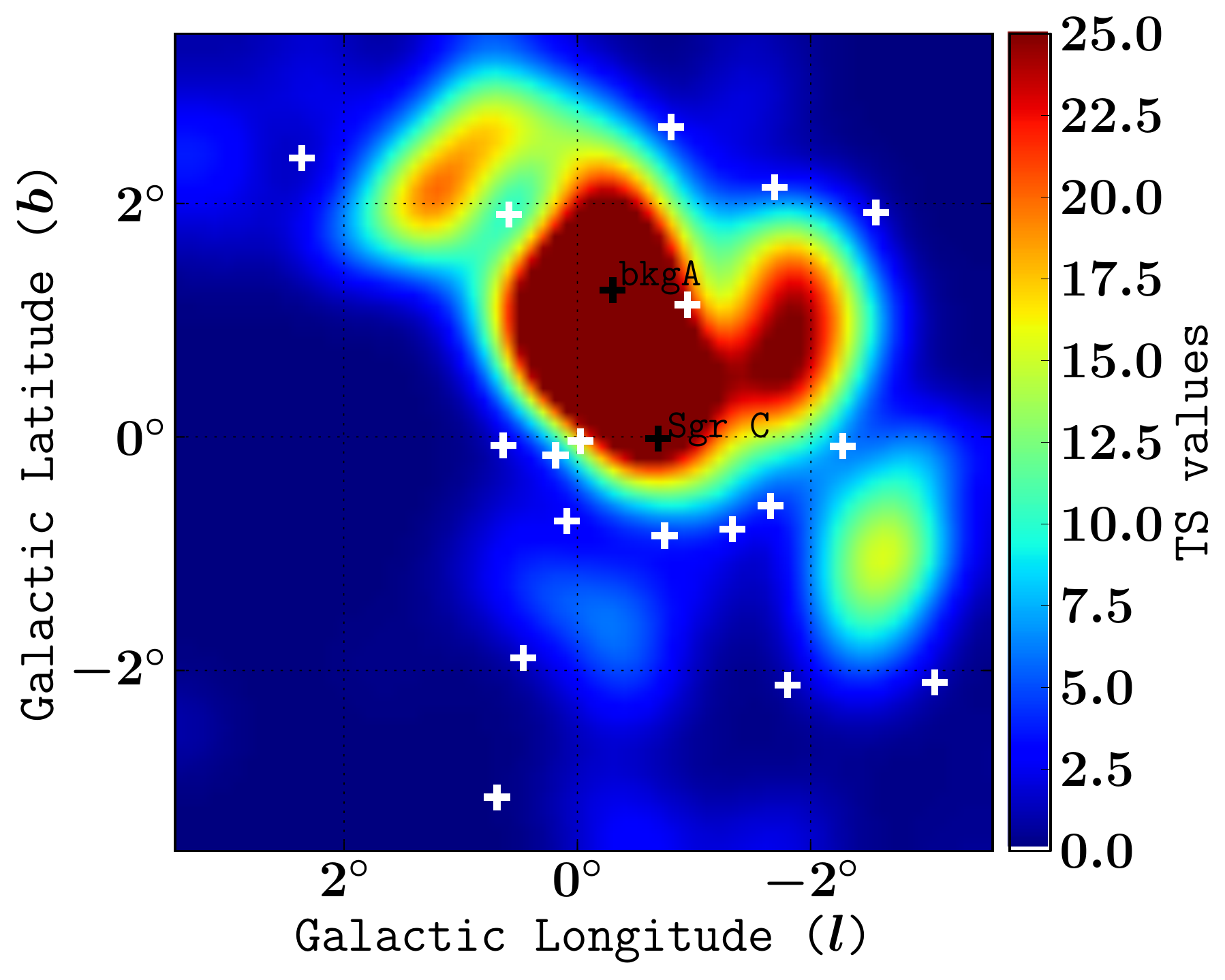} 
\includegraphics[width=1.0\linewidth]{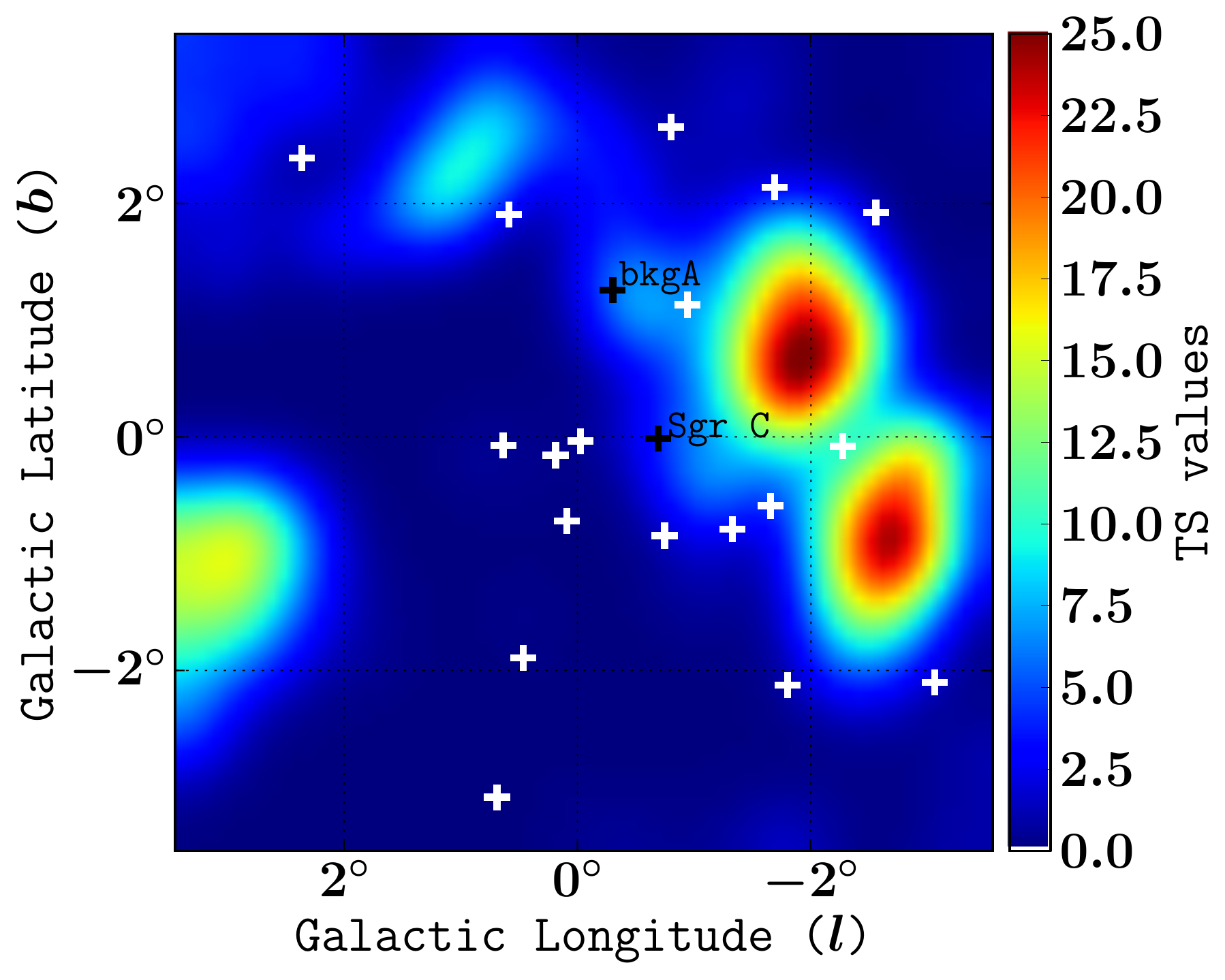} 
 \caption{ \label{fig:tsmaps}Residual test statistics ($TS$) maps in the energy range 300 MeV$-$100 GeV for two different best fit models of the Galactic Center region using: (a) only the known 2FGL point and extended sources (baseline model), highlighted here with white crosses (b) the full set of 2FGL sources plus the best fit spatial and spectral model of an extended source at the Galactic Center (see Sec~\eqref{sec:maps} for details on maps for the extended source). The the two black crosses show the localization of two recently proposed gamma-ray PSs~\cite{yusef-zadehhewittwardle2013}  named \texttt{bkgA} and \texttt{Sgr C}, whose significance drops drastically once the extended source has been taken into consideration. This can be seen in the bottom figure. The maps span a $7^{\circ} \times 7^{\circ}$ region of the sky centered at the Sgr A* position and the extent of every pixel is $0.1^{\circ} \times 0.1^{\circ}$. The residual $TS$ maps have been smoothed for display with a $\sigma=0.3^{\circ}$ Gaussian. For display purposes the images have been thresholded at $TS=25$.}   
\end{figure}

In order to evaluate to what extent the data prefers a model that considers GC excess extended emission instead of the conventional one assumed in the second year Fermi catalogue (2FGL)~\citep{2FGL}, we have constructed two residual test statistics ($TS$) maps shown in Fig.~\eqref{fig:tsmaps}. 
For a given pixel in the map, a trial new PS is added with a power law spectrum and its $TS$ evaluated. 
 The usual convention \cite{2FGL} is to 
investigate the possibility of a new PS if $TS\ge 25$ for PSs far from the galactic plane.
In producing the $TS$ images, we made use of the Fermi Science Tool \texttt{gttsmap} as recommended in the \texttt{Cicerone.}~\footnote{http://fermi.gsfc.nasa.gov/ssc/data/analysis/documentation/Cicerone/}    

We notice that by including the new best fitting spatially extended source Fig.~\eqref{fig:tsmaps}-(b), the ROI integrated $TS$ of the map decreased by $48\%$ relative to a fit with no GC extended source, Fig.~\eqref{fig:tsmaps}-(a). The inclusion of an GC extended source typically has a $TS$ of order 800 and so very significantly favored by the data. In Ref.~\cite{yusef-zadehhewittwardle2013} two new PSs named \texttt{bkgA} and \texttt{Sgr C}, were claimed to have been discovered. In fact, our analysis shows that once the more adequate extended source is included, their significance fades in the Fermi-LAT data. Nevertheless, the incidence that these two new PSs had on the extended source hypothesis was evaluated in Ref~\citep{AK,AKerratum} finding negligible variation on their main conclusions. We therefore do not attempt to model those sources in this article.    

Visual inspection of the $TS$ image shown in Fig.~\eqref{fig:tsmaps}-(b) suggests that the residuals can be further ameliorated by including two new PSs at the coordinates listed in Table~\eqref{tab:1}. However, based on the examination of the sources nearby Cygnus, Orion and molecular clouds, the Fermi collaboration~\citep{2FGL} stipulated that depending on the intensity of the diffuse background, sources near the galactic ridge need to have $TS \gg 25$ to not be considered diffuse features.

We calculated the background photon count per pixel $N_{bkgd}$ by integrating from 589 MeV to 11.4 GeV the diffuse model cube for our ROI and found an average of $N_{bkgd}=42.2$ counts per pixels (where each pixel spans an area of $0.1^{\circ} \times 0.1^{\circ}$). According to this source detection criteria, a new source would need to have a $TS\gtrsim 80$ to be seriously considered for a multi-wavelength search. We therefore do not claim the discovery of new PSs in the field of view.

Interestingly, in a recent study of the Virgo cluster~\citep{han}, it was claimed the detection of extended gamma-ray emission interpreted as Dark Matter annihilation. That hypothesis was later disputed in Ref.~\citep{virgo} arguing that a set of previously unknown PSs or features of the diffuse background could have accounted for the majority of the excess emission. This was later confirmed in Ref.~\citep{han2}. 

We undertook here the same approach as in~\citep{virgo} and evaluated the new significance of the excess emission when the PSs in Table~\eqref{tab:1} are included. However, contrary to what happened in the Virgo case~\citep{virgo} we found that the $TS$ and flux of the extended source at the GC were mildly enhanced (see details in Sec.~\eqref{sec:maps}).               

\begin{table}[H]
\caption{\label{tab:1}Point source candidates found in the GC field of view for almost four years of Fermi-LAT data. The PS detection and localization were carried out following the same approach explained in Ref~\cite{virgo}.}
\begin{ruledtabular}
\begin{tabular}{ccc}
\centering
Right Asc. [deg] & Dec. [deg] & $TS$  \\ \hline 
264.813 & -30.270 & 70.8 \\
265.735& -31.814 & 65.1 \\ 
\end{tabular}
\end{ruledtabular}

\end{table}

\begin{figure*}[ht!]
\begin{center}

\begin{tabular}{cc}
\centering
\includegraphics[width=0.5\linewidth]{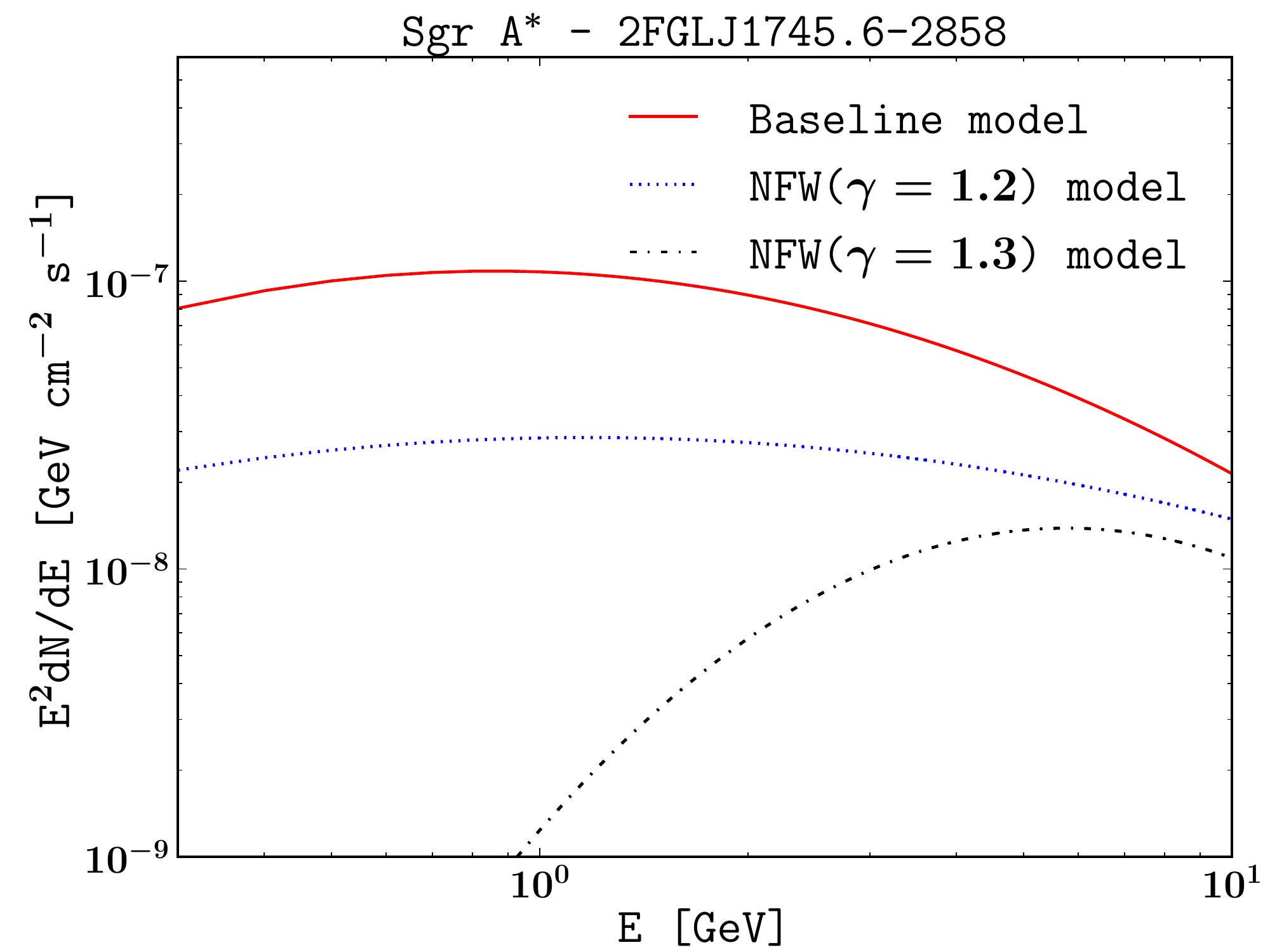} &
\centering
\includegraphics[width=0.5\linewidth]{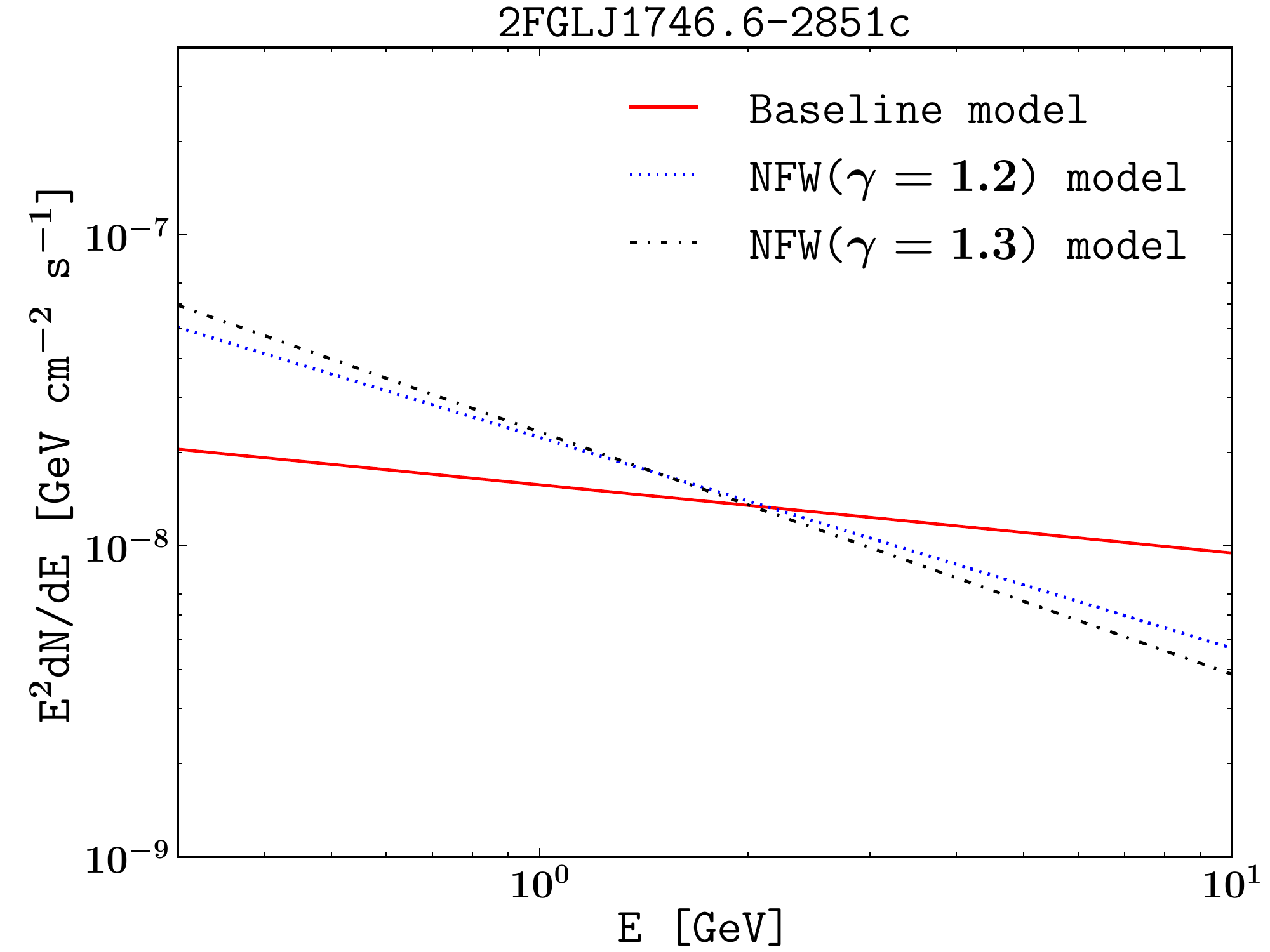} 
\end{tabular}
\begin{tabular}{cc}
\centering
\includegraphics[width=0.5\linewidth]{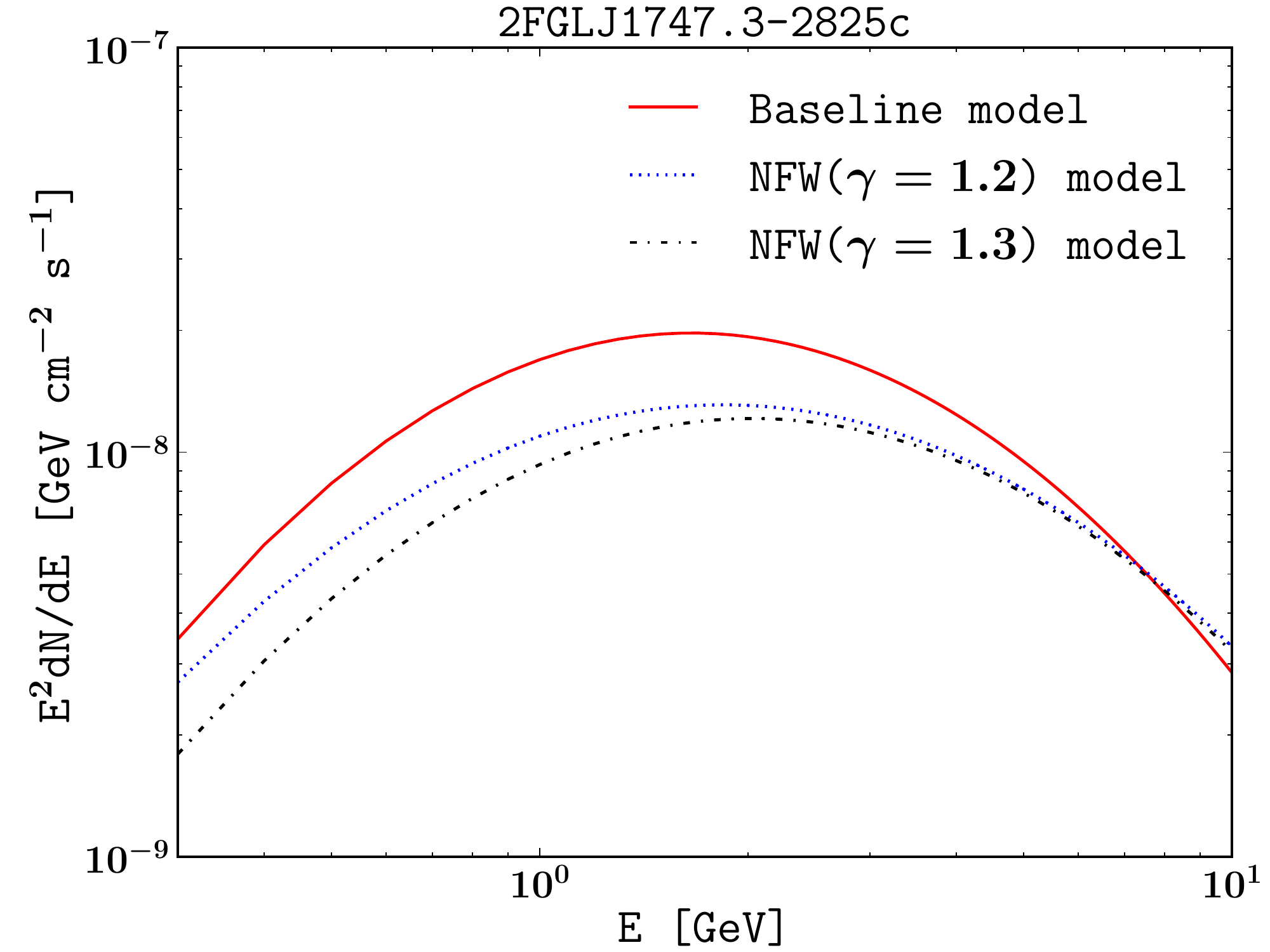} &
\centering
\includegraphics[width=0.5\linewidth]{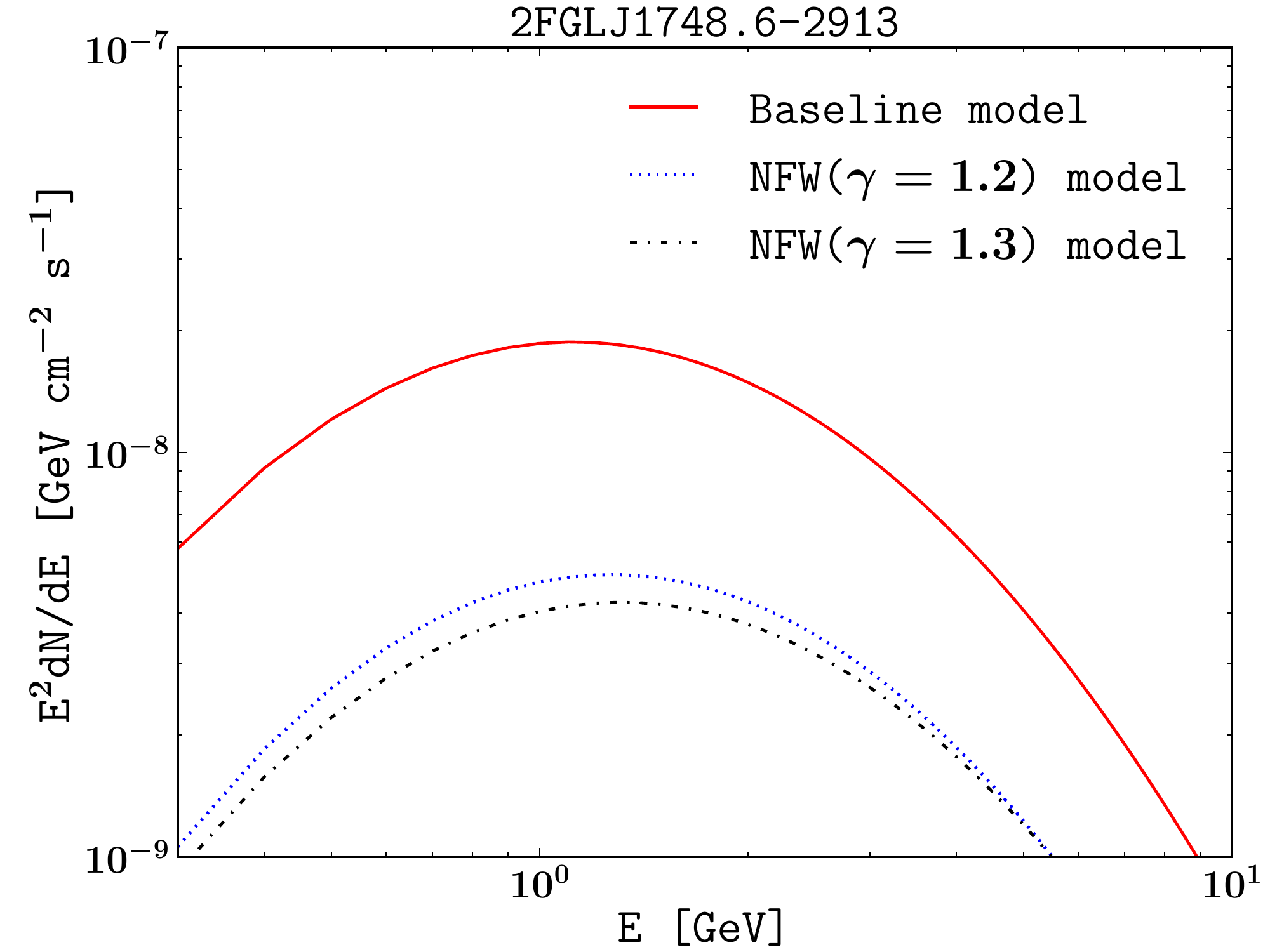} 
\end{tabular}

\end{center}
\caption{ \label{fig:degeneracy} Shown is the spectrum of the four 2FGL PS displaying the largest degeneracy pattern as obtained from three different fits: Continuous red line shows the spectrum for each source that we get from our baseline model (\textit{i.e.} a model that just assumes  the conventional 2FGL sources). Blue dotted and black dash-dotted exhibit the sources spectra when the newly discovered extended source at the GC is included. This extended source is modelled with spatial maps following a universal NFW profile with inner slope $\gamma=1.2$ and $\gamma=1.3$ respectively (see details on maps in Sec.~\eqref{sec:maps}). The spectra of the extended source is modelled with a Log Parabola for both cases. The sources spectra shown here are organized in order of their proximity to the central position from left to right and top to bottom and all of them are located within $1^{\circ}$ of the centre of the ROI.  }
\end{figure*}

In the innermost region of the GC (a circular area with a radius of about $1^{\circ}$ centred on Sgr A*) the spectral parameters describing the gamma-ray sources are degenerate with the extended source parameters \citep{AK,AKerratum}. This means that when the new extended gamma-ray source is not considered in the analysis Fig.~\eqref{fig:tsmaps}-(a), the four nearest sources to the central position are assigned a larger amplitude to account for the excess emission~\citep{AK,AKerratum}. This phenomena can be seen in Fig.~\eqref{fig:degeneracy}, where the behavior of the four sources in the innermost region is depicted.

\section{Morphology of the Extended Source}
\label{sec:maps}

\subsection{Dark Matter and Pulsars Maps}
\label{subsec:DMmaps}    

\begin{figure*}[ht!]
\begin{center}
\centering
\includegraphics[width=0.5\linewidth]{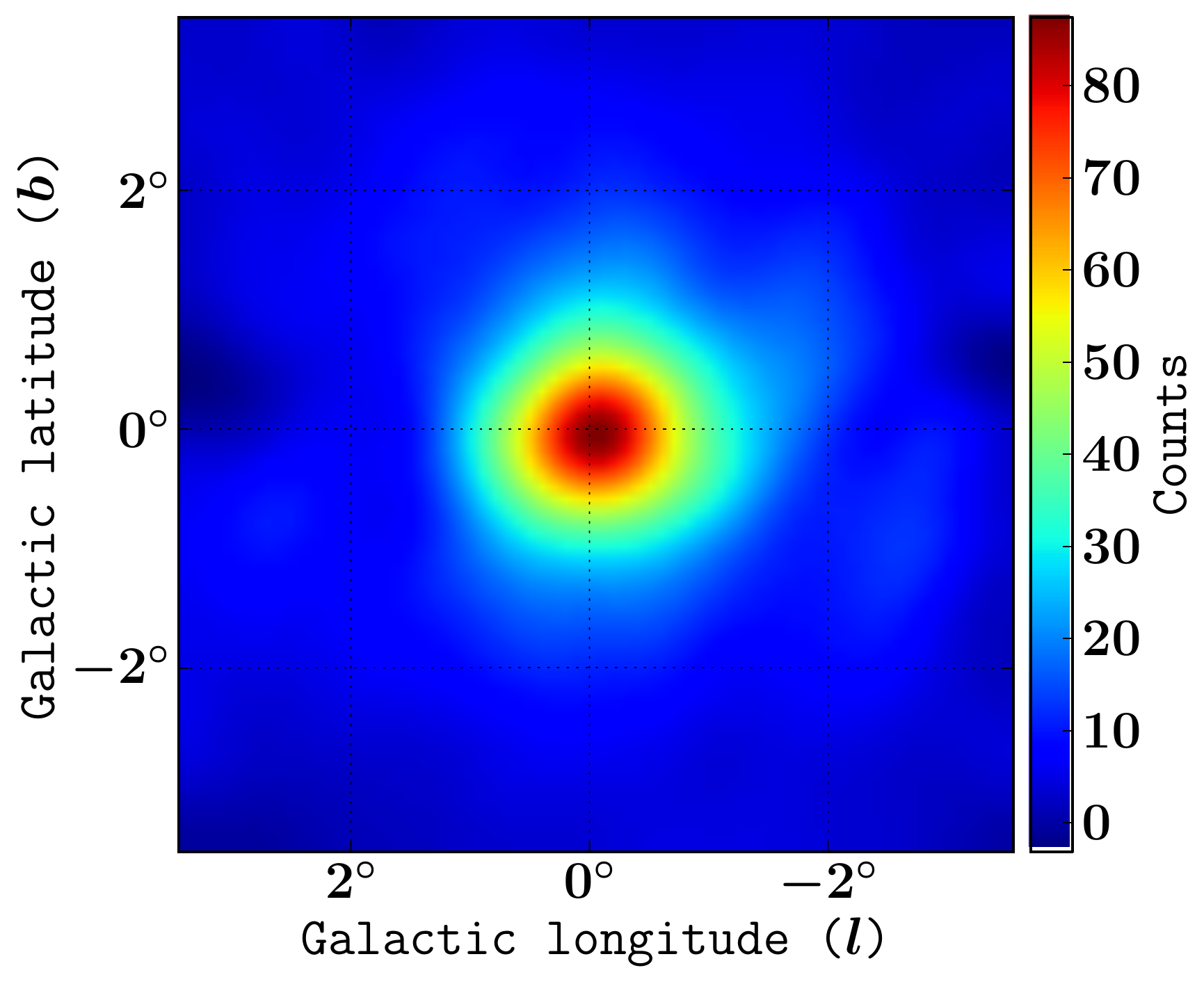} 

\end{center}
\caption{ \label{fig:ExtSrcMorphology1} LAT residual map after subtraction of our best fit model
with an extended GC source, but without subtracting the extended source model component.
 The counts were summed over the energy range 300 MeV$-$10 GeV. The map spans a $7^{\circ} \times 7^{\circ}$ region of the sky centred at the Sgr A* position with pixel size of $0.1^{\circ} \times 0.1^{\circ}$. The residual has been smoothed with a $\sigma=0.3^{\circ}$ Gaussian.   }
\end{figure*}

\begin{figure*}[ht!]
\begin{center}

\begin{tabular}{cc}
\centering
\includegraphics[width=0.4\linewidth]{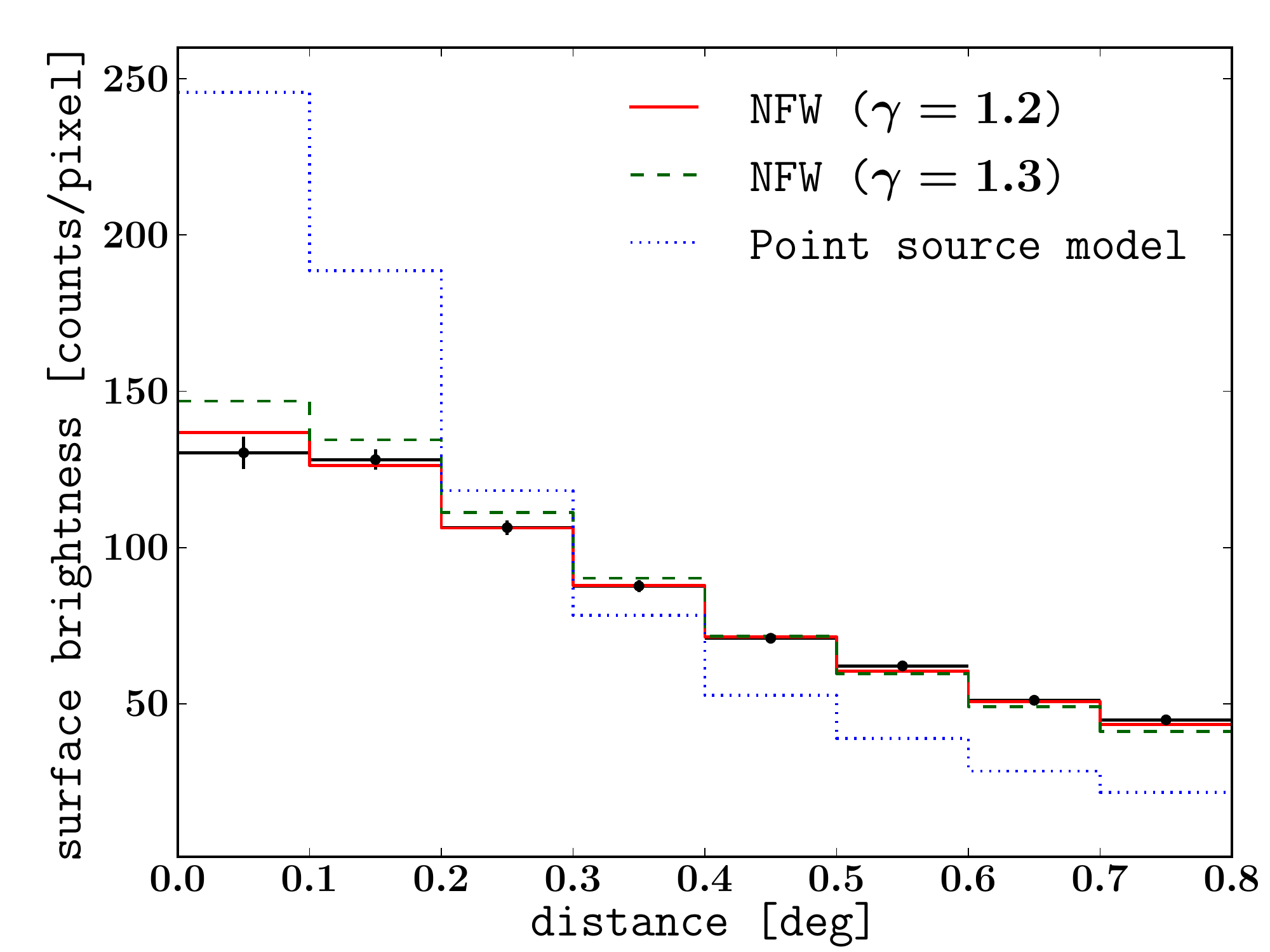} &
\centering
\includegraphics[width=0.4\linewidth]{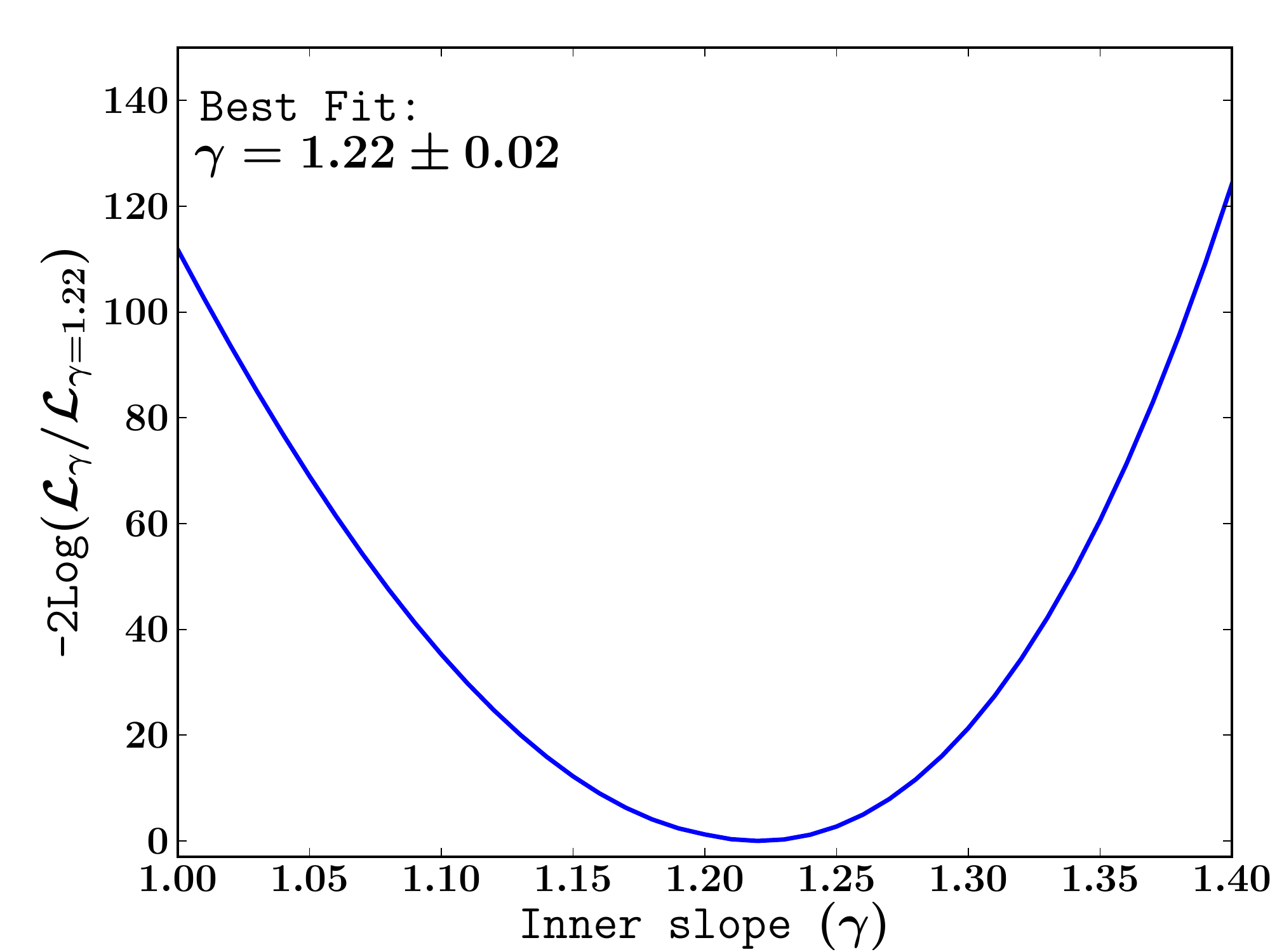} 
\end{tabular}

\end{center}
\caption{ \label{fig:ExtSrcMorphology2}(a) Radial profile of the LAT residuals shown in Fig.~\eqref{fig:ExtSrcMorphology1} as obtained from a ring analysis computed around Sgr A*. The histograms show the effective LAT point spread function (PSF) for three different profile models: (i) NFW with inner slope $\gamma\simeq 1.2$ (red continuous line) for which we get $\chi^2/\mbox{dof}=5.5/7$. (ii) NFW with $\gamma=1.3$ (green dashed line) and $\chi^2/\mbox{dof}=44.6/7$, and lastly (iii) the profile for a PS model (blue dotted line) with $\chi^2/\mbox{dof}=2479.9/7$. For all cases the spectra was modelled with a Log Parabola. (b) Shown is the significance of NFW profiles with varying inner slope, where $\mathcal{L_{\gamma}}$ represents the likelihood function at a given $\gamma$. This was assessed by performing a set Fermi Tools runs where for each case the relaxation method
 was used. The spectra was fitted with a Log Parabola function and only statistical uncertainties were taken into account. }
\end{figure*}

The gamma ray flux emitted by WIMP particle interactions with mass $M_{\rm DM}$ can be factorized~\citep{Baltz, Rott,Bergstrom} in two conceptually distinct terms (i) a ``particle physics factor" $\Phi^{PP}(E_{\gamma})$ that accounts for the number of gamma ray photons  produced per annihilation event at a given photon energy, and (ii) an ``astrophysical factor" $J(b,l)$, which measures the number of dark matter particle pairs producing photons along the line of sight direction. That is
\begin{equation}
\Phi(E_{\gamma},b,l)= \Phi^{PP}(E_{\gamma})\times J(b,l), 
\label{Phi}
\end{equation}
where $b$ and $l$ are the Galactic latitude and longitude respectively. The particle physics contribution is often written as

\begin{equation}
\Phi^{PP}(E_{\gamma})=\frac{1}{2}\frac{\left\langle\sigma v \right\rangle}{4\pi M^2_{DM}}\sum_{f} \frac{dN_{f}}{dE_{\gamma}}B_f,
\label{PhiPP}
\end{equation}        
where $\left\langle \sigma v \right\rangle$ is the annihilation cross-section of two DM particles times their relative velocity, averaged over the velocity distribution. $dN_{f}/dE_{\gamma}$ is the differential gamma ray multiplicity per annihilation, $B_f$ the branching ratio and $f$ stands for the final state particles resulting from the annihilation.

The astrophysical factor in the $(b,l)$ direction is integrated over the line of sight \citep{Bergstrom}
\begin{equation}
J(b,l)=\int^{\infty}_0 ds\; \left.\rho(r)^2\right|_{r=\sqrt{R^2_{\odot}-2sR_{\odot}cos(b)cos(l)+s^2}},
\label{J}
\end{equation}
with $s$ varying in the line-of-sight path and $R_{\odot}=8.25$ kpc is the distance from the solar system to the GC. Since the spatial binning of our Fermi files was $0.1^{\circ}\times 0.1^{\circ}$, we constructed the spatial maps by averaging the astrophysical factor over the corresponding solid angle around the $(b,l)$ coordinates~\citep{Bergstrom}   

\begin{equation}
\left\langle J(b,l) \right\rangle_{\Delta \Omega}=\frac{1}{\Delta\Omega}\int_{\rm pixel} J(b,l)d\Omega,
\end{equation}
where the differential solid angle is given by $d\Omega=db \; dl\cos(b)$.

As in Ref~\citep{AK,AKerratum}, throughout this work we shall use template maps of DM that assume a generalized Navarro-Frenk-White (NFW) profile   \cite{navarrofrenkwhite1996,klypinzhaosomerville2002}
\begin{equation}
\rho(r)=\frac{\rho_s}{\left(\frac{r}{r_s}\right)^{\gamma}\left[1+\left(\frac{r}{r_s}\right)^{\alpha}\right]^{(\beta-\gamma)/\alpha}},
\label{nfw}
\end{equation}
where we fixed $r_s=23.1$ kpc, $\alpha=1$, and $\beta=3$. 

It has been suggested that the excess emission seen in the GC can also be explained by a superposition of unresolved PSs (MSPs) that might be distributed as a mildly contracted NFW profile. We tested this  hypothesis by normalizing to unity the $\left\langle J(b,l) \right\rangle$  maps as explained in the \texttt{Cicerone}.~\footnote{http://fermi.gsfc.nasa.gov/ssc/data/analysis/scitools/extended}

These normalized maps were also used to fit for the inner slope $\gamma$. This was done with two equivalent methods:
\begin{itemize}

\item We first computed the residual emission shown in Fig.~\eqref{fig:ExtSrcMorphology1}. From this we produced a radial profile Fig.~\eqref{fig:ExtSrcMorphology2}-(a) of the photon excess. This was compared with that expected from a PS and also from well motivated spatially extended sources using a $\chi^2$ test. The profiles for extended source shown in the histograms Fig.~\eqref{fig:ExtSrcMorphology2}-(a) were obtained with the \texttt{gtmodel} routine. The models entered to this Tool  were $\left\langle J(b,l) \right\rangle$ maps normalized to unity with $\gamma \in [1.0,1.5]$ and a Log Parabola spectra
\begin{equation}
\frac{dN}{dE}=N_0\left(\frac{E}{E_0}\right)^{-\left(\alpha+\beta \log\left[\frac{E}{E_0}\right]\right)}.
 \label{eq:logparabola}
\end{equation} 
The height of each bin is given by the mean of the residual in a ring of pixels centered around the GC. The error bars were evaluated as the standard deviation of the pixels in the ring divided by the square root of the number of pixels in the ring.

\item Following a more statistically robust approach we proceeded to fit for $\gamma$ with the \textit{pyLikelihood} Tool Fig.~\eqref{fig:ExtSrcMorphology2}-(b). Compared to the previous method, this one has the advantage of carefully considering the energy binning in the likelihood function.  

\end{itemize}
As we are using the profile likelihood approach \cite{rolkelopezconrad2005}, we set $\gamma=1.2$ unless otherwise specified.  Although ideally one should maximize the likelihood for $\gamma$ simultaneously with the other parameters, our initial tests show that the preference for $\gamma=1.2$ is robust to changes in the spectral model.  Also maximizing the likelihood of the microlensing and dynamical data (see Fig.~5 of \cite{ioccopatobertone2011}), $\gamma=1.2$ corresponds to $\rho_0\equiv\rho(R_{\odot})=0.36$~GeV~cm$^{-3}$. From \eqsss{Phi}{PhiPP}{J}{nfw} the annihilation gamma ray flux is $\Phi\propto  \left\langle\sigma v \right\rangle\rho_0^2$ and so $\rho_0$ is not constrained by the Fermi-LAT measurements alone. Also, the microlensing and dynamical data  have a very weak constraint on $\gamma$, compared to our Fermi-LAT analysis and so this justifies using the Fermi-LAT best fit value for $\gamma$ in constraining $\rho_0$. 

The microlensing and dynamical data constrain the scale radius to be
$r_s=20^{+15}_{-10}$ kpc \cite{ioccopatobertone2011}. As this is much larger
than the extent of the excess emission (200 pc),  the gamma ray data is
not able to constrain $r_s$. But, as can be seen from \eqs{J}{nfw},
$r_s$ may effect $J$ and it will also be completely degenerate with
$<\sigma v>$. In line with the profile likelihood approach, we choose
$r_s$ to be consistent with the maximum likelihood value given in Ref.
[34]. It would be better to use the maximum likelihood value of
$r_s$ when $\gamma$ is fixed to 1.2, but the joint confidence intervals for $r_s$ and
$\gamma$ are not given in Ref. \cite{ioccopatobertone2011}. Our current approach should provide a 
reasonable approximation, unless the  microlensing and dynamical data have 
a very high correlation in the joint confidence intervals for $r_s$ and $\gamma$.

As it has been seen in Fig.~\eqref{fig:ExtSrcMorphology2}-(a) and Fig.~\eqref{fig:ExtSrcMorphology2}-(b), comparisons between LAT PSFs and photon distributions indicates that the observed excess emission is consistent with an extended source whose spatial distribution is well described by a mildly contracted NFW profile. Below we outline how we examined its spectral morphology for a DM hypothesis.

We calculated the gamma ray spectra from WIMPs self-annihilations with the \texttt{DMFIT} tool as described in~\citep{profumo1}. This package provides interpolating functions calculated from simulations of DM annihilations with the \texttt{DarkSUSY} software~\citep{darksusy} which in turn interpolates over \texttt{PYTHIA 6.4}~\citep{pythia6} tables.

It has recently been pointed out that there are discrepancies~\citep{Cembranos} between the gamma ray spectra calculated with \texttt{PYTHIA 6.4} (Fortran version) and \texttt{PYTHIA 8.1}~\citep{pythia8} (C++ version), that software analysis using interpolating functions can overestimate the energy cut-off of the gamma-ray spectra~\citep{Cirelli}, and that not considering electroweak corrections can also create deviations between predicted DM annihilitions spectra~\citep{Kachelriess, Ciafaloni}. We therefore looked for a statistical bias in our analysis by producing \texttt{PYTHIA 8.1} tables for a few WIMP masses and found that for the relevant energy scale and annihilation channels used in our work, the discrepancies between the results obtained with DMFIT and \texttt{PYTHIA 8.1} were marginal.       

For the DM spectrum we considered soft gamma ray spectra produced from annihilation into $b\bar{b}$ quarks and hard spectra as produced by annihilations into $\tau^+\tau^-$ or a combination of leptons pairs $e^{+}e^{-}$, $\mu^{+}\mu^{-}$ and $\tau^+\tau^-$. Since the annihilation products are highly model dependent we studied extremes of the possible annihilation channels assuming a branching ratio of 100$\%$ for each of them in turn (except for the case of 100$\%$ $e^{+}e^{-}$), but mixtures of soft and hard spectra were also evaluated in order to fit for the best branching ratio $B_f$. 

\subsection{Examination of Systematics  in the Galactic Diffuse Background Model}
\label{subsec:SysErrors}  

The LAT team developed a model for the Galactic diffuse background map which is an essential input to the analysis for detecting and characterizing gamma ray sources. The model file \texttt{gal$_{-}$2yearp7v6$_{-}$v0.fits} introduced in the 2FGL catalogue~\citep{2FGL} was created by fitting all-sky gamma ray data with a highly sophisticated physical model. In a nutshell; the distribution of interstellar gas and dust was obtained from independent observations, then three-dimensional models of magnetic fields, distributions of optical photons and models of $p$ and $e^{-}$ injections were assumed. By propagating these primary particles through the gas with the  GALPROP\footnote{http://galprop.stanford.edu} software package, the resulting photons from inverse Compton (IC), bremsstrahlung and $\pi^{0}$ decays, were predicted and fitted with gamma ray data.

Since the newly discovered extended source is located in the region where the Galactic diffuse background component largely dominates over any other sources, we therefore expect the uncertainties\footnote{The uncertainties are mainly due to contributions of unresolved PSs and imperfections of the Galactic diffuse background model.} associated with the Galactic diffuse background model to constitute the largest systematic effects for the  analyses in this study. 

In order to estimate the uncertainties of the Galactic diffuse background at the GC, we would like to examine a region of the sky which has  a similar Galactic diffuse background as the GC but does not contain any other sources which then would also contribute to the residuals.
As argued in \cite{ackermannajelloatwood2012}, the Galactic diffuse background has a relatively similar uncertainties within the inner Galaxy ($-80^\circ \geq l\leq-80^\circ$, $-8^\circ\leq b\geq 8$). Based on these considerations, we estimate the percentage uncertainties from nearby regions, along the Galactic plane, which do not have any point sources.

\begin{figure}[!t]
\centering
\includegraphics[width=1.0\linewidth]{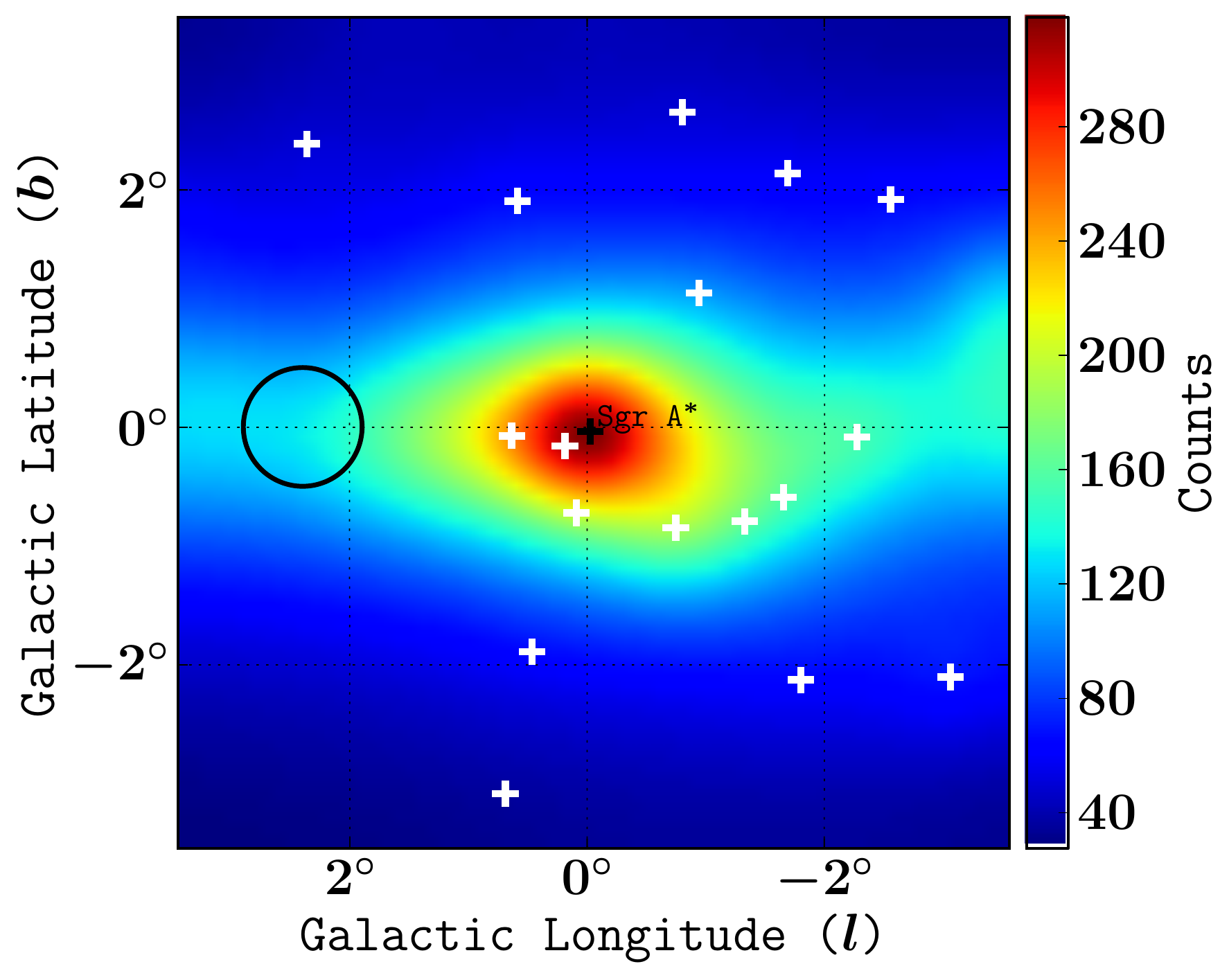} 
\includegraphics[width=1.0\linewidth]{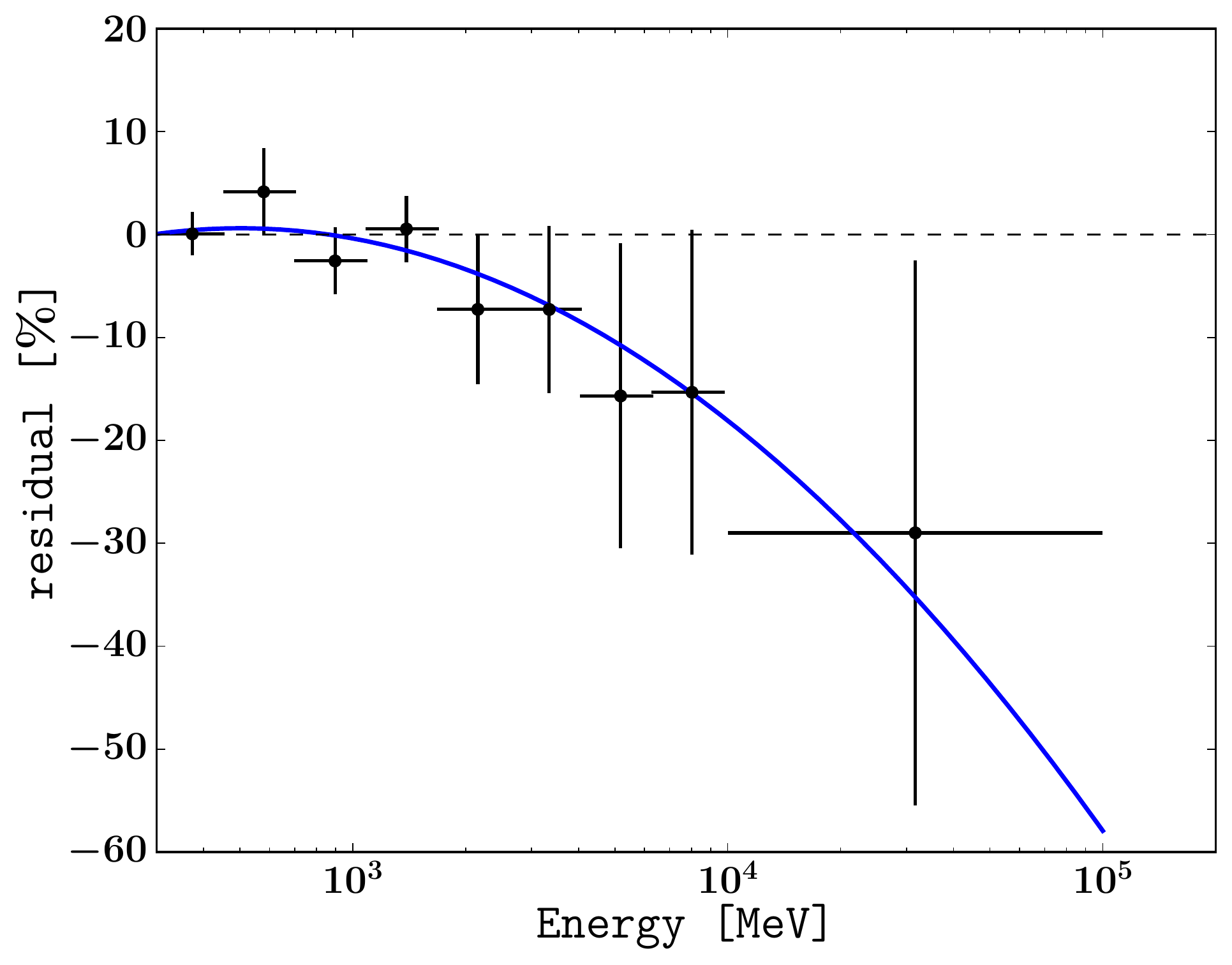} 
 \caption{ \label{fig:Fractionalresidual}(a) Counts map in the 0.3$-$100 GeV energy band of the best fit model for the ROI. This model considered the conventional 2FGL sources plus an additional extended source at the central position (see details in Sec.~\eqref{subsec:DMmaps}). Gaussian smoothing is applied with a kernel size of  $\sigma=0.3^{\circ}$. The black circle superposed on the image shows a region dominated by the Galactic diffuse background that was used to examine the spectral uncertainties. (b) Fractional residuals, that is (observed-model)/model, evaluated at eight energy bins in a circle centered at $(l,b)=(+2.3^{\circ}, 0^{\circ})$ with radius of $0.5^{\circ}$ shown in the above image. The residual data was fitted with a quadratic function in logarithmic scale as described by the blue line.   }   
\end{figure}

We first examined the spectral uncertainties by obtaining the energy dependence of our model residuals. Following a similar approach to that explained in Ref.~\citep{supernovaw49b}, we compared the observed counts with the model counts in a nearby circular region with a radius of $0.5^{\circ}$ centred on $\Delta l\sim +2.3^{\circ}$ and $\Delta b \sim 0^{\circ}$ where the Galactic diffuse background component was found to be dominant, see Fig.~\eqref{fig:Fractionalresidual}-(a). 
The ``model counts" map was computed from our best fit model (\textit{i.e.} the baseline model plus a NFW distributed source with $\gamma=1.2$ and Log Parabola spectra). This step is summarized in Fig.~\eqref{fig:Fractionalresidual}-(b), where the residuals as function of energy are shown.   

\begin{figure*}[ht!]
\begin{center}

\begin{tabular}{cc}
\centering
\includegraphics[width=0.5\linewidth]{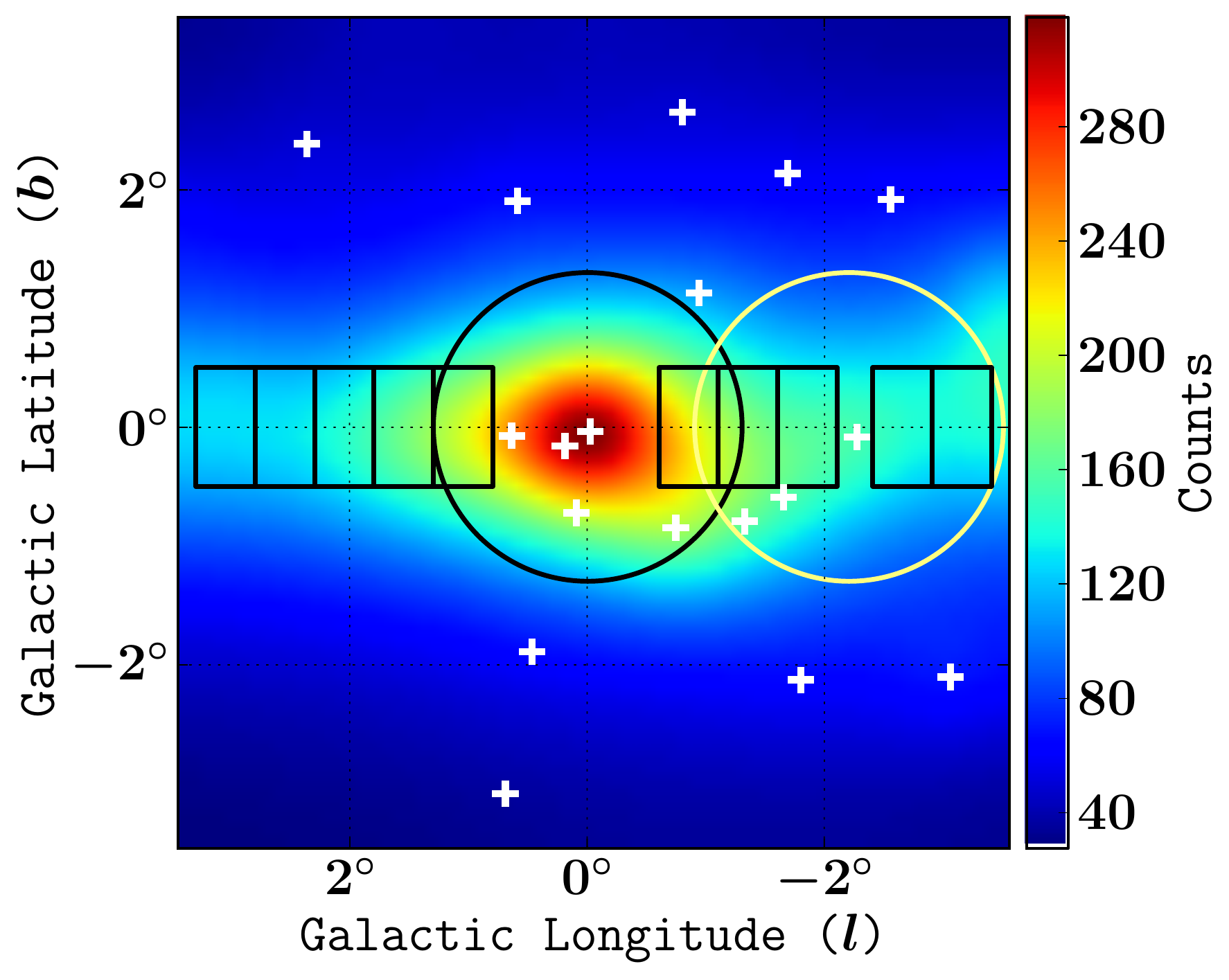} &
\centering
\includegraphics[width=0.55\linewidth]{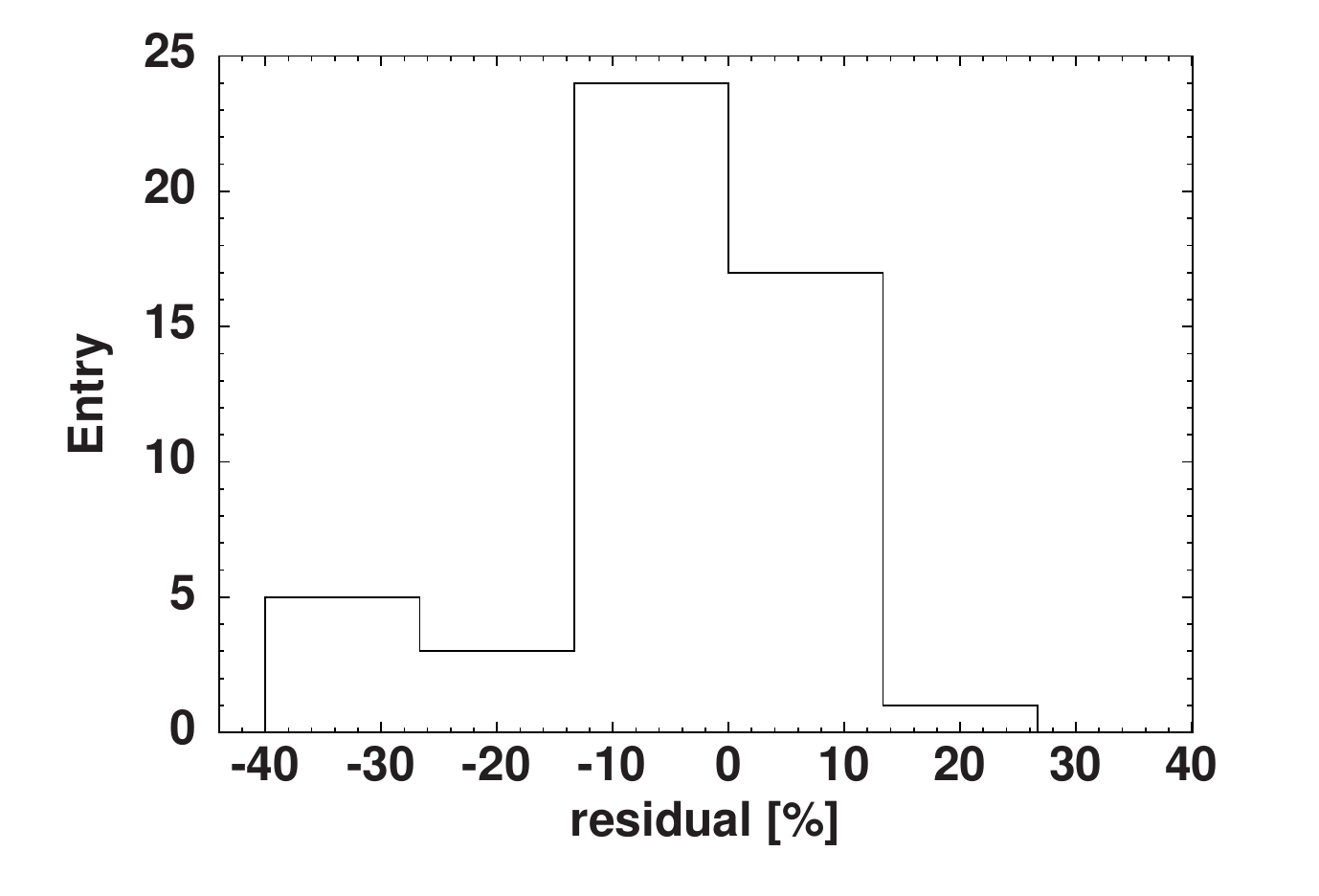} 
\end{tabular}

\end{center}
\caption{ \label{fig:Histogram}(a) Counts map in the 0.3$-$100 GeV energy band smoothed with a Gaussian filter of radius $\sigma=0.3^{\circ}$. The black rectangles ($1.0^{\circ}\times 0.5^{\circ}$) highlight the regions selected for the examination of the spatial uncertainties in the Galactic diffuse background. The black and yellow circles show the regions where the flux of the file \texttt{gal$_{-}$2yearp7v6$_{-}$v0.fits} was varied to evaluate the effects of the spatial dispersion of the model. (b) Histogram of the fractional residuals for ten rectangular regions in five energy bands: 0.30$-$0.50 GeV, 0.50$-$0.80 GeV, 0.80$-$1.30 GeV, 1.3$-$10 GeV and 10$-$100 GeV. The residuals were calculated as (observed-model)/model, where we also subtracted the best fit fluxes of all the sources (except for the Galactic diffuse background source) from the observed counts map.}
\end{figure*}

In order to assess the spatial uncertainties of the Galactic diffuse background component, we quantified the dispersion of the fractional residuals in 10 regions, where the Galactic diffuse background component was found to be dominant. The regions selected are located in the Galactic plane and special attention was put on not considering sectors with known 2FGL PSs within them. The fractional residual for each region was calculated in five energy bands: 0.30$-$0.50 GeV, 0.50$-$0.80 GeV, 0.80$-$1.30 GeV, 1.3$-$10 GeV and 10$-$100 GeV. The results obtained in this step are shown in Fig.~\eqref{fig:Histogram}. It follows that the standard deviation of the fractional residuals is 11$\%$. We thus used this value as an estimate of the uncertainties in the spatial distribution of the Galactic diffuse background component.
 A similar magnitude for the spatial and spectral uncertainties was found in \citep{supernovaw49b} which was also  in the inner Galaxy.

The spectral and spatial uncertainties described above will be used in Sec.~\eqref{subsec:Spectrum} to estimate the systematic error in flux of the extended source. 

\subsection{Spectral Morphology of the Extended Source}
\label{subsec:Spectrum}  

The procedure of obtaining the spectral energy distribution (SED) of the extended source was based on the method used for the flux band analysis
in Ref.~\citep{2FGL}. 
We started by applying the relaxation method (explained in Sec.~\eqref{sec:observations}) to the ROI in the full energy range of 0.3$-$100 GeV.  The extended source was modeled with a NFW($\gamma=1.2$) map normalized to unity and the spectra with a Log Parabola formula, as defined in Eq.~\eqref{eq:logparabola}. 
Once the best fit spectral parameters $\alpha(E_0)$ and $\beta$ have been found, we calculated the spectral slope of the Log Parabola at any given energy as 

\begin{equation}
\alpha(E)=\alpha(E_{0})+2\beta\log\left(\frac{E}{E_0}\right),
\label{eq:slope}
\end{equation}
where $E_0$ is the pivot energy~\citep{2FGL}.

We divided the energy range of the extended source into 12 energy bands evenly separated in the range 0.3$-$10 GeV and one energy band from 10 GeV to 100 GeV.
Next, the extended source photon fluxes in each band were computed  by freezing the spectral indexes of all the 2FGL sources to those obtained in the fit over the full range and fitting the normalizations in each spectral band. Note that the diffuse galactic and extragalactic backgrounds were not frozen and neither where the PS amplitudes. They were optimized along with each band amplitude. In an initial analysis we had also included a  200$-$300 MeV band but we found it had a $TS$ of only 0.4, so we did not include it in our further analysis. Also, the extended source models generally have a negligible amplitude in the 200$-$300 MeV band compared to Sgr A*.
 For each remaining energy band, the GC extended source spectrum was approximated by a power law function
\begin{equation}
\frac{dN}{dE}=N_0\left(\frac{E}{E_0}\right)^{-\Gamma},
\end{equation}
where the spectral index $\Gamma$ in a band was set to the local spectral slope defined in Eq.~\eqref{eq:slope}, at the logarithmic mid-point of the band $\sqrt{E_nE_{n+1}}$, restricted to be in the interval $[0,5]$. We calculated $2\sigma$ upper limits instead of actual fluxes for those bands with either a Test Statistics $TS<10$ or relative uncertainty on the flux $\Delta F_i/F_i>0.5$.  

Systematic errors due to uncertainties in the Galactic diffuse background model were evaluated by modifying the model file \texttt{gal$_{-}$2yearp7v6$_{-}$v0.fits} in the band analysis. This was done differently for spectral and spatial uncertainties:

\begin{itemize}

\item To calculate the spectral uncertainties we performed an additional band analysis where we altered the energy distribution of the Galactic diffuse background model according to the curve in Fig.~\eqref{fig:Fractionalresidual}. We thus compared the fit with and without this modification and set the spectral systematic error to be the difference between the two.

\item Spatial uncertainties were estimated using two modified \texttt{gal$_{-}$2yearp7v6$_{-}$v0.fits} files in the fit. For all energy bins in the model cubes, we varied the fluxes by $11\%$ in first, a disk of radius $1.3^{\circ}$ centred on Sgr A* and then an offset disk at $(b,l)=(0^{\circ},2.1^{\circ})$ with the same dimensions. Again, after a comparison of both fits we chose the one with the largest uncertainties to included in our SED calculation. Both disks are illustrated on Fig.~\eqref{fig:Histogram}-(a).
\end{itemize}

\begin{figure*}[ht!]
\begin{center}

\begin{tabular}{cc}
\centering
\includegraphics[width=0.52\linewidth]{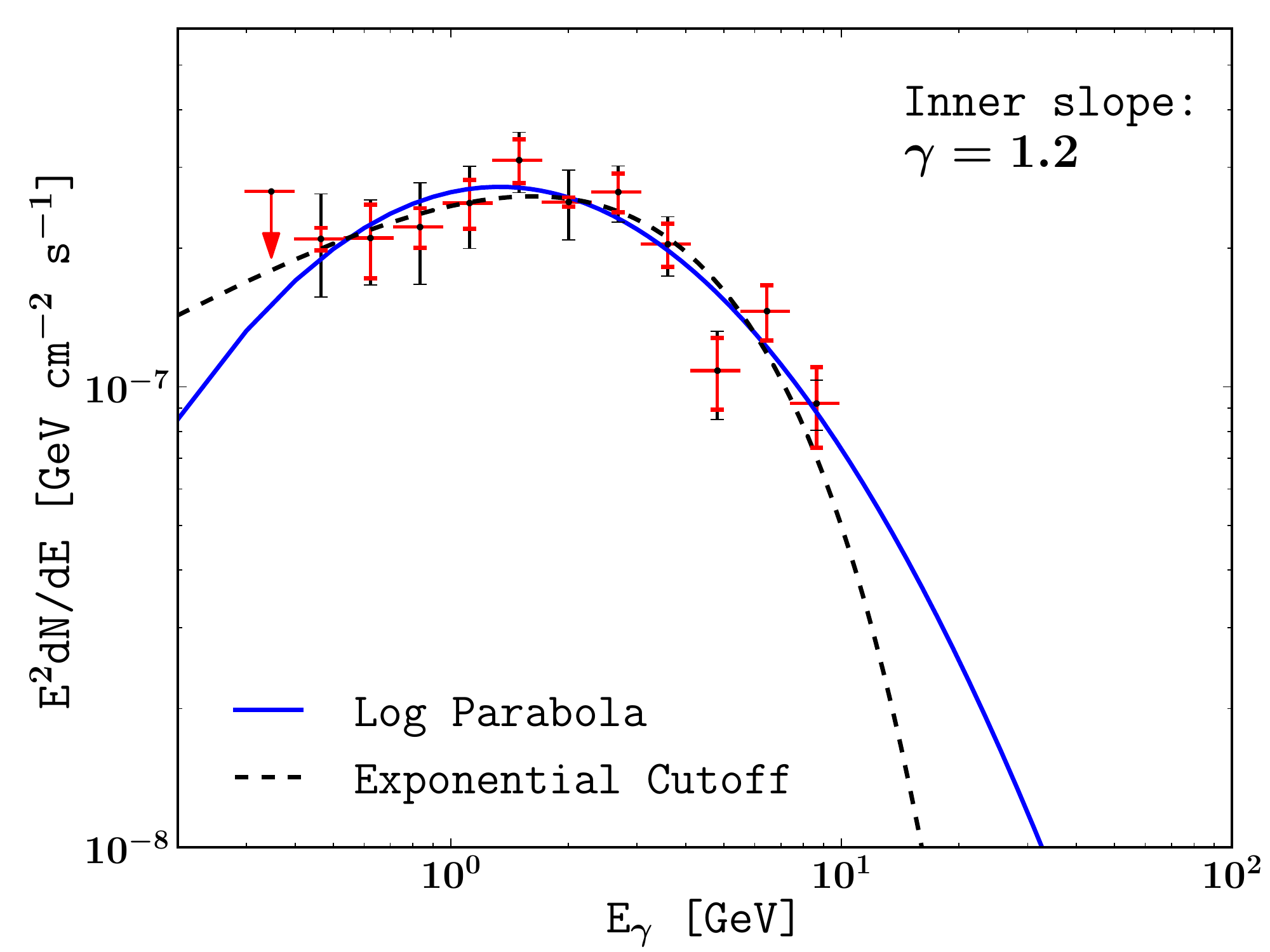} &
\centering
\includegraphics[width=0.52\linewidth]{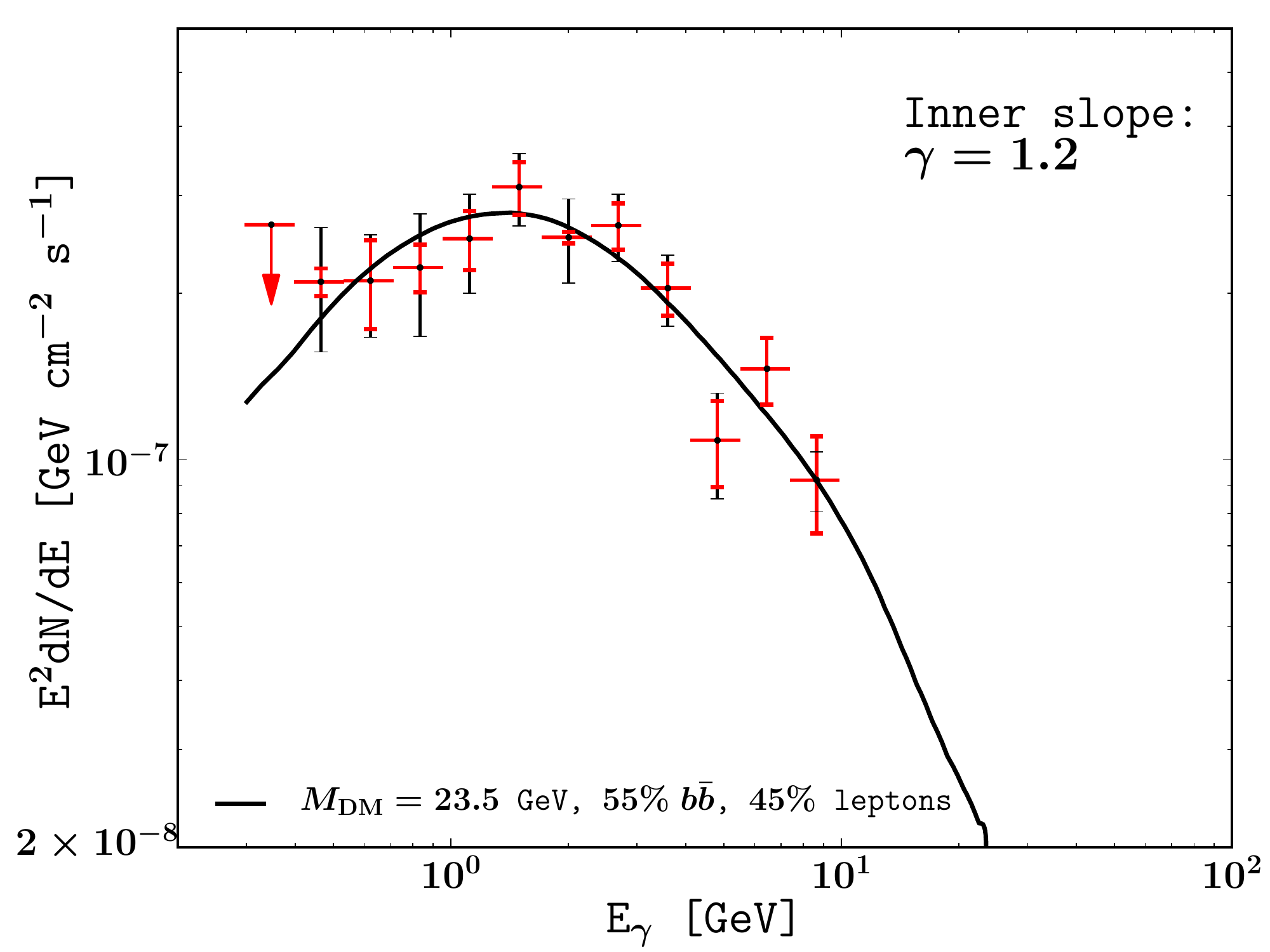} 
\end{tabular}
\includegraphics[width=0.52\linewidth]{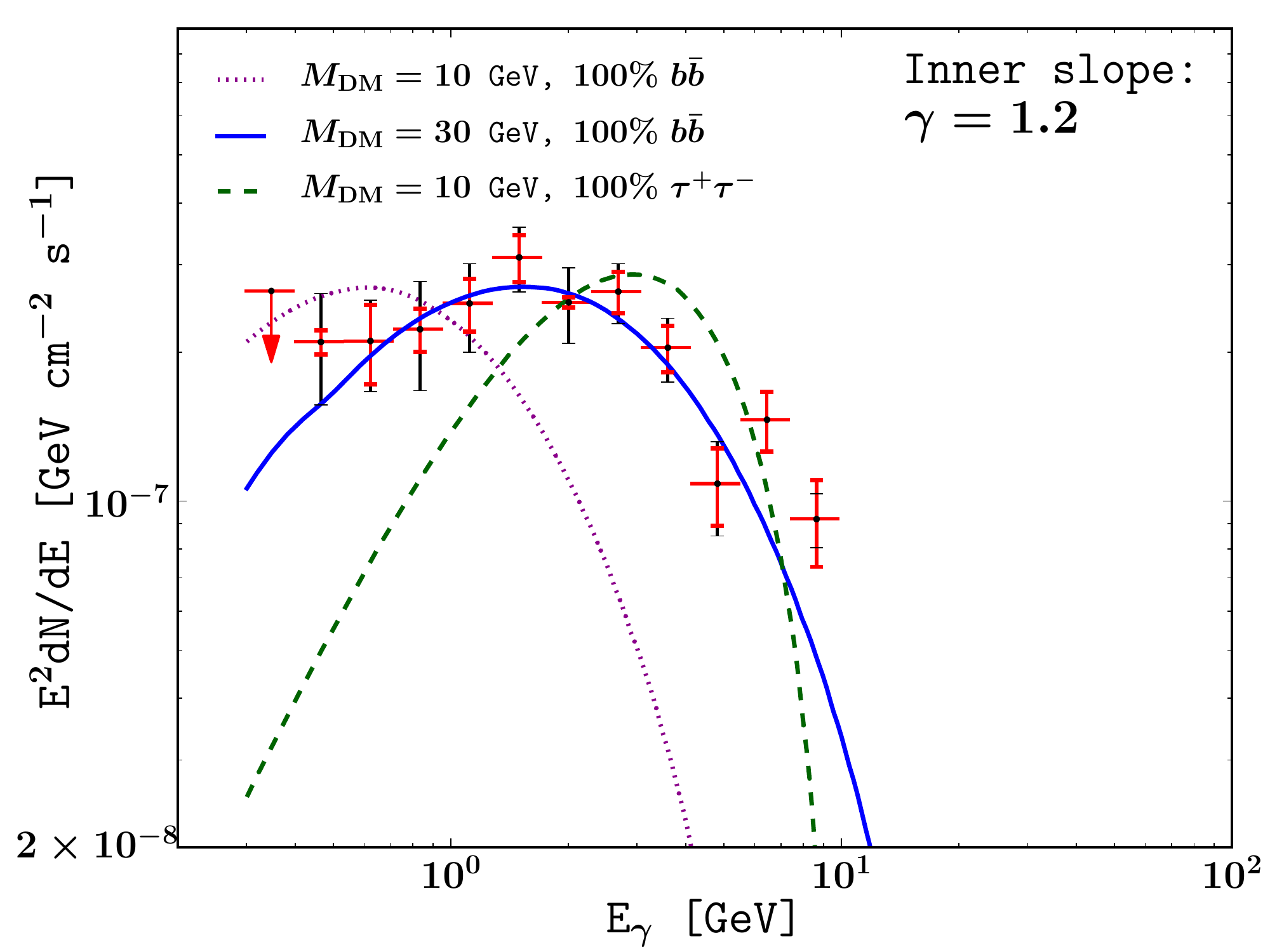} 

\end{center}
\caption{ \label{fig:SED} Spectrum of the extended source measured with the  Fermi-LAT. As shown in the legends, the model for the spatial distribution of the source is a NFW profile with inner slope $\gamma=1.2$. The red and black error bars show the ($1\sigma$) systematic and statistical errors, respectively. The upper limit is $2\sigma$. The fit over the full range is overlaid over the twelve band energy fluxes on each figure as follows: (a) The continuous blue line and dashed black line represent the best fit spectrum for a population of MSPs resembling a NFW spatial distribution, two typical curved spectra of these sources have been used. See text for details on goodness of the fit. (b) Shown is the best fit DM spectrum. $M_{\rm DM}$, $B_f$ and $\langle \sigma v \rangle$ were treated as free parameters in the fit. The black continuous line represents WIMP particles of 23.5 GeV self-annihilating $55\%$ and $45\%$ of the times into quarks $b\bar{b}$ and leptons (here ``leptons" denotes an unweighted mixture of  $e^{+}e^{-}$, $\mu^{+}\mu^{-}$ and $\tau^+\tau^-$), respectively. (c) The figure shows 3 different examples of DM spectra with high $TS$ values as obtained with Fermi Tools, where just $\langle \sigma v \rangle$ was allowed to vary in the fit. Although WIMPs of 10 GeV annihilating all the times into $\tau^+\tau^-$ or $b\bar{b}$ only satisfy the $TS>25$ criteria, they in fact do not pass the goodness of fit threshold, see details in Sec.~\eqref{subsec:darkmatter}. As it can be seen, $M_{\rm DM}=30$ GeV, $100\%$ $b\bar{b}$ exemplifies a good fitting model with significant curved spectra. }
\end{figure*}      

The resulting systematic errors due to uncertainties of the spectral distribution in the Galactic diffuse background model were found to on average be about $2\%$, while for the spatial errors we obtained 
on average about 20\%, both  for the energy ranges 
$\leq$ 10 GeV. For the 10 to 100 GeV band we found the systematic error to be of order 40\%. 
Also, we find that in general the models that fit the $\leq$ 10 GeV range have negligible values in the higher than 10 GeV band. For these reasons we do not use the 10 to 100 GeV energy band.

 Total systematic errors were computed by adding in quadrature the spatial, spectral and effective area systematics which is explained below \eq{eq:Csyst}. In Figure~\eqref{fig:SED} we show the SED of the extended source with the best fit over the full range overlaid. The red error bars indicate the total systematic errors and black error bars the statistical uncertainties.  We also list the SED and errors in Table V of  Appendix~\ref{sec:data} so that the reader may try fit other spectral models.

In order to study the validity of the distinct types of spectral shapes found with high $TS$ values in our Fermi Tools runs,  we used the same spectral fit quality estimator introduced in Ref.~\citep{2FGL} except that we also added our systematic
errors for the diffuse Galactic background

\begin{equation}
\begin{split}
\C &\\
 =&\sum_{i} {(F_i-F_i^{\rm fit})^2 \over \sigma_{i\;\rm stat}^2+\sigma_{i\;\rm spatial}^2+\sigma_{i\;\rm spectral}^2+\sigma_{i\;\rm area}^2}
\end{split}
\label{eq:Csyst}
\end{equation}
where $i$ runs over all bands with $TS>10$,
 $F_i^{\rm fit}$ is the flux predicted in that band from the spectral fit to the full band and the denominator contains a sum of the squares of the statistical error, the Galactic diffuse background spatial systematic error, the Galactic diffuse background spectral systematic error, and the effective area systematic error.
Also, $\sigma_{i\;\rm area}^2 =(f_i^{\rm rel}F_i^{\rm fit})^2$
 where $f_i^{\rm rel}$ represents the systematic uncertainty in the effective area \cite{2FGL}. The $f_i$ were set to 0.05 for the first seven bands and 0.08 from band eight to twelve. The first energy band situated in the range 300 MeV$-$400 MeV was found to have a $TS<10$, therefore it was not included in our analysis.    We will assume that $\C$ has a $\chi^2$ distribution with the number of degrees of freedom equal to the number of bands (11) minus the number of parameters used to determine $F_i^{\rm fit}$.  Assuming that the systematic errors can be treated as independent and Gaussian distributed, this is a good approximation as we have a large number of counts for each band.

The goodness of fit can be evaluated from the p-value which is the probability of $\C$ taken on a value larger than the observed value. We can evaluate the p-value as $\int_{\C}^\infty p(x)\, dx$ where $p(x)$ is a $\chi^2$  distribution with degrees of freedom equal to 11 minus the number of parameters. In Ref.~\cite{2FGL} they take a good fit to be one with a  p-value greater than $10^{-3}$. For a 2 parameter fit with 11 bands this corresponds to $\C<27.9$. 
For the 3 parameter case this corresponds to $\C<26.1$.

In the first row of Fig~\eqref{fig:SED} we show examples of spectra with high $TS$ values and significant curved spectral shapes for two well motivated hypothesis; an unresolved population of MSPs in the GC and dark matter self-annihilating into a mixture of $b\bar{b}$ quarks and leptons. While Figure~\eqref{fig:SED}-(c) shows examples of DM spectra proposed in the literature as good-fitting models for the GC gamma ray excess. However, our analysis demonstrates that DM particles of $M_{\rm DM}=10$ GeV annihilating into $\tau^+\tau^-$ or $b\bar{b}$ only do not fit the LAT data correctly, since they have  $\C\gg 27.9$.

\section{Results}
\label{sec:Results} 
 
\subsection{Millisecond Pulsars}
\label{subsec:Pulsars}

\begin{figure*}[ht!]
\begin{center}

\begin{tabular}{cc}
\centering
\includegraphics[width=0.4\linewidth]{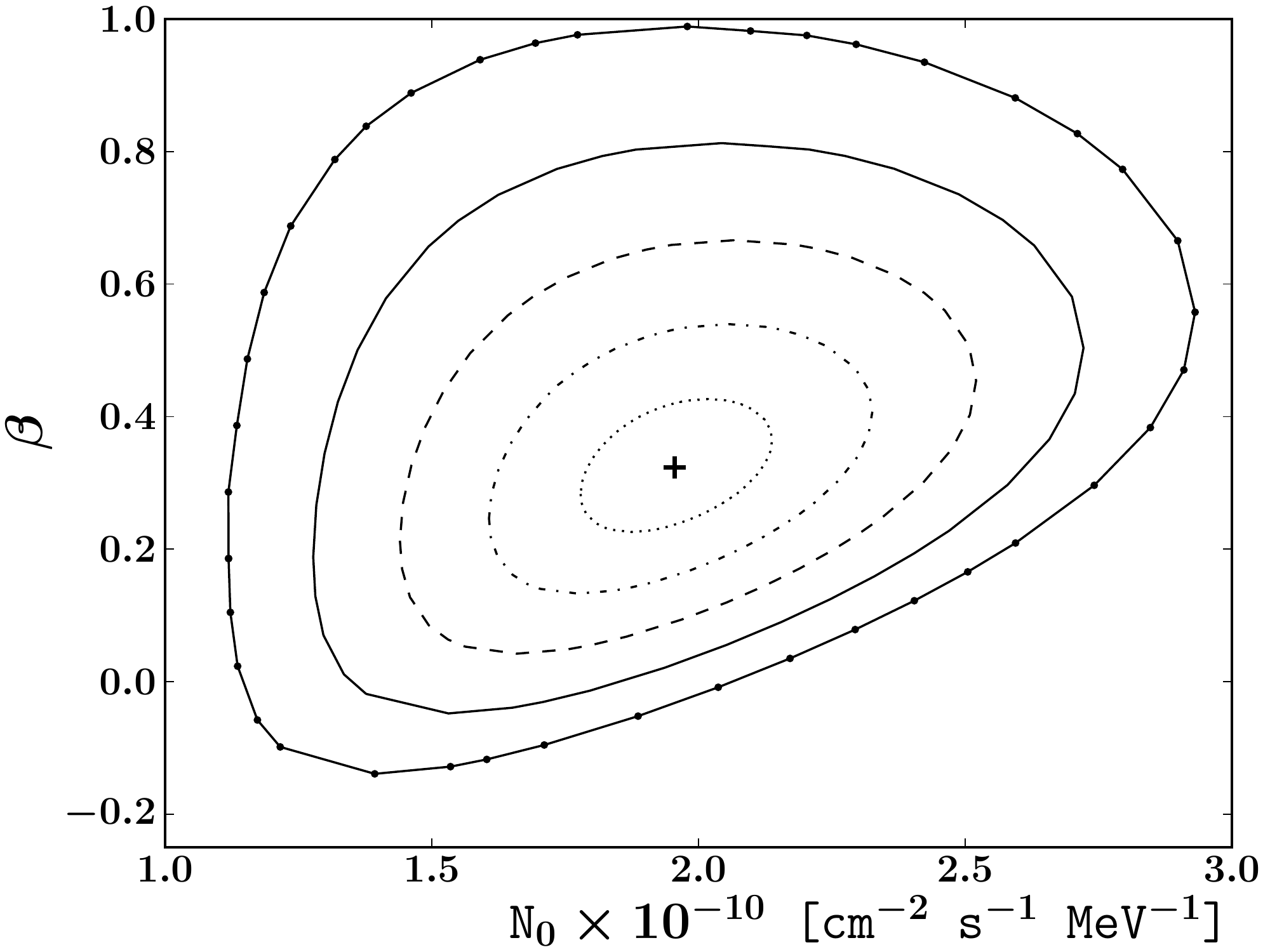} &
\centering
\includegraphics[width=0.4\linewidth]{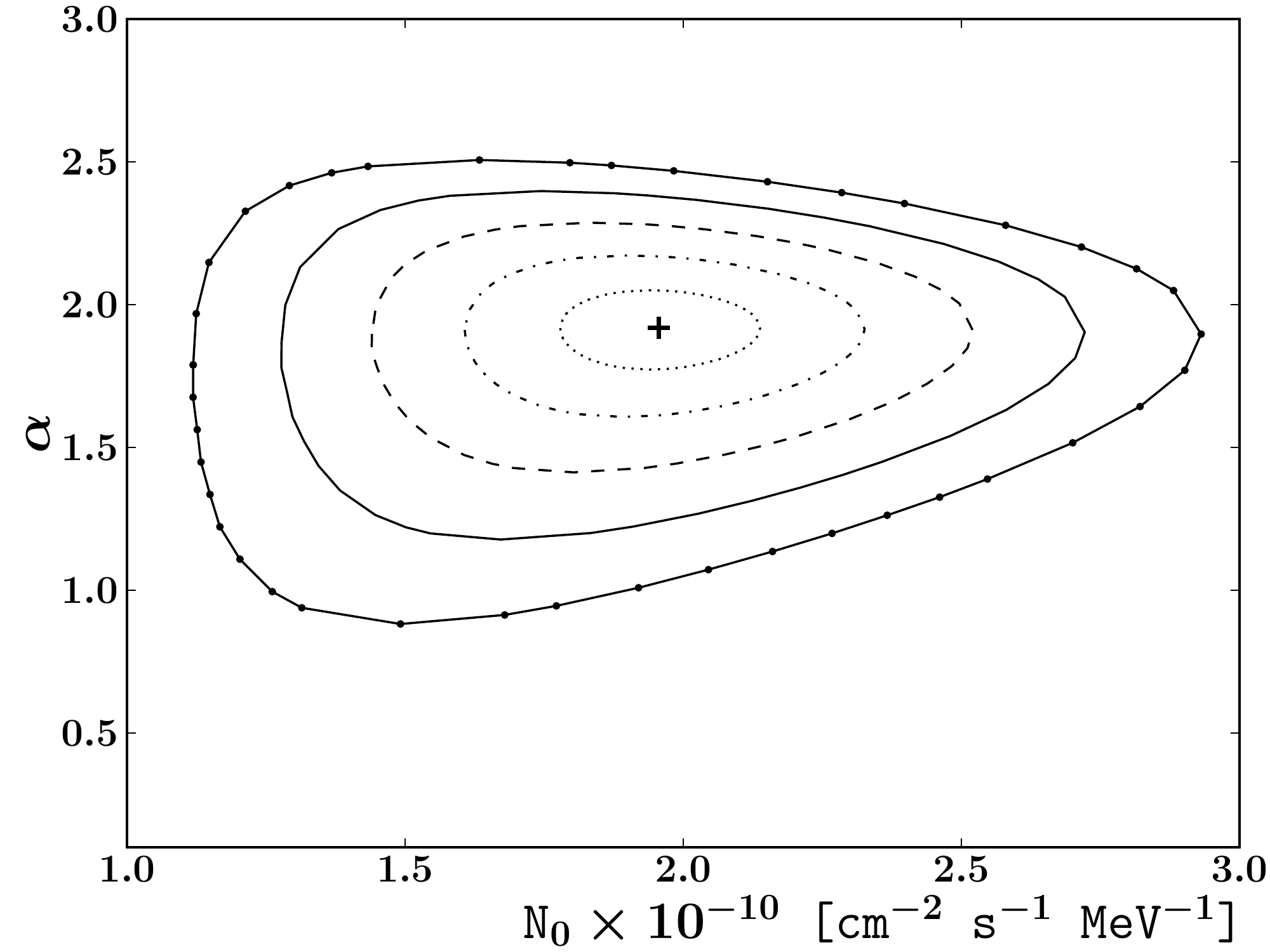} 
\end{tabular}

\begin{tabular}{cc}
\centering
\includegraphics[width=0.4\linewidth]{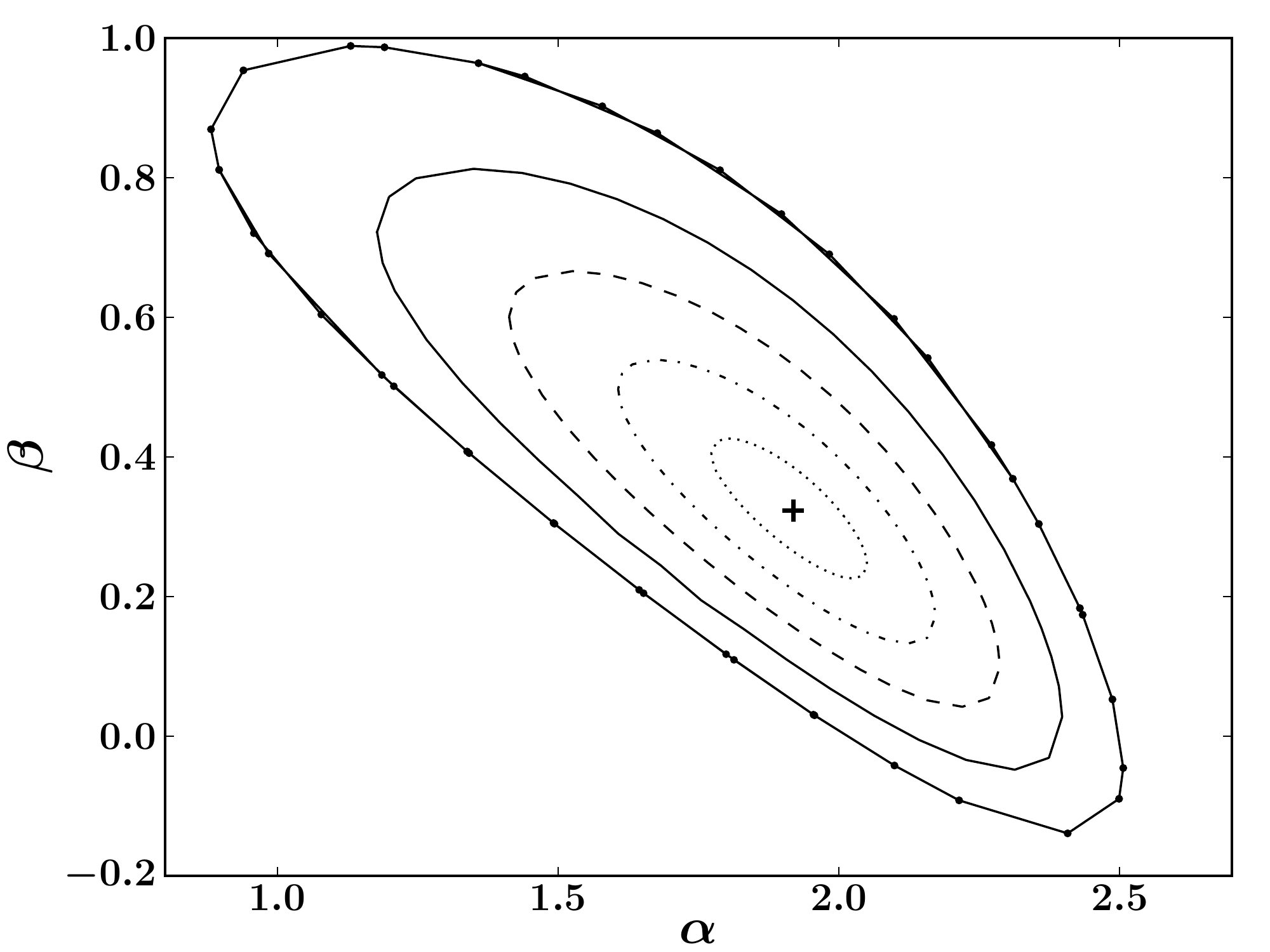} &
\centering
\includegraphics[width=0.4\linewidth]{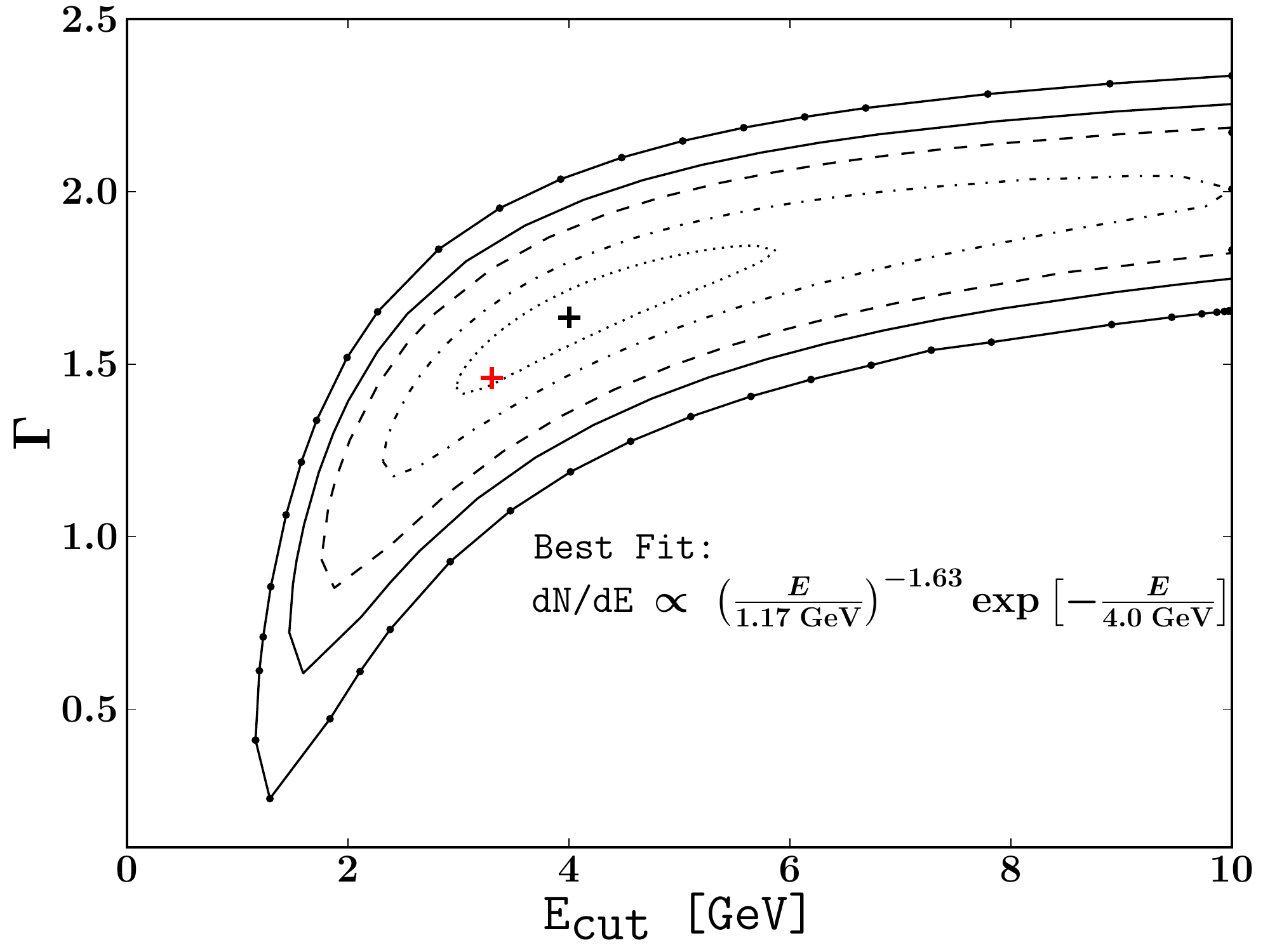} 

\end{tabular}

\end{center}
\caption{ \label{fig:Pulsars} Confidence regions ($1\sigma$, $2\sigma$,...$5\sigma$) for an unresolved population of Millisecond Pulsars using Fermi-LAT data taken from around the GC in the energy range 0.3$-$10 GeV. The spatial distribution of Pulsars follows a normalized NFW profile with inner slope $\gamma=1.2$. The two frames in the upper panel and the first one in the lower panel use a Log Parabola with $E_0=1176$ MeV for spectral shape, but, the second figure in lower row  uses an exponential cut-off as shown in the plot. Best fit parameters are denoted by black crosses. The red cross is the best fit obtained in Ref~\citep{HooperP} as the average best-fit of all the MSPs reported in the 2FGL catalogue.    }

\end{figure*}

\begin{figure*}[hp!]
\begin{center}

\begin{tabular}{cc}
\centering
\includegraphics[width=0.5\linewidth]{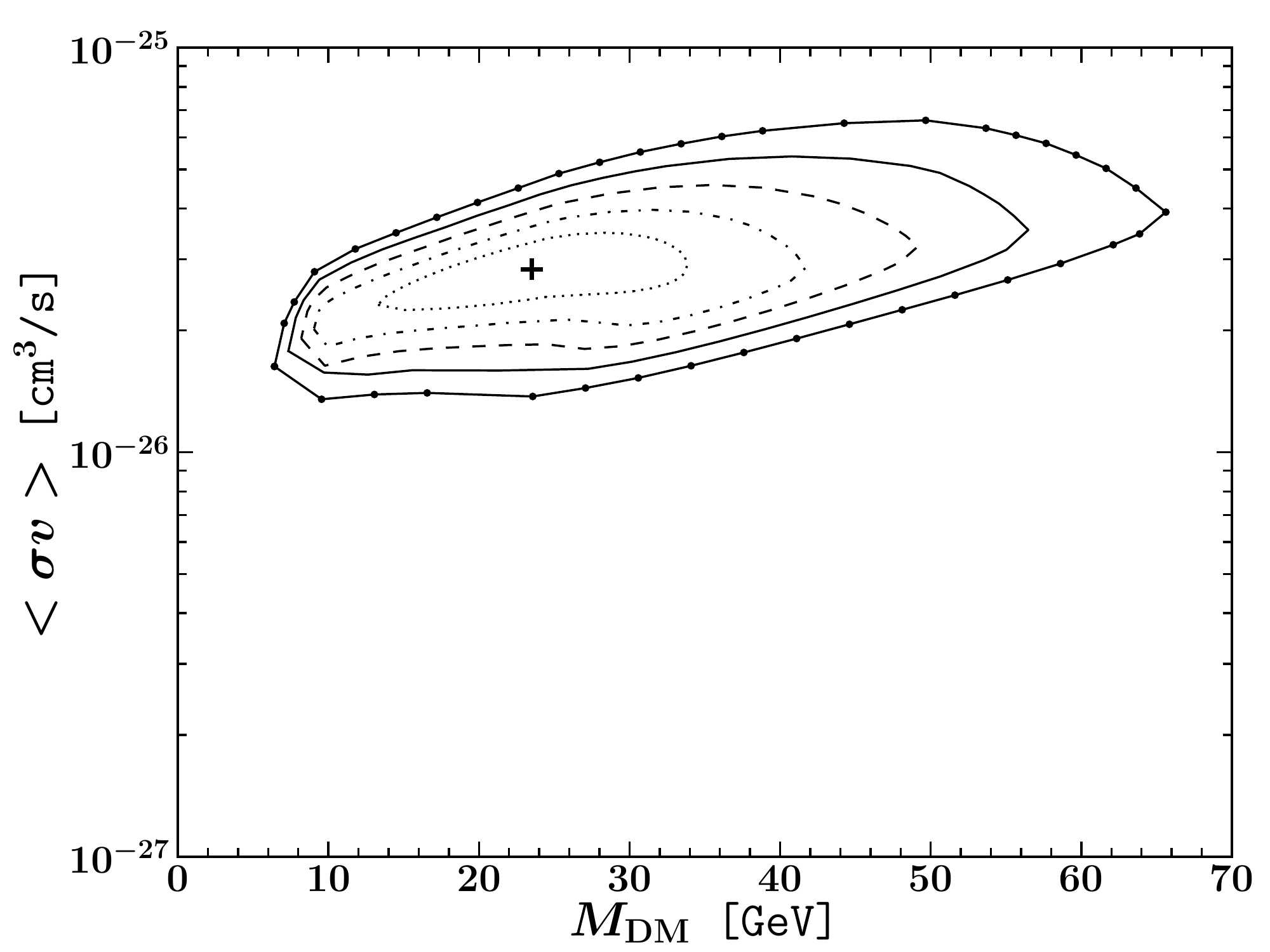} & \includegraphics[width=0.5\linewidth]{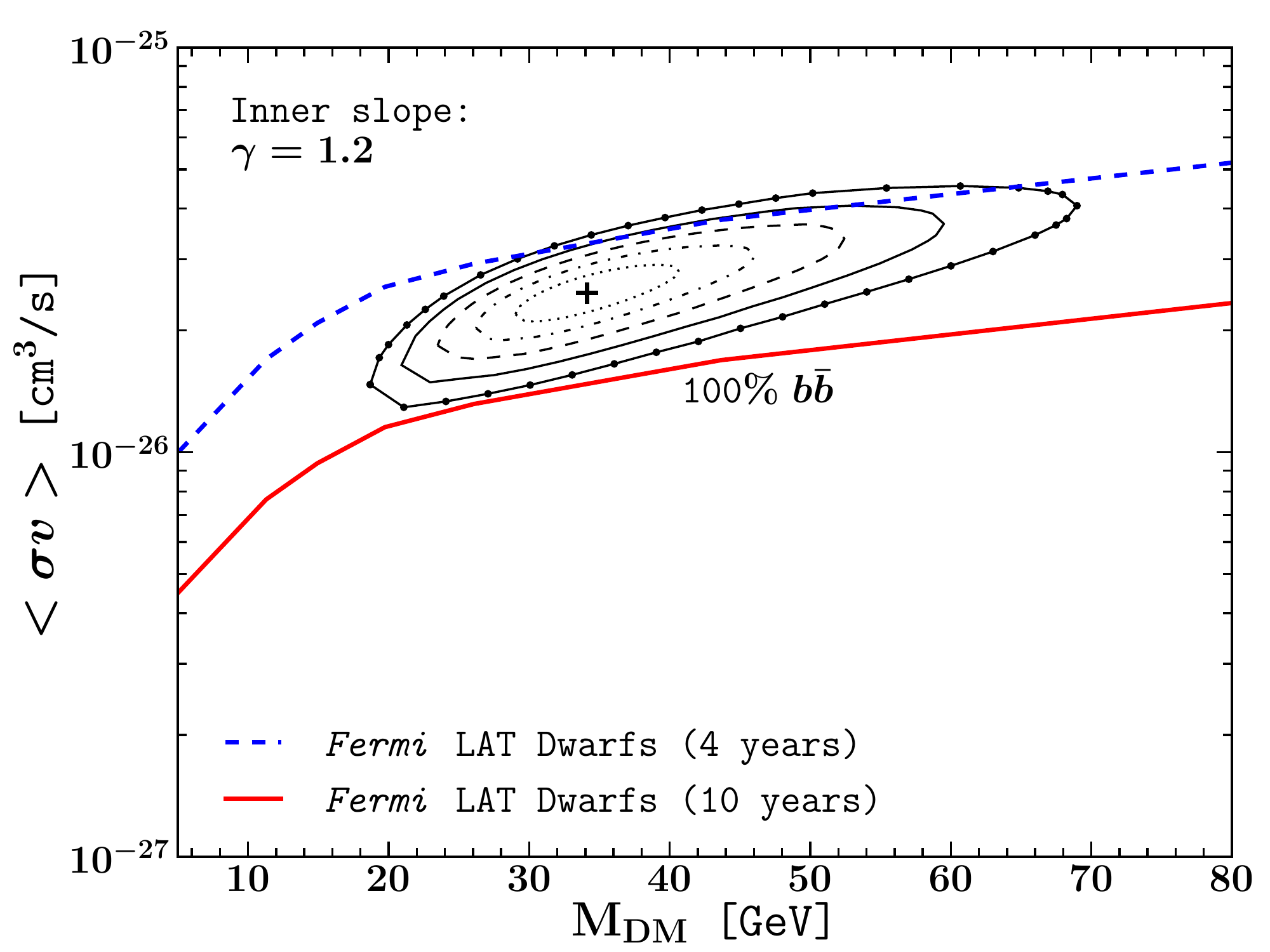}  

\end{tabular}

\begin{tabular}{cc}
\centering
\includegraphics[width=0.5\linewidth]{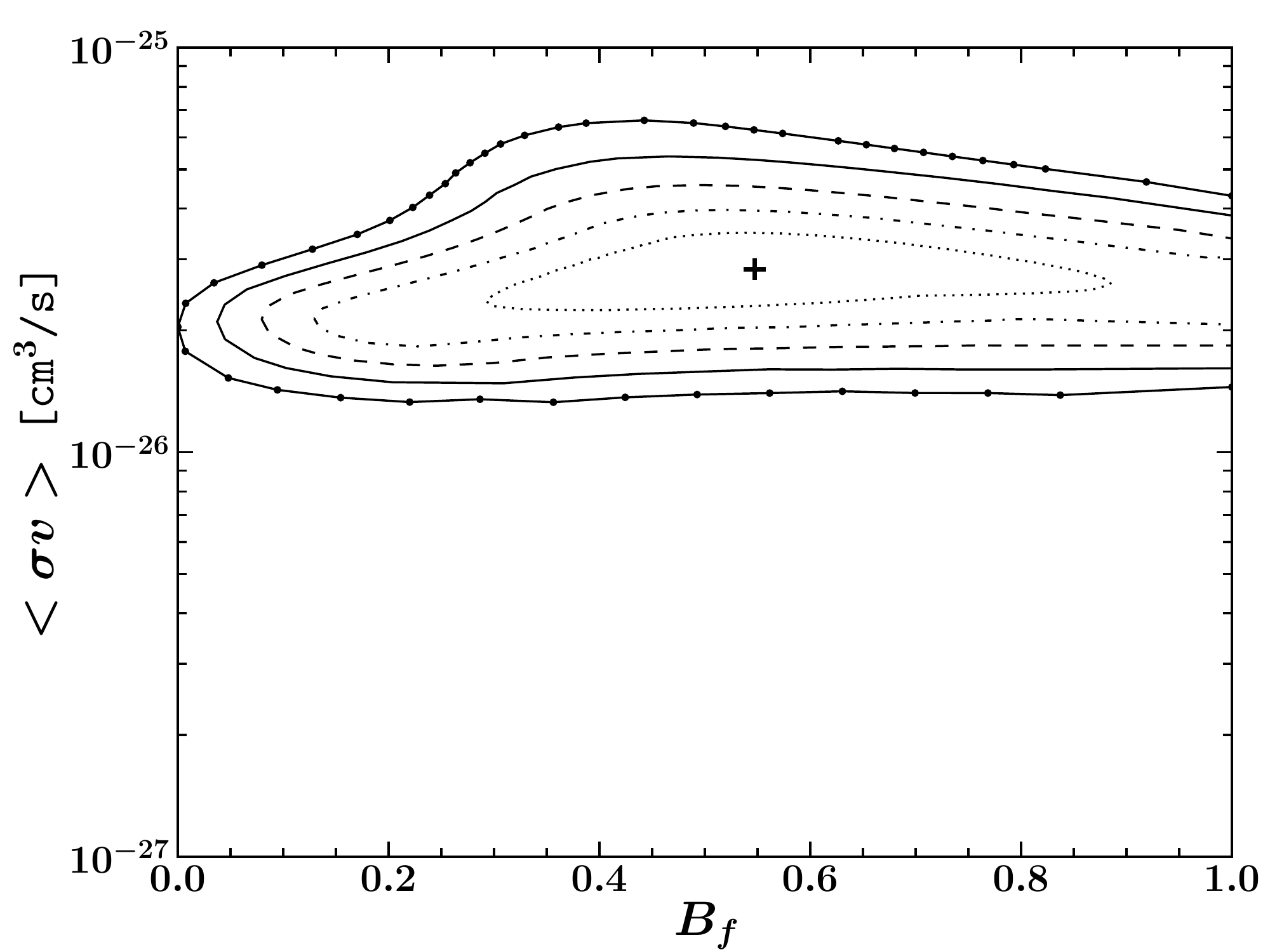} & 
\includegraphics[width=0.5\linewidth]{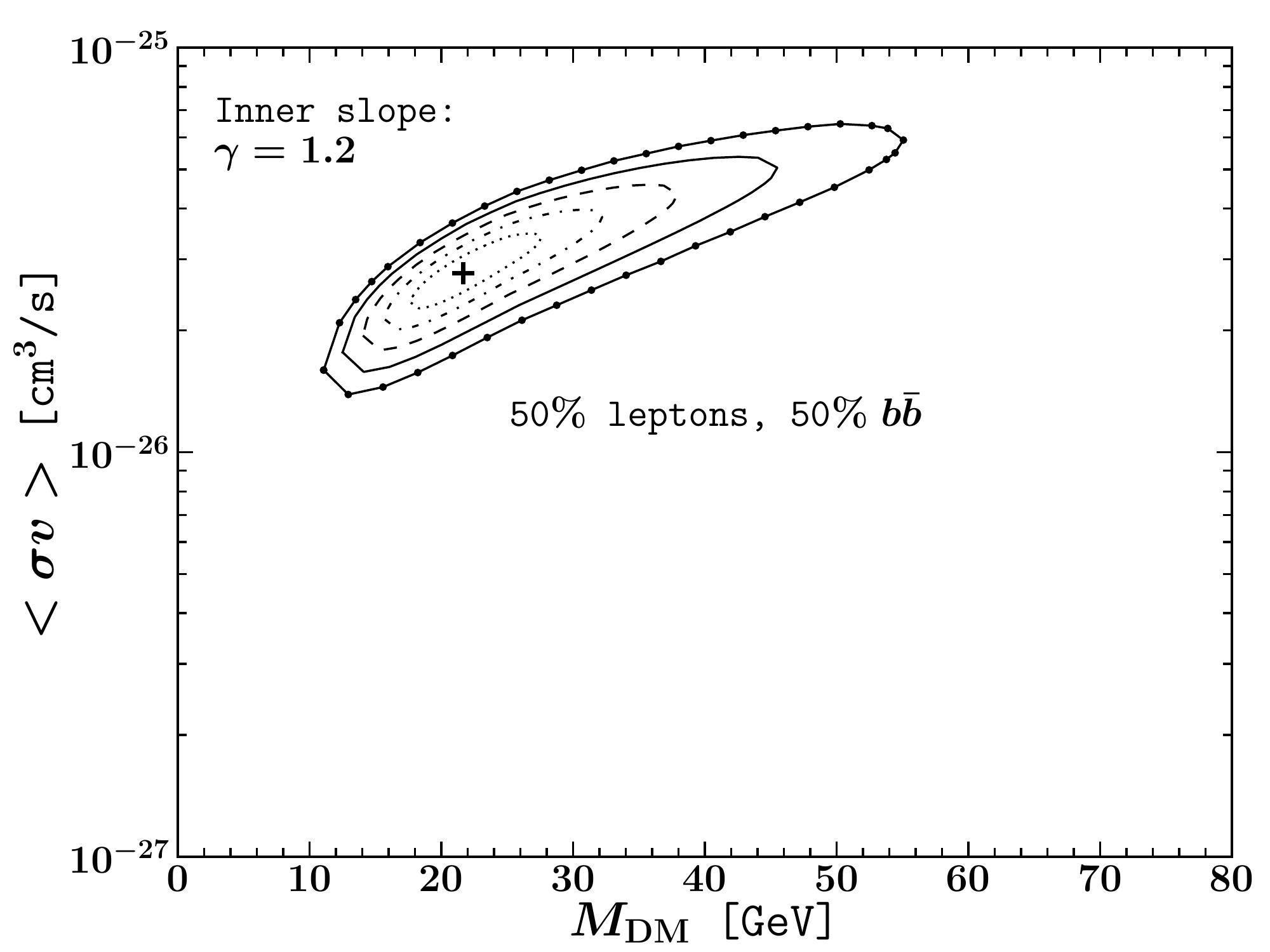} 
\end{tabular}

\begin{tabular}{cc}
\centering
\includegraphics[width=0.5\linewidth]{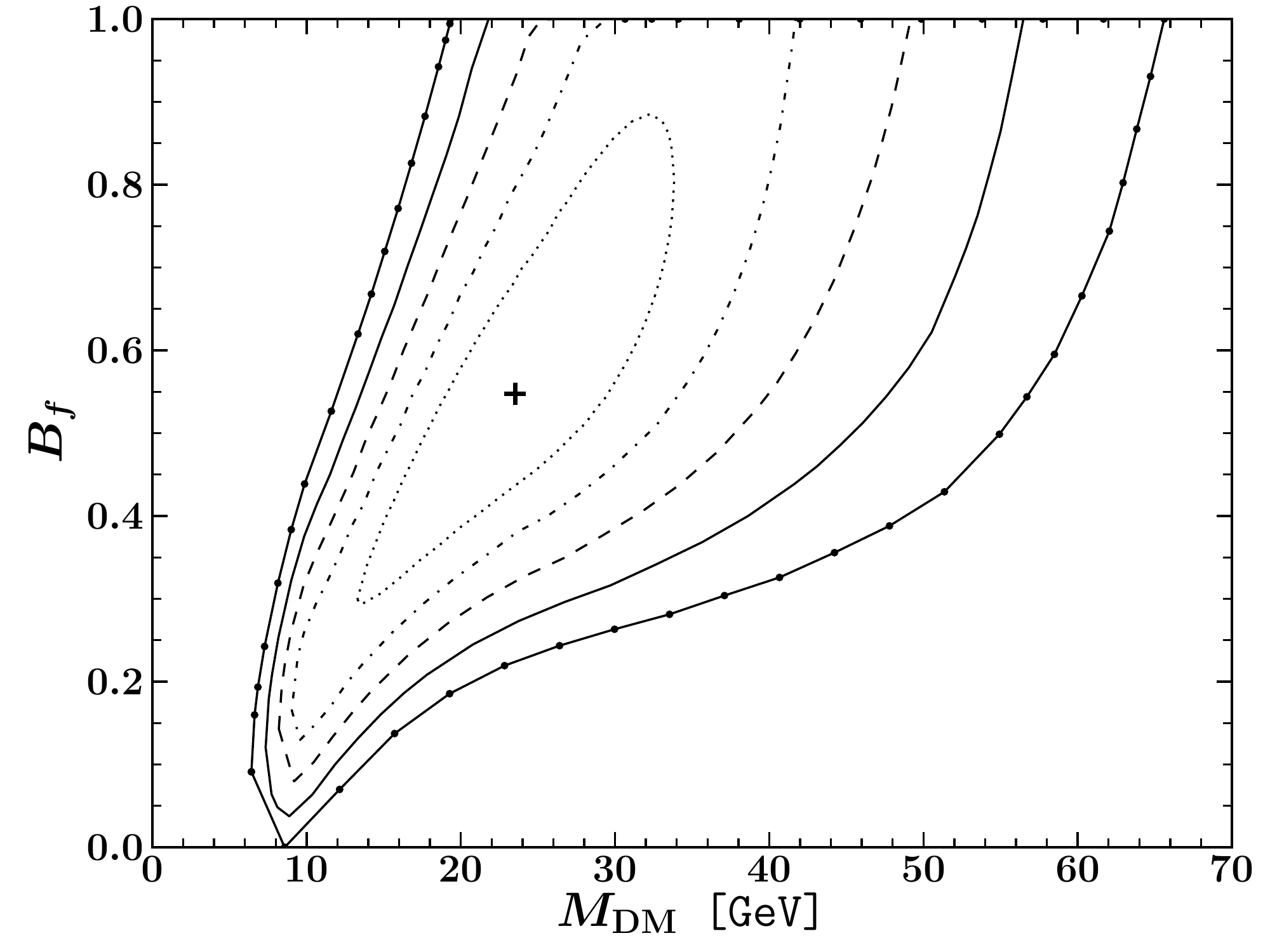} &
\includegraphics[width=0.5\linewidth]{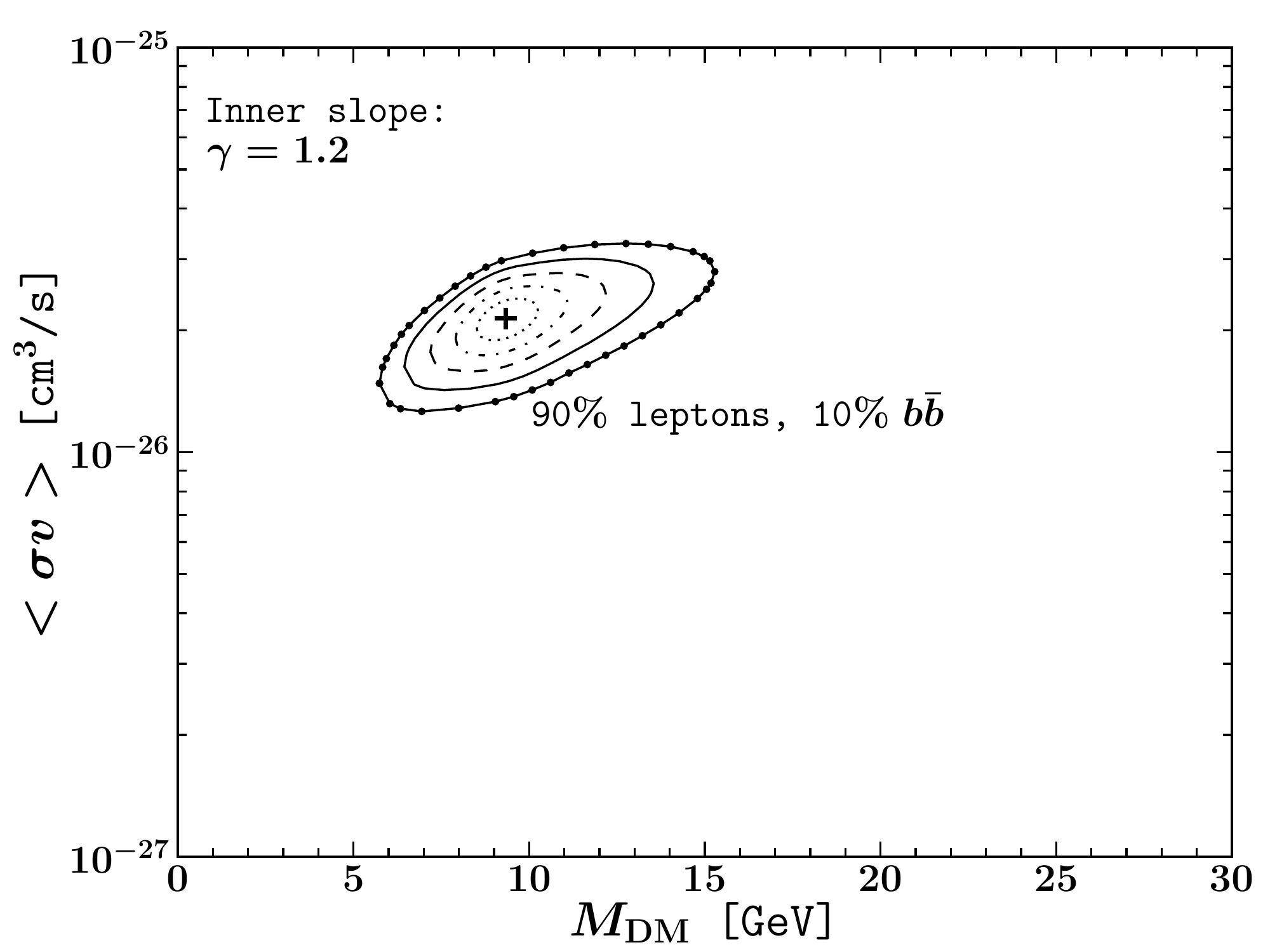} 
\centering
\end{tabular}

\caption{ \label{fig:darkmatterCL} Confidence regions ($1\sigma$, $2\sigma$,...$5\sigma$) for dark matter using Fermi-LAT data taken from around the GC in the range 0.3$-$10 GeV.  \texttt{Left Panel:} Best fit $M_{\rm DM}$, $\left\langle \sigma v\right\rangle$ and $B_f$ and errors, marginalized over the remaining parameter. Where $B_f=1.0$ implies 100\% $b\bar{b}$ and $B_f=0.0$ means 100\% leptons (\textit{i.e} an unweighted combination of $e^{+}e^{-}$, $\mu^{+}\mu^{-}$ and $\tau^+\tau^-$ pairs).   The Dark matter spatial distribution follows a NFW profile with inner slope $\gamma=1.2$.  \texttt{Right Panel:} Best fit $M_{\rm DM}$ and $\left\langle \sigma v\right\rangle$ for several fixed values of $B_f$ as indicated in the figures. The crosses in all frames denote the best-fit point. See Tables~\eqref{tab:bblept1} and~\eqref{tab:bblept2}.     }
\end{center}
\end{figure*}

It has been suggested~\citep{AK,AKerratum} that a population of $\sim10^3$ Millisecond Pulsars (MSPs) constitutes a reasonable explanation for the gamma ray excess seen in the GC. The main physical reasons that support this claim are: MSPs can emit gamma rays over large time scales, their binary companions could prevent them from free-streaming out the GC and estimates of the spatial distribution of M31 low mass X-ray binary population indicate that the number of MSPs located in the GC could scale as steeply as $1/r^{2.4}$ (with $r$ the two-dimensional projected radius).      

To compare the spectral shape of the gamma ray excess seen in the GC with that of typical LAT MSPs in the second year pulsar catalogue~\citep{2FGL}, we fit the LAT spectrum of the GC extended source by a power law with exponential cutoff:

\begin{equation}
\frac{dN}{dE}=K\left({E\over E_0}\right)^{-\Gamma}\exp\left(-\frac{E}{E_{\rm cut}}\right),
\label{eq:expcut}
\end{equation}    
where photon index $\Gamma$, a cut-off energy $E_{\rm cut}$ and a normalization factor $K$ are free parameters. 
The best fit parameters,
with $E_0=1176$ MeV,  were $K=2.5\times 10^{-10} \pm 4\times 10^{-11}$  ph cm$^{-2}$  s$^{-1}$ MeV$^{-1}$,  
$E_{\rm cut}= 4000 \pm 1500 $ MeV, and  $\Gamma=1.6 \pm 0.2$.
The confidence regions
 are shown in the lower right panel of Fig.~\eqref{fig:Pulsars}. It has been found in Ref.~\citep{HooperP} that the sum of the spectra of the 37 MSPs reported in the 2FGL catalogue are well described by Eq.~\eqref{eq:expcut} with $\Gamma=1.46$ and $E_{\rm cut}=3.3$ GeV (see the red cross in Fig.~\eqref{fig:Pulsars}-(d) ). Therefore the LAT spectrum of the extended source in the GC agrees within $1\sigma$ with what has been observed from the 37 resolved MSPs of the 2FGL.

The best fit and confidence intervals were performed with the tool \texttt{Minuit}~\citep{minuit}. 
\eq{eq:Csyst} was used as the goodness of fit statistic.
Note that  the $1\sigma$ contours, for our two dimensional plots,
corresponds to the 68.3\% profile likelihood \cite{rolkelopezconrad2005} confidence region and are defined by all areas of the two dimensional parameter space which
have a $\Delta \C\leq 2.3$ where  $\Delta \C$ is the difference between $\C$ at the best fit point in the plot and $\C$ at the point considered for inclusion within the confidence interval. All other parameters not shown in a plot are chosen to minimize $\C$ at each point in the plot. The corresponding $\Delta \C$ thresholds  for 2, 3, 4, and 5 $\sigma$ are
6.2, 11.8, 19.3, and 28.7 respectively (see for example the Statistics section of the ``The Review of Particle Physics'' \citep{particledatagroup}). For any one parameter confidence intervals, we quote the 68.3\% level which corresponds to a
   $\Delta \C=1$ threshold.

Frames shown in the upper panel and left lower panel of Fig.~\eqref{fig:Pulsars} describe the results of a spectral fit to the LAT data using a Log Parabola formula~\eqref{eq:logparabola} instead of an exponential cut-off. As it can be seen, the full parameter space is shown in three two-dimensional plots. The model parameter $E_0$ in Eq.~\eqref{eq:logparabola}, kept fixed during the fit and set to $E_0=1176$ MeV, was calculated as the energy at which the relative uncertainty on the differential flux $N_0$ was minimal. This was done  with a damping procedure that made use of the covariance matrix between parameters as obtained from the \texttt{MIGRAD} algorithm in \texttt{Minuit}~\citep{minuit}. The best-fit parameters shown with black crosses in the corresponding frames of Fig.~\eqref{fig:Pulsars} are $N_0=1.96^{+0.18}_{-0.17}\times 10^{-10}$ ph cm$^{-2}$ s$^{-1}$ sr$^{-1}$, $\alpha=1.92^{+0.13}_{-0.15}$ and $\beta=0.32^{+0.10}_{-0.09}$ (for completeness, the $\pm 1\sigma$ total errors are included as well).

In Ref.~\citep{AK,AKerratum} the fit to the gamma ray data was performed by considering statistical errors only and fixing $E_0=100$ MeV in the Log Parabola. However, we found that this choice of pivot energy produces a large correlation between the parameters $N_0$, $\alpha$ and $\beta$. We thus notice that this degeneracy can be alleviated by searching for a more adequate value of $E_0$, as outlined above.

\subsection{Self-Annihilating Dark Matter}
\label{subsec:darkmatter}

\begin{figure}[h!]
\begin{center}

\centering
\includegraphics[width=0.97\linewidth]{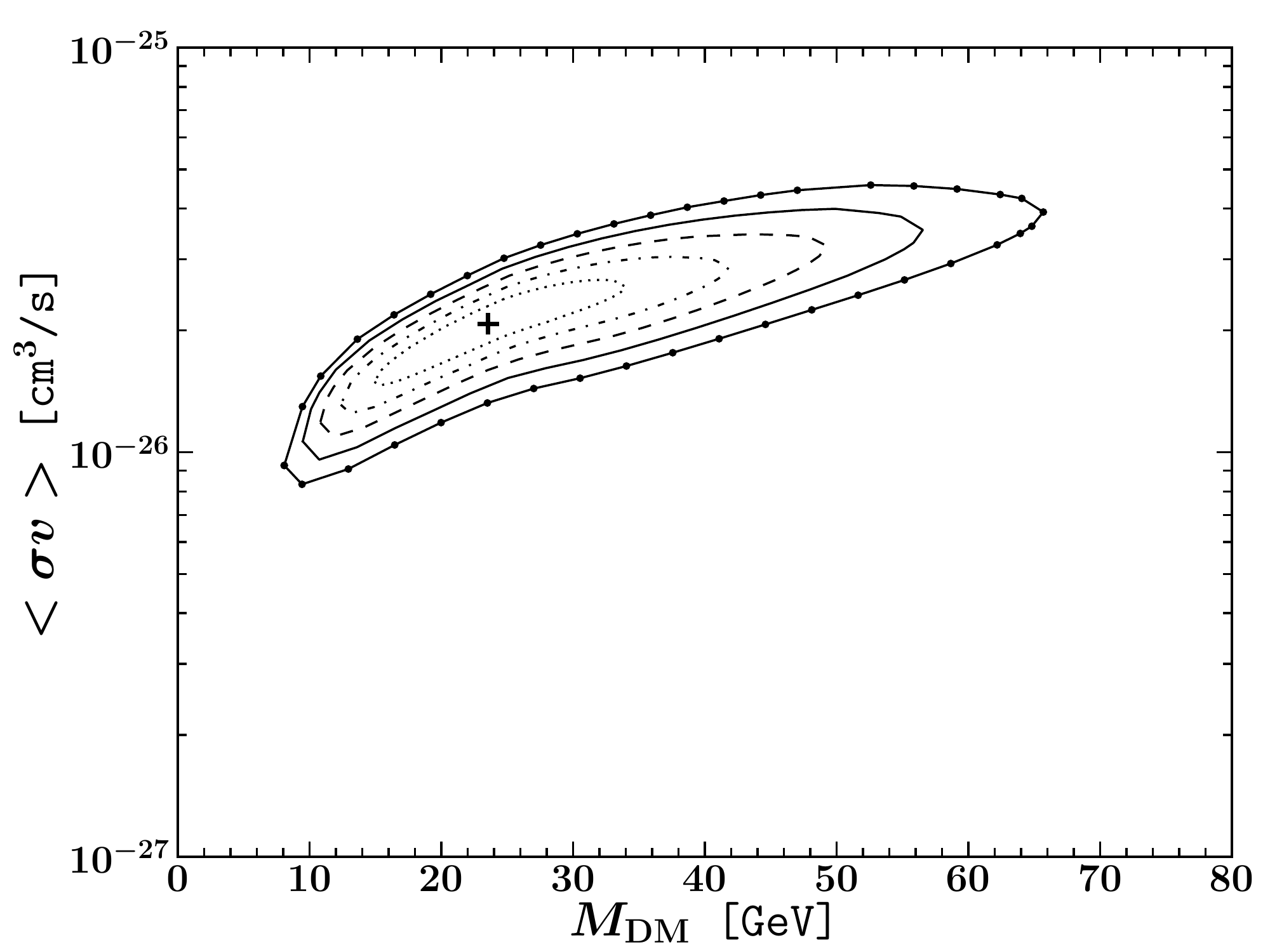}

\centering
\includegraphics[width=0.97\linewidth]{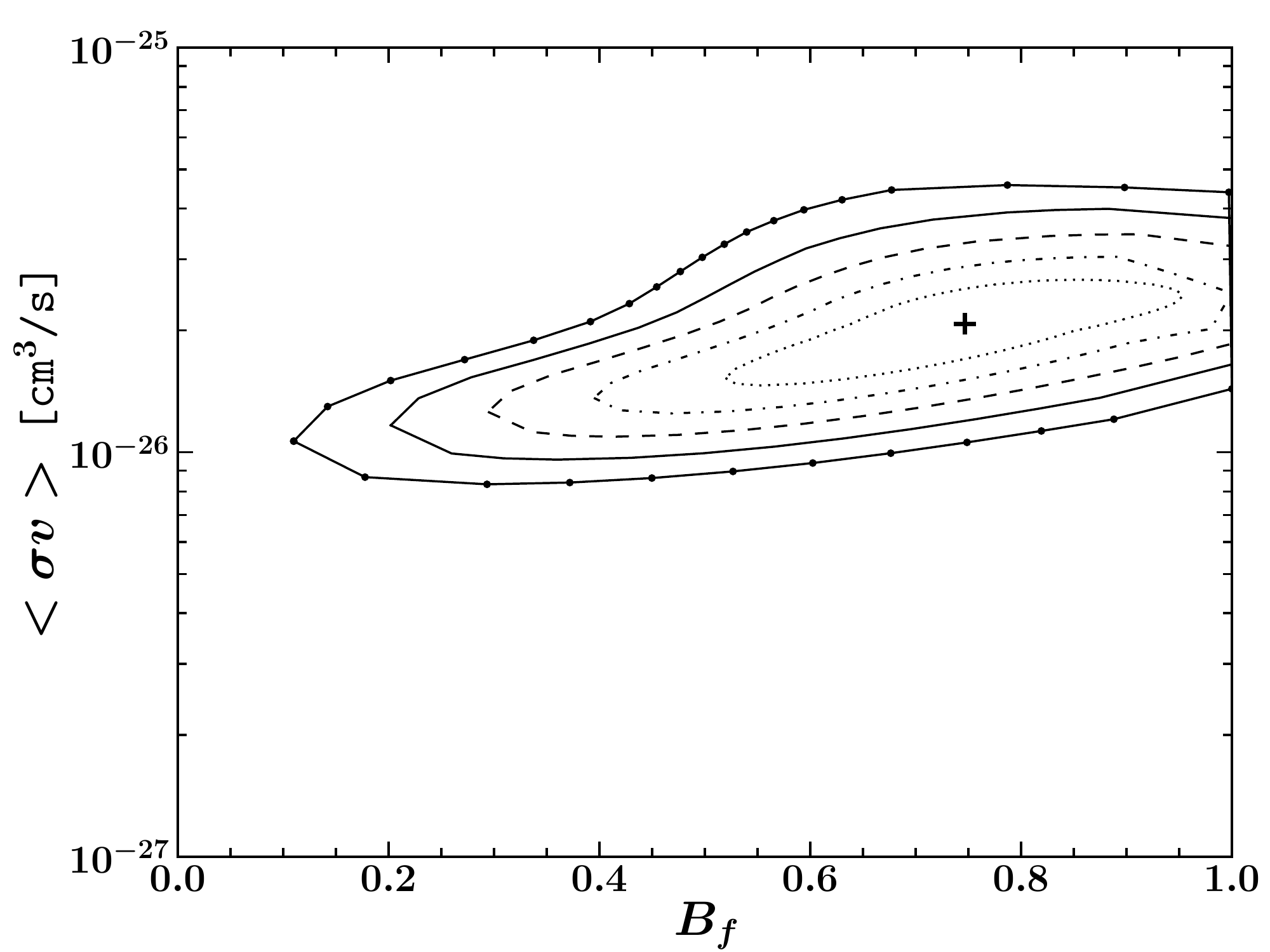} 

\centering
\includegraphics[width=0.97\linewidth]{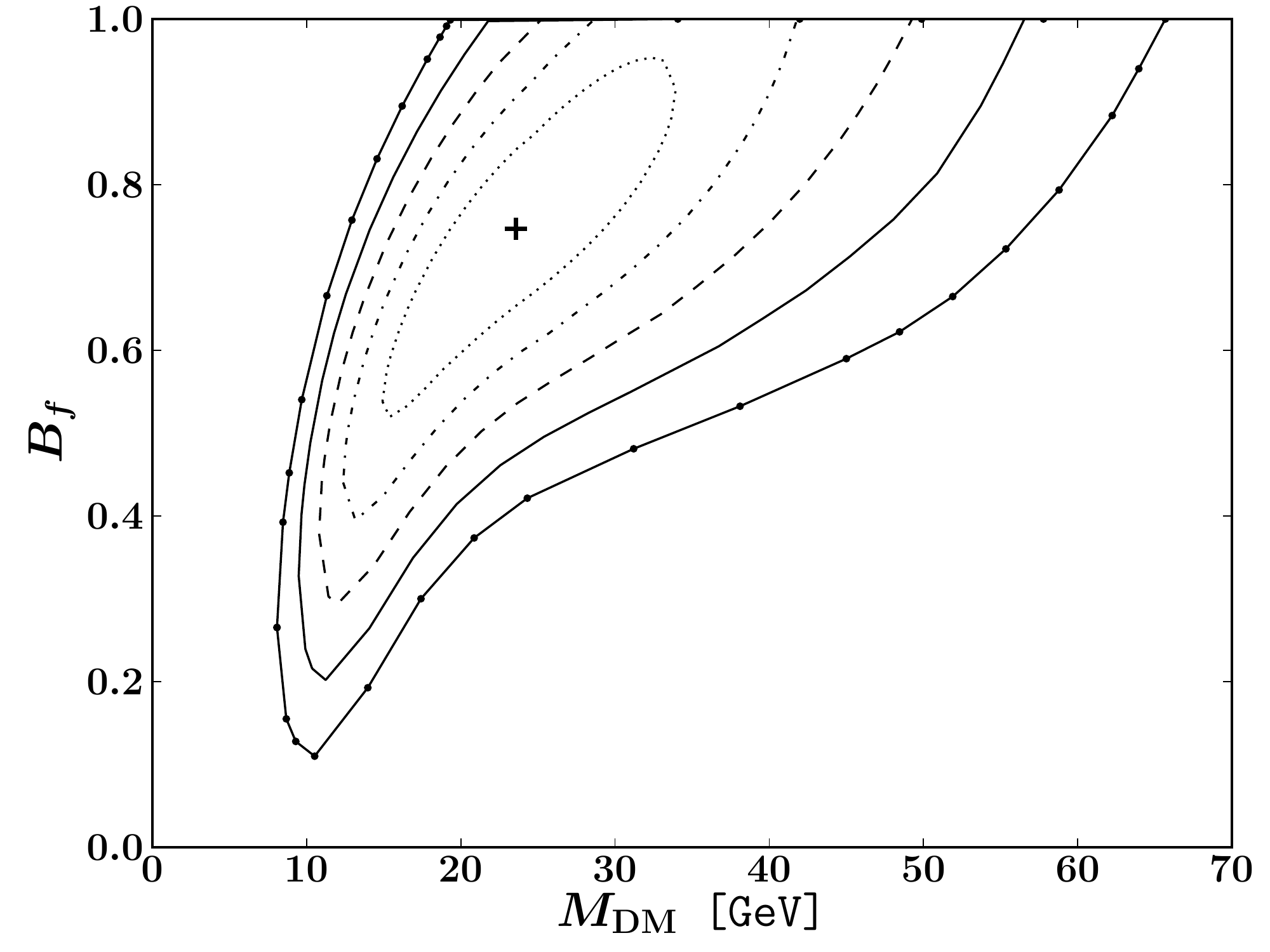} 
\centering

\caption{ \label{fig:darkmatterCLtautau} Confidence regions ($1\sigma$, $2\sigma$,...$5\sigma$) for dark matter using Fermi-LAT data taken from around the GC in the range 0.3$-$10 GeV. Best fit $M_{\rm DM}$, $\left\langle \sigma v\right\rangle$ and $B_f$ and errors, marginalized over the remaining parameter. Where $B_f=1.0$ implies 100\% $b\bar{b}$ and $B_f=0.0$ means 100\% $\tau^+\tau^-$. The Dark matter spatial distribution follows a NFW profile with inner slope $\gamma=1.2$. The crosses in all frames denote the best-fit point. See also Table~\eqref{tab:bbtt}.     }
\end{center}
\end{figure}    

\begin{table}[t!]

\begin{ruledtabular}
\begin{tabular}{ccc}
\centering
Best-fit Branching ratio & $\left\langle\sigma v\right\rangle$ [cm$^3$/s] & $M_{\rm DM}$ [GeV] \\  \hline \\ 
$55^{+18}_{-16}\%$ $b\bar{b}$    & $2.84^{+0.43}_{-0.41}\times 10^{-26}$ & $23.5^{+6.7}_{-6.6}$ \\

\end{tabular}
\end{ruledtabular}

\caption{\label{tab:bblept1} Best fit values on the DM velocity averaged annihilation cross-section, DM mass and branching fraction when the three parameters are varied at a time. The spectra is constructed as an evenly weighted combination of $b\bar{b}$ and leptons pairs. The leptons fraction denotes an unweighted combination of $e^{+}e^{-}$, $\mu^{+}\mu^{-}$ and $\tau^+\tau^-$ pairs. Errors shown here include systematic uncertainties. See left panel of Figure~\eqref{fig:darkmatterCL} for further details.}
\end{table}

\begin{table}[t!]

\begin{ruledtabular}
\begin{tabular}{ccc}
\centering
Branching ratio & $\left\langle\sigma v\right\rangle$ [cm$^3$/s] & $M_{\rm DM}$ [GeV]  \\ \hline  
100\% $b\bar{b}$ & $2.47^{+0.28}_{-0.25}\times 10^{-26}$ & $34.1^{+4.0}_{-3.5}$ \\
50\% $b\bar{b}$, 50\% leptons & $2.77^{+0.47}_{-0.35}\times 10^{-26}$ & $21.7^{+3.8}_{-2.8}$ \\ 
10\% $b\bar{b}$, 90\% leptons & $2.14^{+0.17}_{-0.16}\times 10^{-26}$ & $ 9.3^{+0.6}_{-0.5}$ \\ 
\end{tabular}
\end{ruledtabular}

\caption{\label{tab:bblept2} Best fit $M_{\rm DM}$ and $\left\langle \sigma v\right\rangle$ for several fixed values of $B_f$. The leptons fraction denotes an unweighted combination of $e^{+}e^{-}$, $\mu^{+}\mu^{-}$ and $\tau^+\tau^-$ pairs. Errors shown here include systematic uncertainties. See right panel of Figure~\eqref{fig:darkmatterCL} for further details.}
\end{table}

\begin{table}[t!]

\begin{ruledtabular}
\begin{tabular}{ccc}
\centering
Best-fit Branching ratio & $\left\langle\sigma v\right\rangle$ [cm$^3$/s] & $M_{\rm DM}$ [GeV]  \\ \hline \\ 
$75^{+13}_{-15}\%$ $b\bar{b}$  & $2.1^{+0.27}_{-0.45}\times 10^{-26}$ & $23.6^{+6.7}_{-6.4}$ \\

\end{tabular}
\end{ruledtabular}

\caption{\label{tab:bbtt} Best fit values on the DM velocity averaged annihilation cross-section, DM mass and branching fraction when the three parameters are varied at a time. The spectra is constructed as an unweighted combination of $b\bar{b}$ and $\tau^+\tau^-$ pairs. Errors shown here include systematic uncertainties. See Figure~\eqref{fig:darkmatterCLtautau} for further details.}
\end{table}

We have seen in Figures~\eqref{fig:ExtSrcMorphology1} and \eqref{fig:ExtSrcMorphology2} that there is evidence for a single strong positive residual emission in the Galactic Center with a spatial morphology that agrees well with that of a NFW profile with inner slope $\gamma=1.2$. Also, the evaluation of the systematic uncertainties related to imperfections in the Galactic diffuse background led us to the conclusion that the dark matter signals are much larger in size than the systematic errors. Thus, the next logical step is to calculate the regions of the self-annihilating DM parameter space that provide a good fit to the LAT data. In Figures~\eqref{fig:darkmatterCL} and \eqref{fig:darkmatterCLtautau} we present the main results of this analysis. Contours are shown at $1\sigma$, $2\sigma$,...$5\sigma$ confidence level.

In the right upper panel of Fig.~\eqref{fig:darkmatterCL} we show the preferred regions of the parameter space for 100\% $b\bar{b}$ final states. The 95\% upper limits obtained in the Fermi-LAT analysis of Dwarf Galaxies~\citep{dwarfs} are also shown for comparison. We notice that the  best DM region is not yet in tension with those limits. However, one would expect that the limits obtained from Dwarf Galaxies will be strengthened with larger data sets. We estimated that for 10 years of LAT data obtained from observations of Milky Way Dwarf Galaxies the 95\% upper limits on $\left\langle\sigma v \right\rangle$ can be approximated to two standard deviations of a Gaussian with a mean of zero. As the standard deviation is inversely proportional to the square root of the number of observations
 we can approximate the upper limits for 10 years to be $\sqrt{2/10}=0.45$ of the upper limit of two years (this is plotted in Fig.~\eqref{fig:darkmatterCL} with a red line). We now see that our best fit region would be ruled out by the 10 years data set.
However, this is for our assumed value of $\rho_0=0.36$~GeV~cm$^{-3}$ and so our GC constrained $\left<\sigma v\right>$ contours could move up or down by about 30\%.

\begin{figure*}[ht!]
\begin{center}

\begin{tabular}{cc}
\centering
\includegraphics[width=0.5\linewidth]{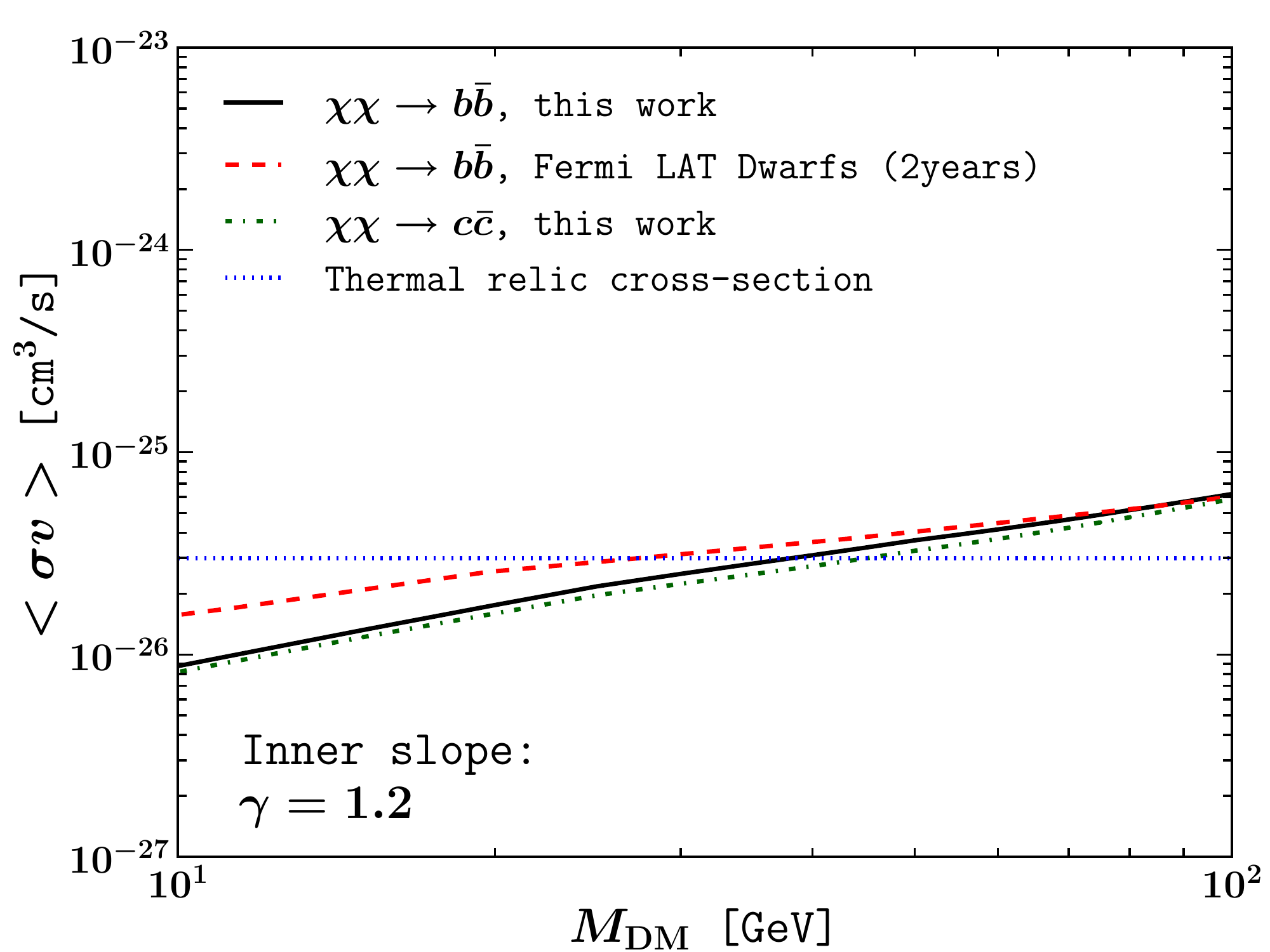} &
\centering
\includegraphics[width=0.5\linewidth]{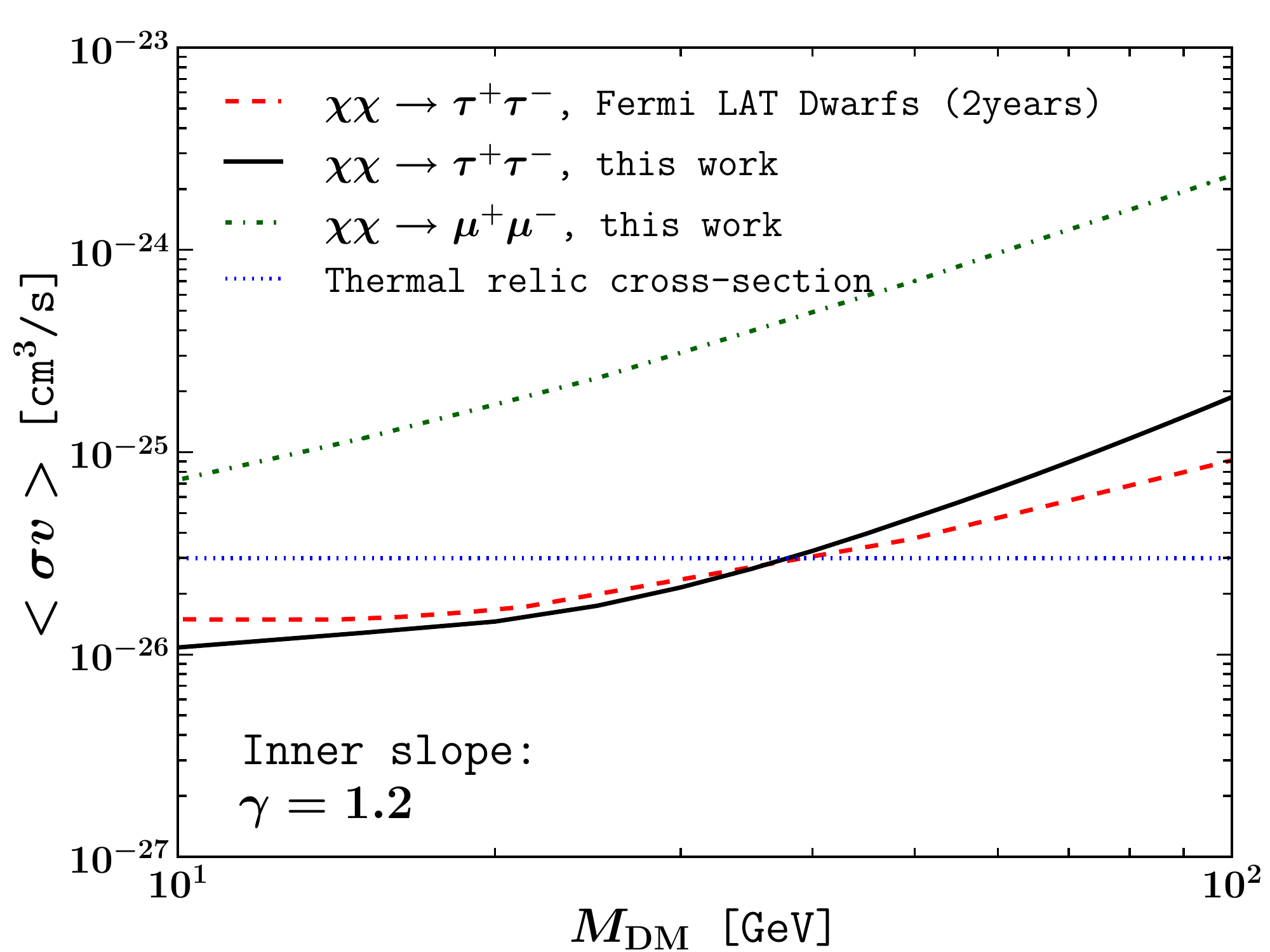} 
\end{tabular}

\begin{tabular}{cc}
\centering
\includegraphics[width=0.5\linewidth]{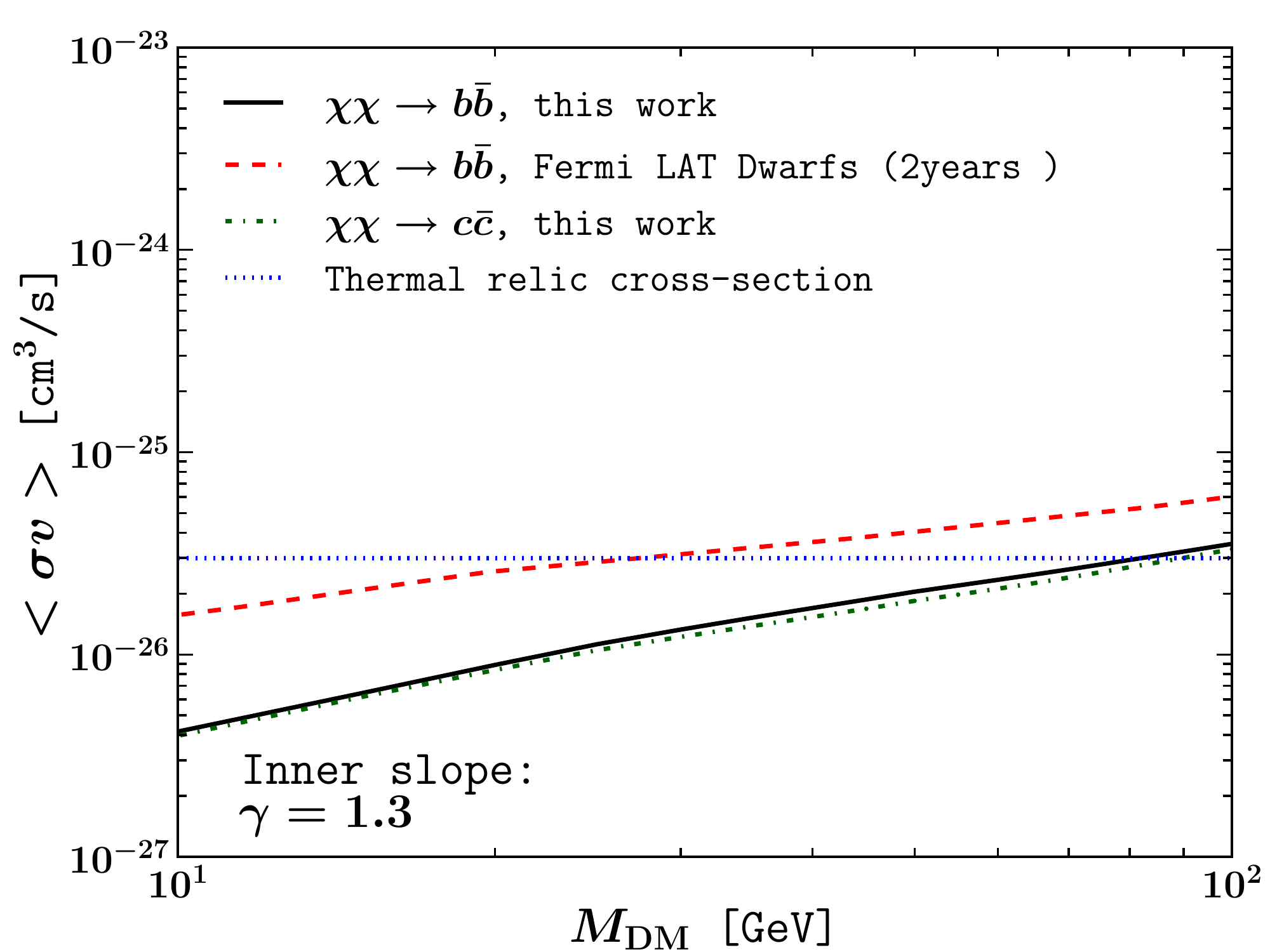} &
\centering
\includegraphics[width=0.5\linewidth]{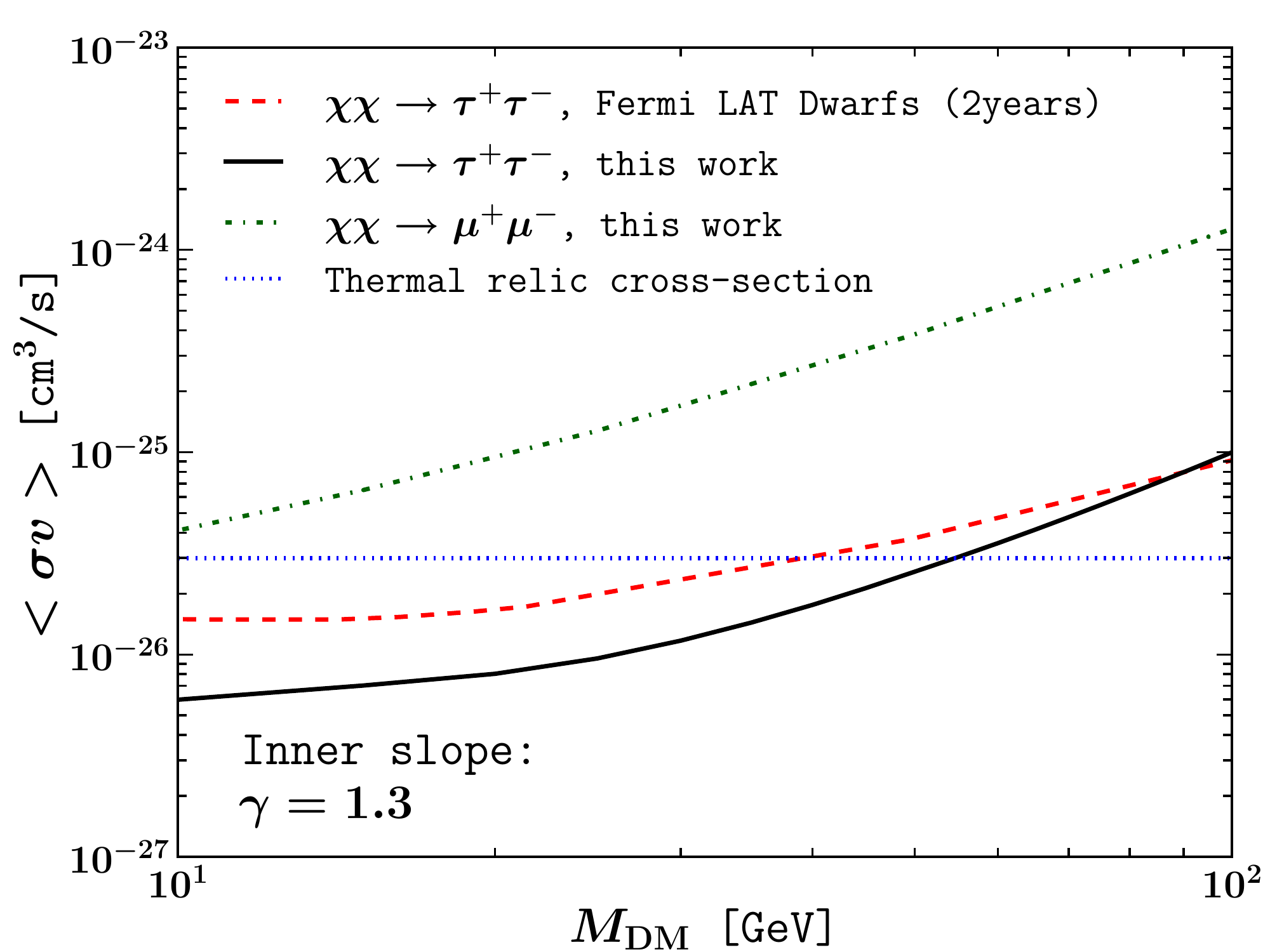} 

\end{tabular}

\end{center}
\caption{ \label{fig:upperlimits} Derived 95\% CL upper limits on the velocity averaged cross-section for various annihilation channels: 100\% $b\bar{b}$, 100\% $c\bar{c}$, 100\% $\tau^+\tau^-$ and 100\% $\mu^+\mu^-$. The horizontal dotted blue line denotes the thermal decoupling cross-section expected for WIMPs particles. Shown for comparison are the upper limits on $\left\langle \sigma v \right\rangle$ obtained from the analysis of Dwarf Galaxies in Ref.~\citep{dwarfs}. Limits are obtained from the analysis of 3.8 years of GC photon data in the energy range 0.3$-$10 GeV. \texttt{Upper panel}: Assumes a DM distribution given by a NFW profile with $\gamma=1.2$ and $\rho(R_{\odot})=3.6$ GeV cm$^{-3}$.  \texttt{Lower panel}: Assumes a DM distribution given by a NFW profile with $\gamma=1.3$ and $\rho(R_{\odot})=3.4$ GeV cm$^{-3}$.    }

\end{figure*}

Finally and for completeness, we present 95\% CL upper limits on $\left\langle\sigma v \right\rangle$ from GC data in Fig.~\eqref{fig:upperlimits}. Since we only used photon data in the energy range 0.3$-$10 GeV, we decided to compute the upper limits up to 100 GeV. We show that our derived limits are competitive with those obtained from Dwarf Galaxies, albeit with more uncertainty in the systematic error.

\begin{figure*}[ht!]
\begin{center}

\begin{tabular}{cc}
\centering
\includegraphics[width=0.5\linewidth]{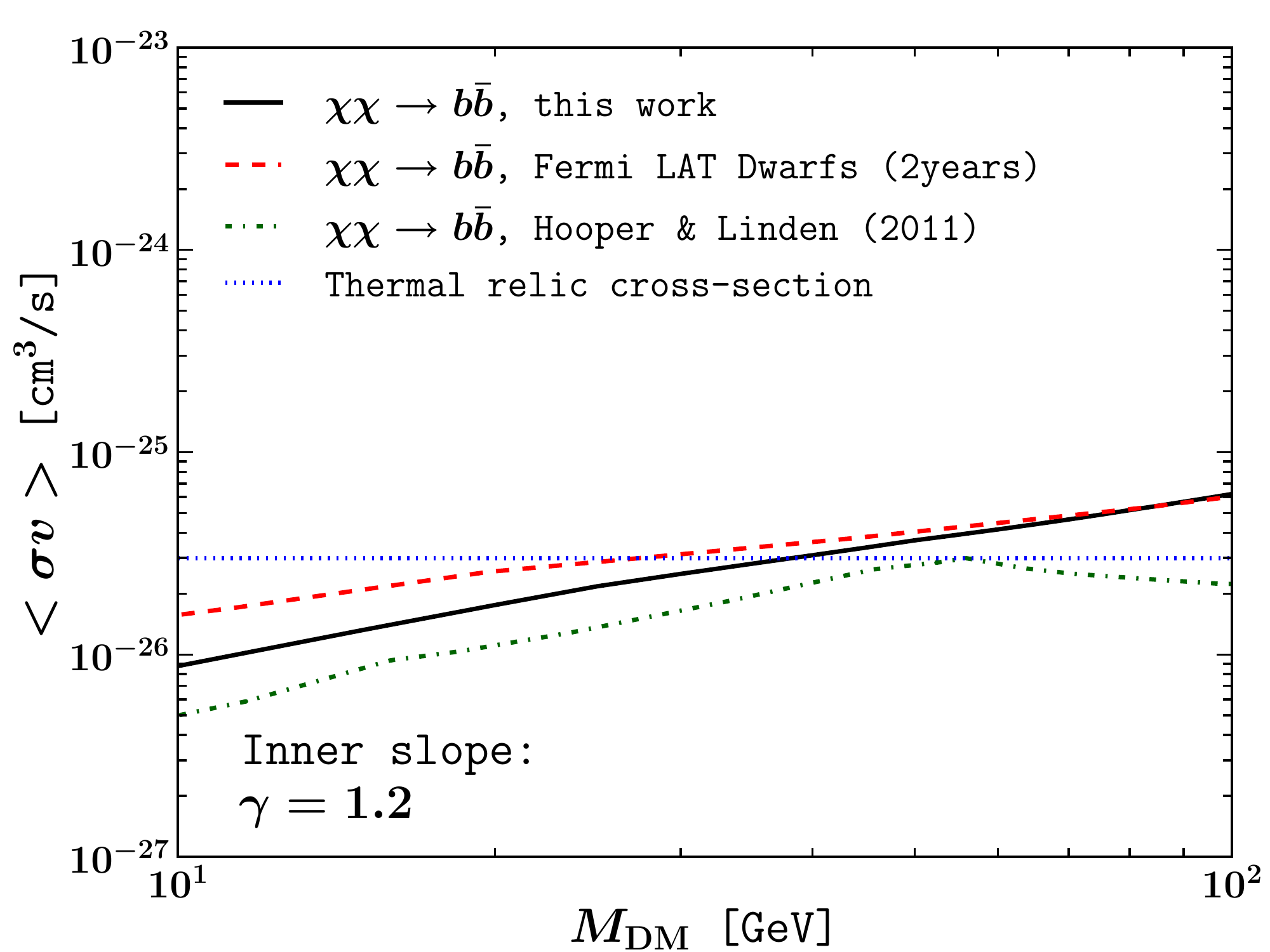} &
\centering
\includegraphics[width=0.5\linewidth]{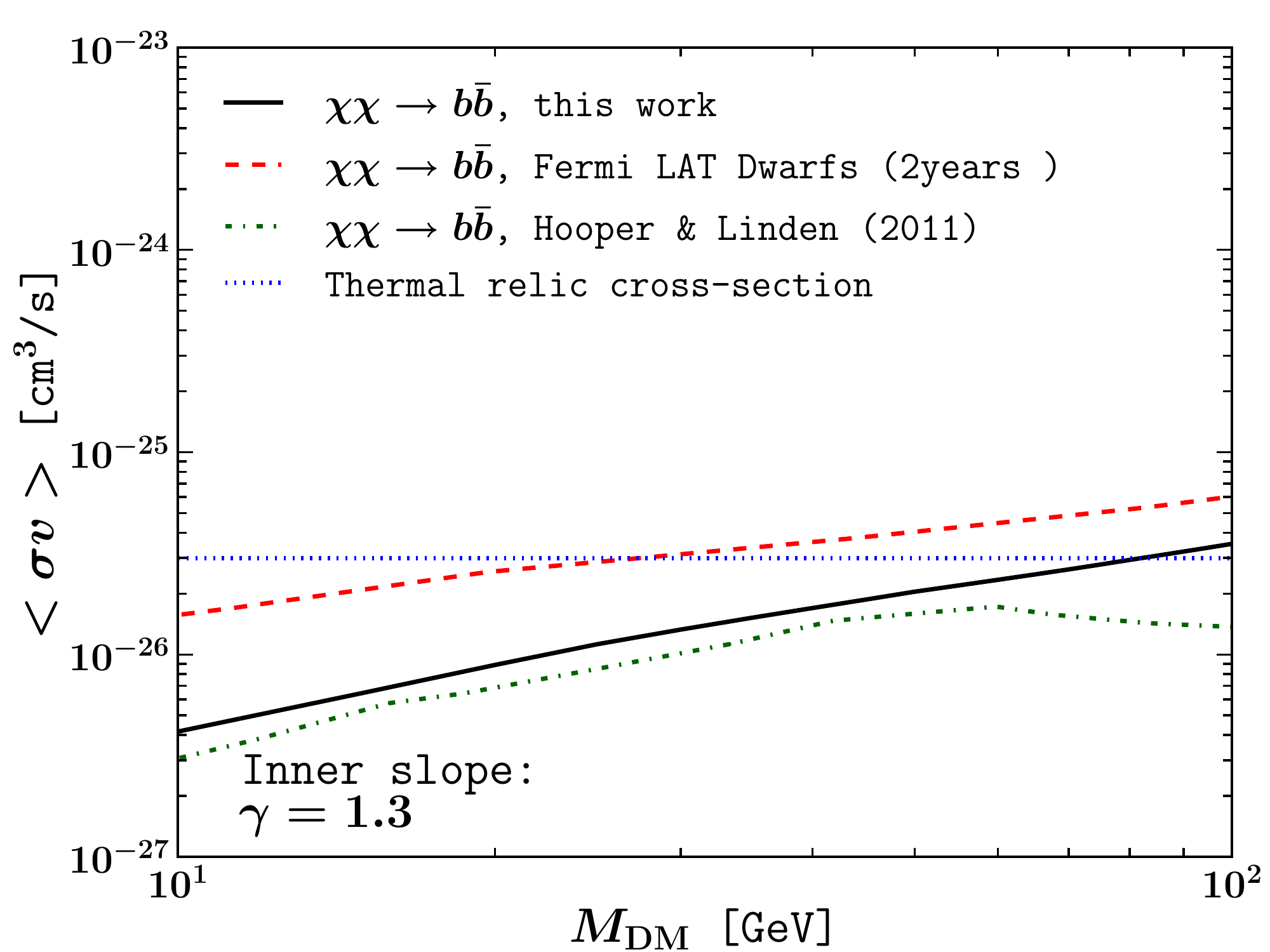} 
\end{tabular}
\includegraphics[width=0.7\linewidth]{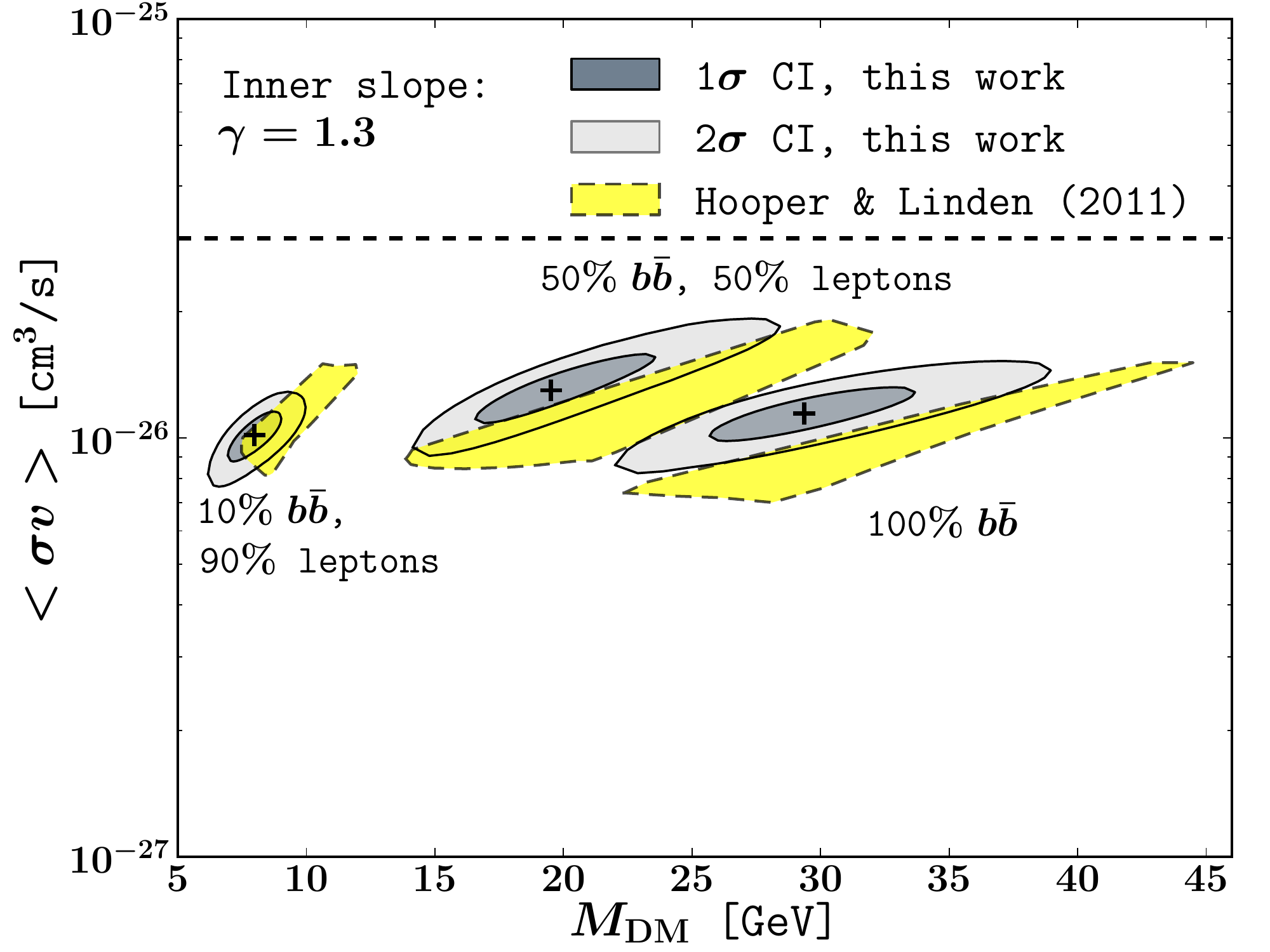} 

\end{center}
\caption{ \label{fig:hooper}\texttt{Upper panel}: Shown are the 95\% CL upper limits on the velocity averaged cross-section for 100\% $b\bar{b}$ final states. The horizontal dotted blue line denotes the thermal decoupling cross-section expected for WIMPs particles. Shown for comparison are the upper limits  obtained from the analysis of Dwarf Galaxies in Ref.~\citep{dwarfs} and GC analysis in Ref.~\citep{hooperlinden2011} (see more details in Fig.~\eqref{fig:upperlimits}). \texttt{Lower panel}: Shown are the regions of the parameter space which provide a good fit to Fermi-LAT data as derived in this work (grey area) and in Hooper et al~\citep{hooperlinden2011} (yellow area).    }
\end{figure*}

\begin{figure}
\centering 
\includegraphics[width=1.0\linewidth]{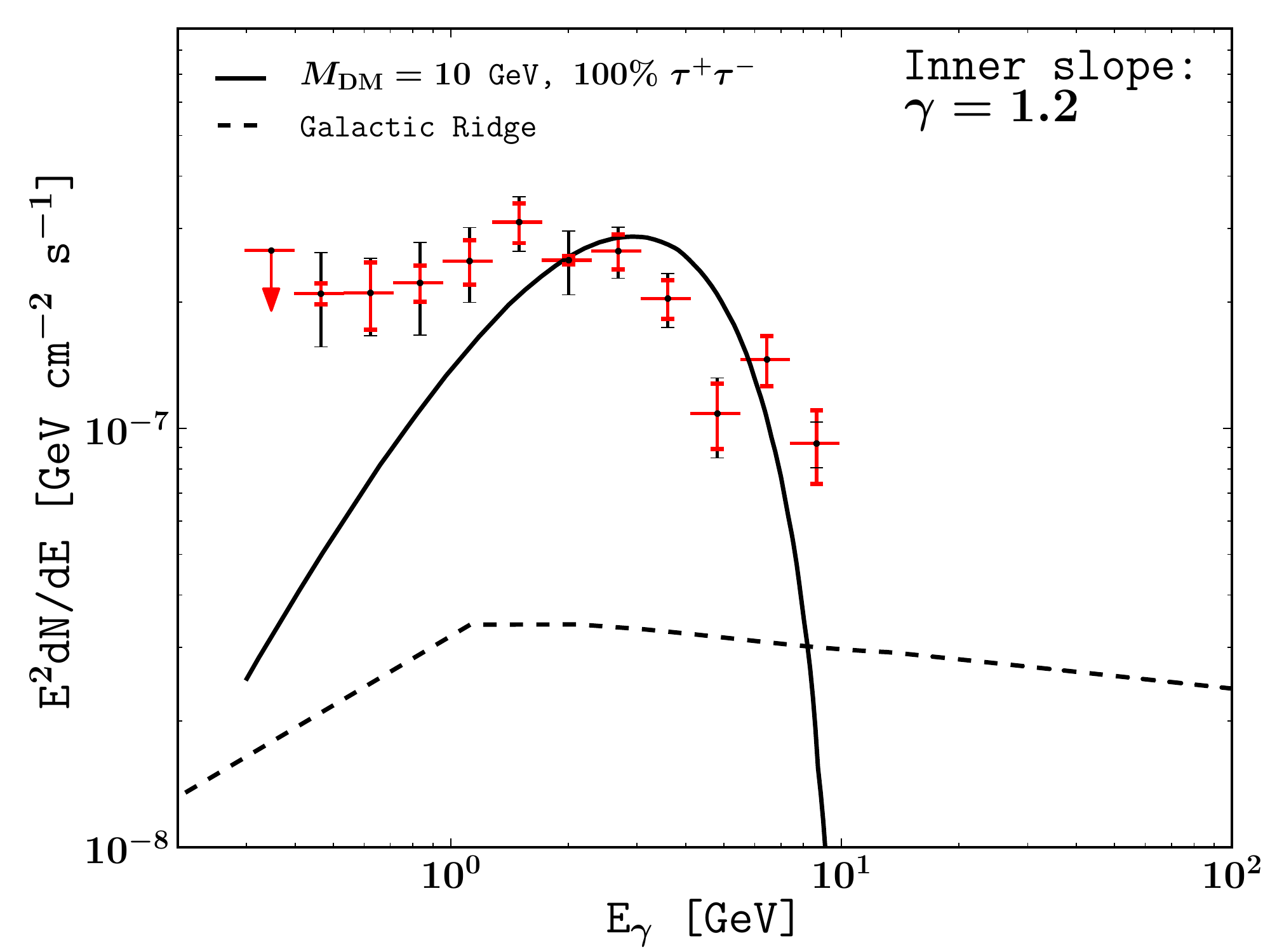}
\caption{ \label{fig:ridge}SED of the extended source assuming a NFW profile with $\gamma=1.2$ and $\rho(R_{\odot})=0.36$ GeV cm$^{-3}$. The best fit spectrum obtained with $M_{\rm DM}=10$ GeV and 100\% $\tau^+\tau^-$ final states is overlaid over the twelve energy fluxes data points. Red error bars represent systematics errors and black error bars statistical errors. For illustration we also plotted the spectra of the Galactic Ridge as obtained in Fig.~7 of Ref.~\citep{hooperlinden2011}, but this source was not considered in our actual fits. See text for a discussion on this.  }
\end{figure}

\section{Discussion}
\label{sec:discussion}
We find that when we include only statistical error bars in our band analysis, we get best fits and errors that are  a good match to using the Fermi Science Tools with the same energy range. This is a good check that our band analysis is providing an accurate representation of the data. 

Using our band analysis, we could evaluate the equivalent of a $TS$ value which includes the $TS$ value by subtracting $\C$ in \eq{eq:Csyst} with $F_i^{\rm fit}$ set to the best fit value from $\C$ with   $F_i^{\rm fit}=0$. Although in this case, as the $F_i^{\rm fit}=0$ is so far from the best fit, the Gaussian approximation, implicit in our use of the band analysis, would be expected to break down. 

Although, the $\tau^+\tau^-$ only case may have $TS\gg 25$, as can be seen from Figs.~\eqref{fig:SED} and \eqref{fig:darkmatterCLtautau}, it does not provide a good fit to the data. 
Ref.~\cite{hooperlinden2011} and  \cite{AK,AKerratum} provide  analysis of the $\tau^+\tau^-$ as an acceptable model. Although, we agree the $\tau^+\tau^-$ with $M\approx 10$~GeV does provide a good $TS$ value, we have shown it provides a poor fit to the data. 
Similarly, as can be seen from Fig.~\eqref{fig:darkmatterCL}, a pure lepton spectrum  does not provide a good fit to the data. However, as can be seen
from Figures~\eqref{fig:darkmatterCL}, \eqref{fig:SED}, and \eqref{fig:darkmatterCLtautau}, a $b\bar{b}$ only spectrum does provide a good fit. 
 In Ref.~\cite{AK,AKerratum}, they fit for a range of masses for a $b{\bar b}$ model.  For their Fermi Science Tools analysis they find models with $10\leq M_{\rm DM}\leq110$ have $TS\ge 25$. Using the same data and method we have reproduced their constraints on $\left<\sigma v\right>$ in Fermi-tools. However, as we see from  Fig.~\ref{fig:darkmatterCL}, only $b{\bar b}$  models with $20\leq M_{\rm DM}\leq 60$ GeV are within the 4$\sigma$ confidence region.  This shows that despite models such as $M_{\rm DM}=10$ GeV providing a good $TS$ value, they do not provide a good fit to the data as can also be seen in Fig.~\eqref{fig:SED}.
 
If the WIMP particles are Majorana fermions, then the pair-annihilation into light fermions is  highly suppressed since the invariant scattering amplitude $|\mathcal{M}|^2\propto m_f^2$~\citep{ProfumoTasi}. Furthermore, if annihilations into gauge bosons are also suppressed and the WIMPs are lighter than the top quark then the prevailing annihilations final states are $b\bar{b}$ and $\tau^+\tau^-$. By virtue of the color charge of the bottom quarks~\citep{ProfumoTasi}, one would expect the production of $b\bar{b}$ pairs to be typically more than three times larger than those of $\tau^+\tau^-$. Thus, we note that one could easily accommodate a theoretical model to these findings.

The best fit DM models, see Tables~\eqref{tab:bblept1}, \eqref{tab:bblept2}, \eqref{tab:bbtt}, Figures \eqref{fig:darkmatterCL}, and \eqref{fig:darkmatterCLtautau} have values for $\left<\sigma v\right>$ intriguingly close to the simple thermal relic value.  
An even closer match is obtained from a more precise WIMP relic abundance cross-section
of $\left<\sigma v\right>=2.2\times10^{-26}$ cm$^3$s$^{-1}$ which has a feeble mass-dependence for masses above 10 GeV \cite{steigmandasguptabeacom2012}.
  
Our SEDs are designed to be of the GC extended emission component only, while those of Ref.~\cite{hooperlinden2011} also include Sgr A* and a component known as the HESS ridge which we will discuss later in this section.
Also, comparing our results with Ref.~\cite{hooperlinden2011} is difficult as they use a profile with a slope $\rho \propto r^{-\gamma}$ rather than a generalized NFW profile as in \eq{nfw}. For a generalized NFW profile the line of sight integral, \eq{J}, formally extends to an infinite distance from the observer. Because of the steep drop off beyond $r_s$, the integral is insensitive to the actual upper bound used provided it is much larger than $r_s$. However, if only the inner slope is used then the $J$ factor depends sensitively on the upper bound assumed and an upper bound of $\infty$ would give much too large an answer. Unfortunately, the range of the line of sight integral used for the galactic center results of \cite{hooperlinden2011} is not provided and so we are unable to reliably compare our constraints with theirs for $\left<\sigma v\right>$. But, interestingly, our constraints for the $\gamma=1.3$ case are a good  match with theirs, see Figs.~\eqref{fig:upperlimits} and \eqref{fig:hooper}. For the $\gamma=1.3$ case, we determined $\rho_0$ from
maximizing the likelihood of the microlensing and dynamical data (see Fig.~5 of \cite{ioccopatobertone2011}) with $\gamma=1.3$ to be $\rho_0=0.34$~GeV~cm$^{-3}$. Our $\rho_0$ for $\gamma=1.2$ and $\gamma=1.3$ match the corresponding $\rho_0$ in Ref.~\cite{hooperlinden2011}. But, without the upper limit for their line of sight integral, it is not clear whether this match is coincidental or not. 
 Note that in the upper-limits plot of 
 Fig.~\ref{fig:hooper}, the match is not as good for $M_{\rm DM}>100$ GeV but this likely due to in their corresponding plot they use their $10$ to 100 GeV bin and for $M_{\rm DM}>100$ GeV the DM spectrum significantly overlaps with that region. 

For $\gamma=1.2$ the match is not as good, see Fig.~\ref{fig:hooper}. As Fig.~\ref{fig:degeneracy} shows the inner PSs are very degenerate with the excess emission component and in the GC analysis of  \cite{hooperlinden2011} they use the 2FGL parameters for all the PSs except Sgr A* which they fit a PS to the data without an GC excess emission component. Their Sgr A* fit (see Fig.~4 of Ref.~\cite{hooperlinden2011} )is very similar to ours
for the baseline model in Fig.~\ref{fig:degeneracy}. They do use a broken power law parametrization rather than a Log Parabola, but that difference has a negligible effect.
So their analysis does not utilize the degeneracy between the PSs, especially Sgr A*, and the GC excess emission component. 
This implies the analysis of Ref.~\cite{hooperlinden2011} will have a suppressed 
dark matter $\left<\sigma v\right>$ when compared to ours.

In Ref.~\citep{hooperkelsoqueiroz2012} they do use a generalized NFW profile, but there they do not account for Sgr A* as they are seeking a robust upper limit to the DM cross-section. They also choose values of $\gamma$ and $\rho_0$ consistent with the microlensing and dynamical data \cite{ioccopatobertone2011} but chosen to be conservative with respect to a potential dark matter annihilation signal. Consistent with this, their upper limits are larger than ours. Also, again the match is more discrepant for $M_{\rm DM}>100$ GeV but this likely due to them using the $10$ to 100 GeV data range and for $M_{\rm DM}>100$ GeV the DM spectrum significantly overlaps with that energy region. 

In the GC analysis of Ref.~\cite{hooperlinden2011}, they investigate adding a HESS Galactic ridge component \cite{Aharonian:2006au}. The 2 by 1 degree HESS Galactic ridge was measured  by HESS over the energy range 0.2 to 10 TeV. It was found to be spatially correlated with the molecular clouds in the central 200 parsec of the Milky Way. Its origin is usually taken to be the decays of neutral pions produced in the interactions of  harder than usual population  of cosmic ray protons and nuclei with the surrounding molecular gas. In Ref.~\cite{hooperlinden2011} they evaluate the spectrum for this model at energies less than 100 GeV and use this in their model fit. 

As can be seen from Fig.~\ref{fig:ridge}, the HESS ridge postulated by \cite{hooperlinden2011} should not significantly effect the fit in our case. This is because we are using $\gamma=1.2$ that leads to a higher inferred DM flux and also we are refitting our PSs which allows the DM flux to be higher by lowering the emission of the PSs close to the GC. 

The GC excess emission results of Ref.~\cite{hooperslatyer2013} are quoted as 
being derived from a generalized NFW which is effectively equivalent to the one we are using for $\gamma=1.2$, albeit with a density of $\rho_0=0.4$ GeV/cm$^3$.
But, as they are not simultaneously fitting their PS and DM models, this is likely to explain why they find a significantly smaller value for $\left<\sigma v\right>=8\times(0.4/0.36)^2\times10^{-27}=9.9\times10^{-27}$ cm$^{3}$ s$^{-1}$ for the $M_{\rm DM}=50$ GeV (100\% $b\bar{b}$)  case for $\gamma=1.2$. Where we have converted their value to the equivalent value for our assumed local density. 
As can be seen from the top left plot in Fig.~\ref{fig:darkmatterCL}, this is outside our 5$\sigma$ confidence region.
However, if instead of taking the best fit $\rho_0$ from the microlensing and dynamical data (see Fig.~5 of \cite{ioccopatobertone2011}), one sees how the result changes if one takes the contour 68\% limits, $\rho_0\in [0.3,0.4]$, the error  for $\left<\sigma v\right>$ becomes of order 20\% taking into account the $\rho_0^2$ dependence of the $J$ factor shown in \eq{J}. However, this is only important in estimating $\left<\sigma v\right>$ and does not affect statements like a 10 GeV DM annihilating only to $\tau^+\tau^-$ does not provide a good fit to the Fermi-LAT data, as the goodness of  fit to the gamma-ray data is independent of $\left<\sigma v\right>$ due to the complete degeneracy between $\left<\sigma v\right>$ and $\rho_0^2$.

In Ref.~\cite{hooperlinden2011}, they state that 
they include the observed spatial variations of the residuals as a systematic error.
 Details are not given on the magnitude. While in an earlier related paper \cite{Hooper:2010mq} a value of 3\% is given.
 By varying the parameters used in GALPROP  for the distribution of cosmic-rays, interstellar gas and radiation fields, the Fermi-LAT team reported a systematic error of order $10\%$ for the inner Galaxy and for energies less than 10 GeV with unresolved point sources being cited as a likely cause \cite{ackermannajelloatwood2012}.
Ref.~\cite{Boyarsky:2010dr} found systematic errors of about 10\% in a $2^\circ$ around the GC by doing Monte Carlo simulations of a model with no GC diffuse source. Thus, overall our estimate of $20\%$ is  higher than other estimates.

Our confidence regions for MSPs are in good agreement with the average pulsar spectrum measured by Fermi-LAT, see Fig.~\ref{fig:Pulsars}. For the pulsar hypothesis, it is interesting to evaluate the number of MSPs needed to account for the excess emission. For an energy range of 100 MeV to 10 GeV, Ref.~\cite{2009Sci...325..845A}  found 47 Tuc had a flux of $2.6(\pm 0.8)\times10^{-8}$ photons cm$^{-2}$ s$^{-1}$. 
The population of MSPs in 47 Tuc is taken to be 30 to 60. 
Following \cite{AK,AKerratum} we then use this to estimate the order of magnitude of the flux of a single MSP to be $\sim 10^{-9}$ ph cm$^{-2}$ s$^{-1}$. 
The flux of our best fit exponential cutoff in the energy range 100 MeV to 10 GeV is obtained by integrating the parametric form \eq{eq:expcut} with the best fit parameters quoted in Sec.~\ref{subsec:Pulsars} and is found to be $1.7\times 10^{-6}$ ph cm$^{-2}$ s$^{-1}$. Therefore we find the number of MSPs needed to explain the GC excess emission to be $\sim 1000$ which is compatible with what \cite{AK,AKerratum} found as our flux estimates of the GC excess are compatible with theirs.

However, if the excess extended emission was also responsible for the bulk of the low latitude, low energy emission of the Fermi Bubbles as suggested in Ref.~\cite{hooperslatyer2013},  the spectral and spatial properties typical of a population of MSPs would not be a good fit to the signal \cite{HooperP}.

Ref.~\cite{AK,AKerratum} also examined a  proposal by Ref.~\cite{yusef-zadehhewittwardle2013} which entails high energy cosmic ray electrons producing bremsstrahlung gamma-rays on molecular gas. This case can have significant extent to the spatial emission. Ref  \cite{yusef-zadehhewittwardle2013} finds that the source electron population is consistent with radio observations of synchrotron emission from the high energy population of electrons, as well as the morphology of the FeI 6.4 keV X-ray emission. Ref.~\cite{AK,AKerratum} find that using the radio emission morphology, tracing the synchrotron emission from the cosmic ray electrons improves the fit over the base model with a $TS$=252 for an energy range of 1 to 100 GeV. But this was significantly smaller than what they obtained for a Log Parabola spectrum  for the same energy range which gave $TS$=412.

This indicates that the bremsstrahlung model may not be providing a good fit, in much the same way as we have found $\tau^+\tau^-$ has a good $TS$ but not as good as $b\bar{b}$ which was a good fit.

\section{Conclusions}
\label{sec:conclusions}
We have found that either a DM annihilation model or unresolved pulsar population is consistent with the observed excess gamma ray emission seen in the GC.
Our analysis marginalized over the PS and diffuse background amplitudes in the region of interest. We included an estimated systematic error for the diffuse galactic background of about 20\%. 
We provide confidence regions for the model parameters. 

We found that a population of 1000-2000 MSPs with 
parameters consistent with the average
 spectral shape of Fermi-LAT measured MSPs was able to fit the GC excess emission.
For DM, we found that a pure  $\tau^+\tau^-$ annihilation channel is not a good fit to the data. But a mixture of  $\tau^+\tau^-$ and $b\bar{b}$ 
with   $\left<\sigma v\right>$ of order the thermal relic value  and a DM mass of around 20 to 60 GeV provides an adequate fit.

\begin{acknowledgements}
We thank  Kev Abazajian, Jenni Adams, Terri Brandt, Tim Cohen,  Dan Hooper,  Junichiro Katsuta, Tim Linden, Gavin Rowell, Tracy Slatyer, and Johann Cohen-Tanugi   for helpful discussions.
We also thank the Fermi Collaboration for making the Fermi-LAT data and Fermi Science Tools so readily available.
O.M. especially expresses his thanks to Stefano Profumo for allowing him to study the original Fortran version of the \texttt{DMFit} package. 
 O.M. is supported by a UC Doctoral Scholarship.
\end{acknowledgements}

\bibliography{references}

\begin{thebibliography}{54}%
\makeatletter
\providecommand \@ifxundefined [1]{%
 \@ifx{#1\undefined}
}%
\providecommand \@ifnum [1]{%
 \ifnum #1\expandafter \@firstoftwo
 \else \expandafter \@secondoftwo
 \fi
}%
\providecommand \@ifx [1]{%
 \ifx #1\expandafter \@firstoftwo
 \else \expandafter \@secondoftwo
 \fi
}%
\providecommand \natexlab [1]{#1}%
\providecommand \enquote  [1]{``#1''}%
\providecommand \bibnamefont  [1]{#1}%
\providecommand \bibfnamefont [1]{#1}%
\providecommand \citenamefont [1]{#1}%
\providecommand \href@noop [0]{\@secondoftwo}%
\providecommand \href [0]{\begingroup \@sanitize@url \@href}%
\providecommand \@href[1]{\@@startlink{#1}\@@href}%
\providecommand \@@href[1]{\endgroup#1\@@endlink}%
\providecommand \@sanitize@url [0]{\catcode `\\12\catcode `\$12\catcode
  `\&12\catcode `\#12\catcode `\^12\catcode `\_12\catcode `\%12\relax}%
\providecommand \@@startlink[1]{}%
\providecommand \@@endlink[0]{}%
\providecommand \url  [0]{\begingroup\@sanitize@url \@url }%
\providecommand \@url [1]{\endgroup\@href {#1}{\urlprefix }}%
\providecommand \urlprefix  [0]{URL }%
\providecommand \Eprint [0]{\href }%
\providecommand \doibase [0]{http://dx.doi.org/}%
\providecommand \selectlanguage [0]{\@gobble}%
\providecommand \bibinfo  [0]{\@secondoftwo}%
\providecommand \bibfield  [0]{\@secondoftwo}%
\providecommand \translation [1]{[#1]}%
\providecommand \BibitemOpen [0]{}%
\providecommand \bibitemStop [0]{}%
\providecommand \bibitemNoStop [0]{.\EOS\space}%
\providecommand \EOS [0]{\spacefactor3000\relax}%
\providecommand \BibitemShut  [1]{\csname bibitem#1\endcsname}%
\let\auto@bib@innerbib\@empty
\bibitem [{\citenamefont {{Bertone}}\ \emph {et~al.}(2005)\citenamefont
  {{Bertone}}, \citenamefont {{Hooper}},\ and\ \citenamefont {{
  Silk}}}]{bertonehoopersilk2005}%
  \BibitemOpen
  \bibfield  {author} {\bibinfo {author} {\bibfnamefont {G.}~\bibnamefont
  {{Bertone}}}, \bibinfo {author} {\bibfnamefont {D.}~\bibnamefont {{Hooper}}},
  \ and\ \bibinfo {author} {\bibfnamefont {{J.}}~\bibnamefont {{ Silk}}},\
  }\bibfield  {title} {\enquote {\bibinfo {title} {{Particle dark matter:
  evidence, candidates and constraints}},}\ }\href {\doibase
  10.1016/j.physrep.2004.08.031} {\bibfield  {journal} {\bibinfo  {journal}
  {\physrep}\ }\textbf {\bibinfo {volume} {405}},\ \bibinfo {pages} {279--390}
  (\bibinfo {year} {2005})},\ \Eprint
  {http://arxiv.org/abs/arXiv:hep-ph/0404175} {arXiv:hep-ph/0404175}
  \BibitemShut {NoStop}%
\bibitem [{\citenamefont {Cirelli}(2012)}]{cirelli2}%
  \BibitemOpen
  \bibfield  {author} {\bibinfo {author} {\bibfnamefont {Marco}\ \bibnamefont
  {Cirelli}},\ }\bibfield  {title} {\enquote {\bibinfo {title} {{Indirect
  Searches for Dark Matter: a status review}},}\ }\href@noop {} {\  (\bibinfo
  {year} {2012})},\ \Eprint {http://arxiv.org/abs/1202.1454} {arXiv:1202.1454
  [hep-ph]} \BibitemShut {NoStop}%
\bibitem [{\citenamefont {Clowe}\ \emph {et~al.}(2006)\citenamefont {Clowe},
  \citenamefont {Bradac}, \citenamefont {Gonzalez}, \citenamefont {Markevitch},
  \citenamefont {Randall} \emph {et~al.}}]{clowebradacgonzalez2006}%
  \BibitemOpen
  \bibfield  {author} {\bibinfo {author} {\bibfnamefont {Douglas}\ \bibnamefont
  {Clowe}}, \bibinfo {author} {\bibfnamefont {Marusa}\ \bibnamefont {Bradac}},
  \bibinfo {author} {\bibfnamefont {Anthony~H.}\ \bibnamefont {Gonzalez}},
  \bibinfo {author} {\bibfnamefont {Maxim}\ \bibnamefont {Markevitch}},
  \bibinfo {author} {\bibfnamefont {Scott~W.}\ \bibnamefont {Randall}},  \emph
  {et~al.},\ }\bibfield  {title} {\enquote {\bibinfo {title} {{A direct
  empirical proof of the existence of dark matter}},}\ }\href {\doibase
  10.1086/508162} {\bibfield  {journal} {\bibinfo  {journal} {Astrophys.J.}\
  }\textbf {\bibinfo {volume} {648}},\ \bibinfo {pages} {L109--L113} (\bibinfo
  {year} {2006})},\ \Eprint {http://arxiv.org/abs/astro-ph/0608407}
  {arXiv:astro-ph/0608407 [astro-ph]} \BibitemShut {NoStop}%
\bibitem [{\citenamefont {Bringmann}\ and\ \citenamefont
  {Weniger}(2012)}]{Bringmann}%
  \BibitemOpen
  \bibfield  {author} {\bibinfo {author} {\bibfnamefont {Torsten}\ \bibnamefont
  {Bringmann}}\ and\ \bibinfo {author} {\bibfnamefont {Christoph}\ \bibnamefont
  {Weniger}},\ }\bibfield  {title} {\enquote {\bibinfo {title} {{Gamma Ray
  Signals from Dark Matter: Concepts, Status and Prospects}},}\ }\href
  {\doibase 10.1016/j.dark.2012.10.005} {\bibfield  {journal} {\bibinfo
  {journal} {Phys.Dark Univ.}\ }\textbf {\bibinfo {volume} {1}},\ \bibinfo
  {pages} {194--217} (\bibinfo {year} {2012})},\ \Eprint
  {http://arxiv.org/abs/1208.5481} {arXiv:1208.5481 [hep-ph]} \BibitemShut
  {NoStop}%
\bibitem [{\citenamefont {Baltz}\ \emph
  {et~al.}(2008{\natexlab{a}})\citenamefont {Baltz} \emph
  {et~al.}}]{Baltz:2008wd}%
  \BibitemOpen
  \bibfield  {author} {\bibinfo {author} {\bibfnamefont {E.A.}\ \bibnamefont
  {Baltz}} \emph {et~al.},\ }\bibfield  {title} {\enquote {\bibinfo {title}
  {{Pre-launch estimates for GLAST sensitivity to Dark Matter annihilation
  signals}},}\ }\href {\doibase 10.1088/1475-7516/2008/07/013} {\bibfield
  {journal} {\bibinfo  {journal} {JCAP}\ }\textbf {\bibinfo {volume} {0807}},\
  \bibinfo {pages} {013} (\bibinfo {year} {2008}{\natexlab{a}})},\ \Eprint
  {http://arxiv.org/abs/0806.2911} {arXiv:0806.2911 [astro-ph]} \BibitemShut
  {NoStop}%
\bibitem [{\citenamefont {Nolan}\ \emph {et~al.}(2012)\citenamefont {Nolan}
  \emph {et~al.}}]{2FGL}%
  \BibitemOpen
  \bibfield  {author} {\bibinfo {author} {\bibfnamefont {P.~L.}\ \bibnamefont
  {Nolan}} \emph {et~al.},\ }\bibfield  {title} {\enquote {\bibinfo {title}
  {{Fermi Large Area Telescope Second Source Catalog}},}\ }\href {\doibase
  10.1088/0067-0049/199/2/31} {\bibfield  {journal} {\bibinfo  {journal}
  {Astrophys.J.Suppl.}\ }\textbf {\bibinfo {volume} {199}},\ \bibinfo {pages}
  {31} (\bibinfo {year} {2012})},\ \Eprint {http://arxiv.org/abs/1108.1435}
  {arXiv:1108.1435 [astro-ph.HE]} \BibitemShut {NoStop}%
\bibitem [{\citenamefont {Linden}\ \emph {et~al.}(2012)\citenamefont {Linden},
  \citenamefont {Lovegrove},\ and\ \citenamefont {Profumo}}]{Linden:2012iv}%
  \BibitemOpen
  \bibfield  {author} {\bibinfo {author} {\bibfnamefont {Tim}\ \bibnamefont
  {Linden}}, \bibinfo {author} {\bibfnamefont {Elizabeth}\ \bibnamefont
  {Lovegrove}}, \ and\ \bibinfo {author} {\bibfnamefont {Stefano}\ \bibnamefont
  {Profumo}},\ }\bibfield  {title} {\enquote {\bibinfo {title} {{The Morphology
  of Hadronic Emission Models for the Gamma-Ray Source at the Galactic
  Center}},}\ }\href {\doibase 10.1088/0004-637X/753/1/41} {\bibfield
  {journal} {\bibinfo  {journal} {Astrophys.J.}\ }\textbf {\bibinfo {volume}
  {753}},\ \bibinfo {pages} {41} (\bibinfo {year} {2012})},\ \Eprint
  {http://arxiv.org/abs/1203.3539} {arXiv:1203.3539 [astro-ph.HE]} \BibitemShut
  {NoStop}%
\bibitem [{\citenamefont {Geringer-Sameth}\ and\ \citenamefont
  {Koushiappas}(2011)}]{GeringerSameth:2011iw}%
  \BibitemOpen
  \bibfield  {author} {\bibinfo {author} {\bibfnamefont {Alex}\ \bibnamefont
  {Geringer-Sameth}}\ and\ \bibinfo {author} {\bibfnamefont {Savvas~M.}\
  \bibnamefont {Koushiappas}},\ }\bibfield  {title} {\enquote {\bibinfo {title}
  {{Exclusion of canonical WIMPs by the joint analysis of Milky Way dwarfs with
  Fermi}},}\ }\href {\doibase 10.1103/PhysRevLett.107.241303} {\bibfield
  {journal} {\bibinfo  {journal} {Phys.Rev.Lett.}\ }\textbf {\bibinfo {volume}
  {107}},\ \bibinfo {pages} {241303} (\bibinfo {year} {2011})},\ \Eprint
  {http://arxiv.org/abs/1108.2914} {arXiv:1108.2914 [astro-ph.CO]} \BibitemShut
  {NoStop}%
\bibitem [{\citenamefont {Ackermann}\ \emph
  {et~al.}(2011{\natexlab{a}})\citenamefont {Ackermann} \emph
  {et~al.}}]{Ackermann:2011wa}%
  \BibitemOpen
  \bibfield  {author} {\bibinfo {author} {\bibfnamefont {M.}~\bibnamefont
  {Ackermann}} \emph {et~al.} (\bibinfo {collaboration} {Fermi-LAT
  collaboration}),\ }\bibfield  {title} {\enquote {\bibinfo {title}
  {{Constraining Dark Matter Models from a Combined Analysis of Milky Way
  Satellites with the Fermi Large Area Telescope}},}\ }\href {\doibase
  10.1103/PhysRevLett.107.241302} {\bibfield  {journal} {\bibinfo  {journal}
  {Phys.Rev.Lett.}\ }\textbf {\bibinfo {volume} {107}},\ \bibinfo {pages}
  {241302} (\bibinfo {year} {2011}{\natexlab{a}})},\ \Eprint
  {http://arxiv.org/abs/1108.3546} {arXiv:1108.3546 [astro-ph.HE]} \BibitemShut
  {NoStop}%
\bibitem [{\citenamefont {Ando}\ and\ \citenamefont
  {Nagai}(2012)}]{Ando:2012vu}%
  \BibitemOpen
  \bibfield  {author} {\bibinfo {author} {\bibfnamefont {Shinichiro}\
  \bibnamefont {Ando}}\ and\ \bibinfo {author} {\bibfnamefont {Daisuke}\
  \bibnamefont {Nagai}},\ }\bibfield  {title} {\enquote {\bibinfo {title}
  {{Fermi-LAT constraints on dark matter annihilation cross section from
  observations of the Fornax cluster}},}\ }\href {\doibase
  10.1088/1475-7516/2012/07/017} {\bibfield  {journal} {\bibinfo  {journal}
  {JCAP}\ }\textbf {\bibinfo {volume} {1207}},\ \bibinfo {pages} {017}
  (\bibinfo {year} {2012})},\ \Eprint {http://arxiv.org/abs/1201.0753}
  {arXiv:1201.0753 [astro-ph.HE]} \BibitemShut {NoStop}%
\bibitem [{\citenamefont {Han}\ \emph {et~al.}(2012{\natexlab{a}})\citenamefont
  {Han} \emph {et~al.}}]{Han:2012uw}%
  \BibitemOpen
  \bibfield  {author} {\bibinfo {author} {\bibfnamefont {Jiaxin}\ \bibnamefont
  {Han}} \emph {et~al.},\ }\bibfield  {title} {\enquote {\bibinfo {title}
  {{Constraining Extended Gamma-ray Emission from Galaxy Clusters}},}\ }\href
  {\doibase 10.1111/j.1365-2966.2012.22080.x} {\bibfield  {journal} {\bibinfo
  {journal} {Mon.Not.Roy.Astron.Soc.}\ }\textbf {\bibinfo {volume} {427}},\
  \bibinfo {pages} {1651--1665} (\bibinfo {year} {2012}{\natexlab{a}})},\
  \Eprint {http://arxiv.org/abs/1207.6749} {arXiv:1207.6749 [astro-ph.CO]}
  \BibitemShut {NoStop}%
\bibitem [{\citenamefont {Mac\'ias}\ \emph {et~al.}(2012)\citenamefont
  {Mac\'ias}, \citenamefont {Gordon}, \citenamefont {Brown},\ and\
  \citenamefont {Adams}}]{virgo}%
  \BibitemOpen
  \bibfield  {author} {\bibinfo {author} {\bibfnamefont {Oscar}\ \bibnamefont
  {Mac\'ias}}, \bibinfo {author} {\bibfnamefont {Chris}\ \bibnamefont
  {Gordon}}, \bibinfo {author} {\bibfnamefont {Anthony~M.}\ \bibnamefont
  {Brown}}, \ and\ \bibinfo {author} {\bibfnamefont {Jenni}\ \bibnamefont
  {Adams}},\ }\bibfield  {title} {\enquote {\bibinfo {title} {{Evaluating the
  Gamma-Ray Evidence for Self-Annihilating Dark Matter from the Virgo
  Cluster}},}\ }\href {\doibase 10.1103/PhysRevD.86.076004} {\bibfield
  {journal} {\bibinfo  {journal} {Phys.Rev.}\ }\textbf {\bibinfo {volume}
  {D86}},\ \bibinfo {pages} {076004} (\bibinfo {year} {2012})},\ \Eprint
  {http://arxiv.org/abs/1207.6257} {arXiv:1207.6257 [astro-ph.HE]} \BibitemShut
  {NoStop}%
\bibitem [{\citenamefont {Goodenough}\ and\ \citenamefont
  {Hooper}(2009)}]{Goodenough:2009gk}%
  \BibitemOpen
  \bibfield  {author} {\bibinfo {author} {\bibfnamefont {Lisa}\ \bibnamefont
  {Goodenough}}\ and\ \bibinfo {author} {\bibfnamefont {Dan}\ \bibnamefont
  {Hooper}},\ }\bibfield  {title} {\enquote {\bibinfo {title} {{Possible
  Evidence For Dark Matter Annihilation In The Inner Milky Way From The Fermi
  Gamma Ray Space Telescope}},}\ }\href@noop {} {\  (\bibinfo {year} {2009})},\
  \Eprint {http://arxiv.org/abs/0910.2998} {arXiv:0910.2998 [hep-ph]}
  \BibitemShut {NoStop}%
\bibitem [{\citenamefont {Hooper}\ and\ \citenamefont
  {Goodenough}(2011)}]{Hooper:2010mq}%
  \BibitemOpen
  \bibfield  {author} {\bibinfo {author} {\bibfnamefont {D.}~\bibnamefont
  {Hooper}}\ and\ \bibinfo {author} {\bibfnamefont {L.}~\bibnamefont
  {Goodenough}},\ }\bibfield  {title} {\enquote {\bibinfo {title} {{Dark Matter
  Annihilation in The Galactic Center As Seen by the Fermi Gamma Ray Space
  Telescope}},}\ }\href {\doibase 10.1016/j.physletb.2011.02.029} {\bibfield
  {journal} {\bibinfo  {journal} {Phys.Lett.}\ }\textbf {\bibinfo {volume}
  {B697}},\ \bibinfo {pages} {412--428} (\bibinfo {year} {2011})},\ \Eprint
  {http://arxiv.org/abs/1010.2752} {arXiv:1010.2752 [hep-ph]} \BibitemShut
  {NoStop}%
\bibitem [{\citenamefont {Boyarsky}\ \emph {et~al.}(2011)\citenamefont
  {Boyarsky}, \citenamefont {Malyshev},\ and\ \citenamefont
  {Ruchayskiy}}]{Boyarsky:2010dr}%
  \BibitemOpen
  \bibfield  {author} {\bibinfo {author} {\bibfnamefont {Alexey}\ \bibnamefont
  {Boyarsky}}, \bibinfo {author} {\bibfnamefont {Denys}\ \bibnamefont
  {Malyshev}}, \ and\ \bibinfo {author} {\bibfnamefont {Oleg}\ \bibnamefont
  {Ruchayskiy}},\ }\bibfield  {title} {\enquote {\bibinfo {title} {{A comment
  on the emission from the Galactic Center as seen by the Fermi telescope}},}\
  }\href {\doibase 10.1016/j.physletb.2011.10.014} {\bibfield  {journal}
  {\bibinfo  {journal} {Phys.Lett.}\ }\textbf {\bibinfo {volume} {B705}},\
  \bibinfo {pages} {165--169} (\bibinfo {year} {2011})},\ \Eprint
  {http://arxiv.org/abs/1012.5839} {arXiv:1012.5839 [hep-ph]} \BibitemShut
  {NoStop}%
\bibitem [{\citenamefont {Hooper}\ and\ \citenamefont
  {Linden}(2011)}]{hooperlinden2011}%
  \BibitemOpen
  \bibfield  {author} {\bibinfo {author} {\bibfnamefont {D.}~\bibnamefont
  {Hooper}}\ and\ \bibinfo {author} {\bibfnamefont {T.}~\bibnamefont
  {Linden}},\ }\bibfield  {title} {\enquote {\bibinfo {title} {{On The Origin
  Of The Gamma Rays From The Galactic Center}},}\ }\href {\doibase
  10.1103/PhysRevD.84.123005} {\bibfield  {journal} {\bibinfo  {journal}
  {Phys.Rev.}\ }\textbf {\bibinfo {volume} {D84}},\ \bibinfo {pages} {123005}
  (\bibinfo {year} {2011})},\ \Eprint {http://arxiv.org/abs/1110.0006}
  {arXiv:1110.0006 [astro-ph.HE]} \BibitemShut {NoStop}%
\bibitem [{\citenamefont {Abazajian}\ and\ \citenamefont
  {Kaplinghat}(2012)}]{AK}%
  \BibitemOpen
  \bibfield  {author} {\bibinfo {author} {\bibfnamefont {Kevork~N.}\
  \bibnamefont {Abazajian}}\ and\ \bibinfo {author} {\bibfnamefont {Manoj}\
  \bibnamefont {Kaplinghat}},\ }\bibfield  {title} {\enquote {\bibinfo {title}
  {{Detection of a Gamma-Ray Source in the Galactic Center Consistent with
  Extended Emission from Dark Matter Annihilation and Concentrated
  Astrophysical Emission}},}\ }\href {\doibase 10.1103/PhysRevD.86.083511}
  {\bibfield  {journal} {\bibinfo  {journal} {Phys.Rev.}\ }\textbf {\bibinfo
  {volume} {D86}},\ \bibinfo {pages} {083511} (\bibinfo {year} {2012})},\
  \Eprint {http://arxiv.org/abs/1207.6047} {arXiv:1207.6047 [astro-ph.HE]}
  \BibitemShut {NoStop}%
\bibitem [{\citenamefont {Abazajian}\ and\ \citenamefont
  {Kaplinghat}(2013)}]{AKerratum}%
  \BibitemOpen
  \bibfield  {author} {\bibinfo {author} {\bibfnamefont {Kevork~N.}\
  \bibnamefont {Abazajian}}\ and\ \bibinfo {author} {\bibfnamefont {Manoj}\
  \bibnamefont {Kaplinghat}},\ }\bibfield  {title} {\enquote {\bibinfo {title}
  {Erratum: Detection of a gamma-ray source in the galactic center consistent
  with extended emission from dark matter annihilation and concentrated
  astrophysical emission [phys. rev. d 86, 083511 (2012)]},}\ }\href {\doibase
  10.1103/PhysRevD.87.129902} {\bibfield  {journal} {\bibinfo  {journal} {Phys.
  Rev. D}\ }\textbf {\bibinfo {volume} {87}},\ \bibinfo {pages} {129902}
  (\bibinfo {year} {2013})}\BibitemShut {NoStop}%
\bibitem [{\citenamefont {Vitale}\ and\ \citenamefont
  {Morselli}(2009)}]{Vitale:2009hr}%
  \BibitemOpen
  \bibfield  {author} {\bibinfo {author} {\bibfnamefont {Vincenzo}\
  \bibnamefont {Vitale}}\ and\ \bibinfo {author} {\bibfnamefont {Aldo}\
  \bibnamefont {Morselli}} (\bibinfo {collaboration} {Fermi/LAT
  Collaboration}),\ }\bibfield  {title} {\enquote {\bibinfo {title} {{Indirect
  Search for Dark Matter from the center of the Milky Way with the Fermi-Large
  Area Telescope}},}\ }\href@noop {} {\  (\bibinfo {year} {2009})},\ \Eprint
  {http://arxiv.org/abs/0912.3828} {arXiv:0912.3828 [astro-ph.HE]} \BibitemShut
  {NoStop}%
\bibitem [{\citenamefont {{Vitale}}\ \emph {et~al.}(2011)\citenamefont
  {{Vitale}}, \citenamefont {{Morselli}},\ and\ \citenamefont {{Fermi/LAT
  Collaboration}}}]{2011NIMPA.630..147V}%
  \BibitemOpen
  \bibfield  {author} {\bibinfo {author} {\bibfnamefont {V.}~\bibnamefont
  {{Vitale}}}, \bibinfo {author} {\bibfnamefont {A.}~\bibnamefont
  {{Morselli}}}, \ and\ \bibinfo {author} {\bibnamefont {{Fermi/LAT
  Collaboration}}},\ }\bibfield  {title} {\enquote {\bibinfo {title} {{Search
  for Dark Matter with Fermi Large Area Telescope: The Galactic Center}},}\
  }\href {\doibase 10.1016/j.nima.2010.06.048} {\bibfield  {journal} {\bibinfo
  {journal} {Nuclear Instruments and Methods in Physics Research A}\ }\textbf
  {\bibinfo {volume} {630}},\ \bibinfo {pages} {147--150} (\bibinfo {year}
  {2011})}\BibitemShut {NoStop}%
\bibitem [{\citenamefont {Abazajian}(2011)}]{Abazajian:2010zy}%
  \BibitemOpen
  \bibfield  {author} {\bibinfo {author} {\bibfnamefont {Kevork~N.}\
  \bibnamefont {Abazajian}},\ }\bibfield  {title} {\enquote {\bibinfo {title}
  {{The Consistency of Fermi-LAT Observations of the Galactic Center with a
  Millisecond Pulsar Population in the Central Stellar Cluster}},}\ }\href
  {\doibase 10.1088/1475-7516/2011/03/010} {\bibfield  {journal} {\bibinfo
  {journal} {JCAP}\ }\textbf {\bibinfo {volume} {1103}},\ \bibinfo {pages}
  {010} (\bibinfo {year} {2011})},\ \Eprint {http://arxiv.org/abs/1011.4275}
  {arXiv:1011.4275 [astro-ph.HE]} \BibitemShut {NoStop}%
\bibitem [{\citenamefont {Yusef-Zadeh}\ \emph {et~al.}(2013)\citenamefont
  {Yusef-Zadeh} \emph {et~al.}}]{yusef-zadehhewittwardle2013}%
  \BibitemOpen
  \bibfield  {author} {\bibinfo {author} {\bibfnamefont {F.}~\bibnamefont
  {Yusef-Zadeh}} \emph {et~al.},\ }\bibfield  {title} {\enquote {\bibinfo
  {title} {{Interacting Cosmic Rays with Molecular Clouds: A Bremsstrahlung
  Origin of Diffuse High Energy Emission from the Inner 2deg by 1deg of the
  Galactic Center}},}\ }\href {\doibase 10.1088/0004-637X/762/1/33} {\bibfield
  {journal} {\bibinfo  {journal} {Astrophys.J.}\ }\textbf {\bibinfo {volume}
  {762}},\ \bibinfo {pages} {33} (\bibinfo {year} {2013})},\ \Eprint
  {http://arxiv.org/abs/1206.6882} {arXiv:1206.6882 [astro-ph.HE]} \BibitemShut
  {NoStop}%
\bibitem [{\citenamefont {Ackermann}\ \emph
  {et~al.}(2012{\natexlab{a}})\citenamefont {Ackermann} \emph
  {et~al.}}]{FermiInstrument}%
  \BibitemOpen
  \bibfield  {author} {\bibinfo {author} {\bibfnamefont {M.}~\bibnamefont
  {Ackermann}} \emph {et~al.} (\bibinfo {collaboration} {Fermi-LAT
  Collaboration}),\ }\bibfield  {title} {\enquote {\bibinfo {title} {{The Fermi
  Large Area Telescope On Orbit: Event Classification, Instrument Response
  Functions, and Calibration}},}\ }\href {\doibase 10.1088/0067-0049/203/1/4}
  {\bibfield  {journal} {\bibinfo  {journal} {Astrophys.J.Suppl.}\ }\textbf
  {\bibinfo {volume} {203}},\ \bibinfo {pages} {4} (\bibinfo {year}
  {2012}{\natexlab{a}})},\ \Eprint {http://arxiv.org/abs/1206.1896}
  {arXiv:1206.1896 [astro-ph.IM]} \BibitemShut {NoStop}%
\bibitem [{\citenamefont {Mattox}\ \emph {et~al.}(1996)\citenamefont {Mattox},
  \citenamefont {Bertsch}, \citenamefont {Chiang}, \citenamefont {Dingus},
  \citenamefont {Digel} \emph {et~al.}}]{mattox}%
  \BibitemOpen
  \bibfield  {author} {\bibinfo {author} {\bibfnamefont {J.R.}\ \bibnamefont
  {Mattox}}, \bibinfo {author} {\bibfnamefont {D.L.}\ \bibnamefont {Bertsch}},
  \bibinfo {author} {\bibfnamefont {J.}~\bibnamefont {Chiang}}, \bibinfo
  {author} {\bibfnamefont {B.L.}\ \bibnamefont {Dingus}}, \bibinfo {author}
  {\bibfnamefont {S.W.}\ \bibnamefont {Digel}},  \emph {et~al.},\ }\bibfield
  {title} {\enquote {\bibinfo {title} {{The Likelihood Analysis of EGRET
  Data}},}\ }\href@noop {} {\bibfield  {journal} {\bibinfo  {journal}
  {Astrophys.J.}\ }\textbf {\bibinfo {volume} {461}},\ \bibinfo {pages} {396}
  (\bibinfo {year} {1996})}\BibitemShut {NoStop}%
\bibitem [{\citenamefont {Wilks}(1938)}]{wilks}%
  \BibitemOpen
  \bibfield  {author} {\bibinfo {author} {\bibfnamefont {S.~S.}\ \bibnamefont
  {Wilks}},\ }\bibfield  {title} {\enquote {\bibinfo {title} {{The Large-Sample
  Distribution of the Likelihood Ratio for Testing Composite Hypotheses}},}\
  }\href {\doibase 10.1214/aoms/1177732360} {\bibfield  {journal} {\bibinfo
  {journal} {Ann.Math.Stat.}\ }\textbf {\bibinfo {volume} {9}},\ \bibinfo
  {pages} {60} (\bibinfo {year} {1938})}\BibitemShut {NoStop}%
\bibitem [{\citenamefont {Han}\ \emph {et~al.}(2012{\natexlab{b}})\citenamefont
  {Han}, \citenamefont {Frenk}, \citenamefont {Eke}, \citenamefont {Gao},\ and\
  \citenamefont {White}}]{han}%
  \BibitemOpen
  \bibfield  {author} {\bibinfo {author} {\bibfnamefont {Jiaxin}\ \bibnamefont
  {Han}}, \bibinfo {author} {\bibfnamefont {Carlos~S.}\ \bibnamefont {Frenk}},
  \bibinfo {author} {\bibfnamefont {Vincent~R.}\ \bibnamefont {Eke}}, \bibinfo
  {author} {\bibfnamefont {Liang}\ \bibnamefont {Gao}}, \ and\ \bibinfo
  {author} {\bibfnamefont {Simon~D.M.}\ \bibnamefont {White}},\ }\bibfield
  {title} {\enquote {\bibinfo {title} {{Evidence for extended gamma-ray
  emission from galaxy clusters}},}\ }\href@noop {} {\  (\bibinfo {year}
  {2012}{\natexlab{b}})},\ \Eprint {http://arxiv.org/abs/1201.1003}
  {arXiv:1201.1003 [astro-ph.HE]} \BibitemShut {NoStop}%
\bibitem [{\citenamefont {Han}\ \emph {et~al.}(2012{\natexlab{c}})\citenamefont
  {Han}, \citenamefont {Frenk}, \citenamefont {Eke}, \citenamefont {Gao},
  \citenamefont {White} \emph {et~al.}}]{han2}%
  \BibitemOpen
  \bibfield  {author} {\bibinfo {author} {\bibfnamefont {Jiaxin}\ \bibnamefont
  {Han}}, \bibinfo {author} {\bibfnamefont {Carlos~S.}\ \bibnamefont {Frenk}},
  \bibinfo {author} {\bibfnamefont {Vincent~R.}\ \bibnamefont {Eke}}, \bibinfo
  {author} {\bibfnamefont {Liang}\ \bibnamefont {Gao}}, \bibinfo {author}
  {\bibfnamefont {Simon~D.M.}\ \bibnamefont {White}},  \emph {et~al.},\
  }\bibfield  {title} {\enquote {\bibinfo {title} {{Constraining Extended
  Gamma-ray Emission from Galaxy Clusters}},}\ }\href {\doibase
  10.1111/j.1365-2966.2012.22080.x} {\bibfield  {journal} {\bibinfo  {journal}
  {Mon.Not.Roy.Astron.Soc.}\ }\textbf {\bibinfo {volume} {427}},\ \bibinfo
  {pages} {1651--1665} (\bibinfo {year} {2012}{\natexlab{c}})},\ \Eprint
  {http://arxiv.org/abs/1207.6749} {arXiv:1207.6749 [astro-ph.CO]} \BibitemShut
  {NoStop}%
\bibitem [{\citenamefont {Baltz}\ \emph
  {et~al.}(2008{\natexlab{b}})\citenamefont {Baltz}, \citenamefont {Berenji},
  \citenamefont {Bertone}, \citenamefont {Bergstrom}, \citenamefont {Bloom}
  \emph {et~al.}}]{Baltz}%
  \BibitemOpen
  \bibfield  {author} {\bibinfo {author} {\bibfnamefont {E.A.}\ \bibnamefont
  {Baltz}}, \bibinfo {author} {\bibfnamefont {B.}~\bibnamefont {Berenji}},
  \bibinfo {author} {\bibfnamefont {G.}~\bibnamefont {Bertone}}, \bibinfo
  {author} {\bibfnamefont {L.}~\bibnamefont {Bergstrom}}, \bibinfo {author}
  {\bibfnamefont {E.}~\bibnamefont {Bloom}},  \emph {et~al.},\ }\bibfield
  {title} {\enquote {\bibinfo {title} {{Pre-launch estimates for GLAST
  sensitivity to Dark Matter annihilation signals}},}\ }\href {\doibase
  10.1088/1475-7516/2008/07/013} {\bibfield  {journal} {\bibinfo  {journal}
  {JCAP}\ }\textbf {\bibinfo {volume} {0807}},\ \bibinfo {pages} {013}
  (\bibinfo {year} {2008}{\natexlab{b}})},\ \Eprint
  {http://arxiv.org/abs/0806.2911} {arXiv:0806.2911 [astro-ph]} \BibitemShut
  {NoStop}%
\bibitem [{\citenamefont {Rott}(2013)}]{Rott}%
  \BibitemOpen
  \bibfield  {author} {\bibinfo {author} {\bibfnamefont {Carsten}\ \bibnamefont
  {Rott}},\ }\bibfield  {title} {\enquote {\bibinfo {title} {{Review of
  Indirect WIMP Search Experiments}},}\ }\href {\doibase
  10.1016/j.nuclphysbps.2013.04.040} {\bibfield  {journal} {\bibinfo  {journal}
  {Nucl.Phys.Proc.Suppl.}\ }\textbf {\bibinfo {volume} {235-236}},\ \bibinfo
  {pages} {413--420} (\bibinfo {year} {2013})},\ \Eprint
  {http://arxiv.org/abs/1210.4161} {arXiv:1210.4161 [astro-ph.HE]} \BibitemShut
  {NoStop}%
\bibitem [{\citenamefont {Bergstrom}\ \emph {et~al.}(1998)\citenamefont
  {Bergstrom}, \citenamefont {Ullio},\ and\ \citenamefont
  {Buckley}}]{Bergstrom}%
  \BibitemOpen
  \bibfield  {author} {\bibinfo {author} {\bibfnamefont {Lars}\ \bibnamefont
  {Bergstrom}}, \bibinfo {author} {\bibfnamefont {Piero}\ \bibnamefont
  {Ullio}}, \ and\ \bibinfo {author} {\bibfnamefont {James~H.}\ \bibnamefont
  {Buckley}},\ }\bibfield  {title} {\enquote {\bibinfo {title} {{Observability
  of gamma-rays from dark matter neutralino annihilations in the Milky Way
  halo}},}\ }\href {\doibase 10.1016/S0927-6505(98)00015-2} {\bibfield
  {journal} {\bibinfo  {journal} {Astropart.Phys.}\ }\textbf {\bibinfo {volume}
  {9}},\ \bibinfo {pages} {137--162} (\bibinfo {year} {1998})},\ \Eprint
  {http://arxiv.org/abs/astro-ph/9712318} {arXiv:astro-ph/9712318 [astro-ph]}
  \BibitemShut {NoStop}%
\bibitem [{\citenamefont {Navarro}\ \emph {et~al.}(1996)\citenamefont
  {Navarro}, \citenamefont {Frenk},\ and\ \citenamefont
  {White}}]{navarrofrenkwhite1996}%
  \BibitemOpen
  \bibfield  {author} {\bibinfo {author} {\bibfnamefont {Julio~F.}\
  \bibnamefont {Navarro}}, \bibinfo {author} {\bibfnamefont {Carlos~S.}\
  \bibnamefont {Frenk}}, \ and\ \bibinfo {author} {\bibfnamefont {Simon~D.M.}\
  \bibnamefont {White}},\ }\bibfield  {title} {\enquote {\bibinfo {title} {{The
  Structure of cold dark matter halos}},}\ }\href {\doibase 10.1086/177173}
  {\bibfield  {journal} {\bibinfo  {journal} {Astrophys.J.}\ }\textbf {\bibinfo
  {volume} {462}},\ \bibinfo {pages} {563--575} (\bibinfo {year} {1996})},\
  \Eprint {http://arxiv.org/abs/astro-ph/9508025} {arXiv:astro-ph/9508025
  [astro-ph]} \BibitemShut {NoStop}%
\bibitem [{\citenamefont {{Klypin}}\ \emph {et~al.}(2002)\citenamefont
  {{Klypin}}, \citenamefont {{Zhao}},\ and\ \citenamefont
  {{Somerville}}}]{klypinzhaosomerville2002}%
  \BibitemOpen
  \bibfield  {author} {\bibinfo {author} {\bibfnamefont {A.}~\bibnamefont
  {{Klypin}}}, \bibinfo {author} {\bibfnamefont {H.}~\bibnamefont {{Zhao}}}, \
  and\ \bibinfo {author} {\bibfnamefont {R.~S.}\ \bibnamefont {{Somerville}}},\
  }\bibfield  {title} {\enquote {\bibinfo {title} {{{$\Lambda$}CDM-based Models
  for the Milky Way and M31. I. Dynamical Models}},}\ }\href {\doibase
  10.1086/340656} {\bibfield  {journal} {\bibinfo  {journal} {\apj}\ }\textbf
  {\bibinfo {volume} {573}},\ \bibinfo {pages} {597--613} (\bibinfo {year}
  {2002})},\ \Eprint {http://arxiv.org/abs/arXiv:astro-ph/0110390}
  {arXiv:astro-ph/0110390} \BibitemShut {NoStop}%
\bibitem [{\citenamefont {{Rolke}}\ \emph {et~al.}(2005)\citenamefont
  {{Rolke}}, \citenamefont {{L{\'o}pez}},\ and\ \citenamefont
  {{Conrad}}}]{rolkelopezconrad2005}%
  \BibitemOpen
  \bibfield  {author} {\bibinfo {author} {\bibfnamefont {W.~A.}\ \bibnamefont
  {{Rolke}}}, \bibinfo {author} {\bibfnamefont {A.~M.}\ \bibnamefont
  {{L{\'o}pez}}}, \ and\ \bibinfo {author} {\bibfnamefont {J.}~\bibnamefont
  {{Conrad}}},\ }\bibfield  {title} {\enquote {\bibinfo {title} {{Limits and
  confidence intervals in the presence of nuisance parameters}},}\ }\href
  {\doibase 10.1016/j.nima.2005.05.068} {\bibfield  {journal} {\bibinfo
  {journal} {Nuclear Instruments and Methods in Physics Research A}\ }\textbf
  {\bibinfo {volume} {551}},\ \bibinfo {pages} {493--503} (\bibinfo {year}
  {2005})},\ \Eprint {http://arxiv.org/abs/arXiv:physics/0403059}
  {arXiv:physics/0403059} \BibitemShut {NoStop}%
\bibitem [{\citenamefont {Iocco}\ \emph {et~al.}(2011)\citenamefont {Iocco},
  \citenamefont {Pato}, \citenamefont {Bertone},\ and\ \citenamefont
  {Jetzer}}]{ioccopatobertone2011}%
  \BibitemOpen
  \bibfield  {author} {\bibinfo {author} {\bibfnamefont {Fabio}\ \bibnamefont
  {Iocco}}, \bibinfo {author} {\bibfnamefont {Miguel}\ \bibnamefont {Pato}},
  \bibinfo {author} {\bibfnamefont {Gianfranco}\ \bibnamefont {Bertone}}, \
  and\ \bibinfo {author} {\bibfnamefont {Philippe}\ \bibnamefont {Jetzer}},\
  }\bibfield  {title} {\enquote {\bibinfo {title} {{Dark Matter distribution in
  the Milky Way: microlensing and dynamical constraints}},}\ }\href {\doibase
  10.1088/1475-7516/2011/11/029} {\bibfield  {journal} {\bibinfo  {journal}
  {JCAP}\ }\textbf {\bibinfo {volume} {1111}},\ \bibinfo {pages} {029}
  (\bibinfo {year} {2011})},\ \Eprint {http://arxiv.org/abs/1107.5810}
  {arXiv:1107.5810 [astro-ph.GA]} \BibitemShut {NoStop}%
\bibitem [{\citenamefont {Jeltema}\ and\ \citenamefont
  {Profumo}(2008)}]{profumo1}%
  \BibitemOpen
  \bibfield  {author} {\bibinfo {author} {\bibfnamefont {Tesla~E.}\
  \bibnamefont {Jeltema}}\ and\ \bibinfo {author} {\bibfnamefont {Stefano}\
  \bibnamefont {Profumo}},\ }\bibfield  {title} {\enquote {\bibinfo {title}
  {{Fitting the Gamma-Ray Spectrum from Dark Matter with DMFIT: GLAST and the
  Galactic Center Region}},}\ }\href {\doibase 10.1088/1475-7516/2008/11/003}
  {\bibfield  {journal} {\bibinfo  {journal} {JCAP}\ }\textbf {\bibinfo
  {volume} {0811}},\ \bibinfo {pages} {003} (\bibinfo {year} {2008})},\ \Eprint
  {http://arxiv.org/abs/0808.2641} {arXiv:0808.2641 [astro-ph]} \BibitemShut
  {NoStop}%
\bibitem [{\citenamefont {Gondolo}\ \emph {et~al.}(2004)\citenamefont
  {Gondolo}, \citenamefont {Edsjo}, \citenamefont {Ullio}, \citenamefont
  {Bergstrom}, \citenamefont {Schelke} \emph {et~al.}}]{darksusy}%
  \BibitemOpen
  \bibfield  {author} {\bibinfo {author} {\bibfnamefont {P.}~\bibnamefont
  {Gondolo}}, \bibinfo {author} {\bibfnamefont {J.}~\bibnamefont {Edsjo}},
  \bibinfo {author} {\bibfnamefont {P.}~\bibnamefont {Ullio}}, \bibinfo
  {author} {\bibfnamefont {L.}~\bibnamefont {Bergstrom}}, \bibinfo {author}
  {\bibfnamefont {Mia}\ \bibnamefont {Schelke}},  \emph {et~al.},\ }\bibfield
  {title} {\enquote {\bibinfo {title} {{DarkSUSY: Computing supersymmetric dark
  matter properties numerically}},}\ }\href {\doibase
  10.1088/1475-7516/2004/07/008} {\bibfield  {journal} {\bibinfo  {journal}
  {JCAP}\ }\textbf {\bibinfo {volume} {0407}},\ \bibinfo {pages} {008}
  (\bibinfo {year} {2004})},\ \Eprint {http://arxiv.org/abs/astro-ph/0406204}
  {arXiv:astro-ph/0406204 [astro-ph]} \BibitemShut {NoStop}%
\bibitem [{\citenamefont {Sjostrand}\ \emph {et~al.}(2006)\citenamefont
  {Sjostrand}, \citenamefont {Mrenna},\ and\ \citenamefont {Skands}}]{pythia6}%
  \BibitemOpen
  \bibfield  {author} {\bibinfo {author} {\bibfnamefont {Torbjorn}\
  \bibnamefont {Sjostrand}}, \bibinfo {author} {\bibfnamefont {Stephen}\
  \bibnamefont {Mrenna}}, \ and\ \bibinfo {author} {\bibfnamefont {Peter~Z.}\
  \bibnamefont {Skands}},\ }\bibfield  {title} {\enquote {\bibinfo {title}
  {{PYTHIA 6.4 Physics and Manual}},}\ }\href {\doibase
  10.1088/1126-6708/2006/05/026} {\bibfield  {journal} {\bibinfo  {journal}
  {JHEP}\ }\textbf {\bibinfo {volume} {0605}},\ \bibinfo {pages} {026}
  (\bibinfo {year} {2006})},\ \Eprint {http://arxiv.org/abs/hep-ph/0603175}
  {arXiv:hep-ph/0603175 [hep-ph]} \BibitemShut {NoStop}%
\bibitem [{\citenamefont {Cembranos}\ \emph {et~al.}(2013)\citenamefont
  {Cembranos}, \citenamefont {de~la Cruz-Dombriz}, \citenamefont {Gammaldi},
  \citenamefont {Lineros},\ and\ \citenamefont {Maroto}}]{Cembranos}%
  \BibitemOpen
  \bibfield  {author} {\bibinfo {author} {\bibfnamefont {J.A.R.}\ \bibnamefont
  {Cembranos}}, \bibinfo {author} {\bibfnamefont {A.}~\bibnamefont {de~la
  Cruz-Dombriz}}, \bibinfo {author} {\bibfnamefont {V.}~\bibnamefont
  {Gammaldi}}, \bibinfo {author} {\bibfnamefont {R.A.}\ \bibnamefont
  {Lineros}}, \ and\ \bibinfo {author} {\bibfnamefont {A.L.}\ \bibnamefont
  {Maroto}},\ }\bibfield  {title} {\enquote {\bibinfo {title} {{Reliability of
  Monte Carlo event generators for gamma ray dark matter searches}},}\
  }\href@noop {} {\  (\bibinfo {year} {2013})},\ \Eprint
  {http://arxiv.org/abs/1305.2124} {arXiv:1305.2124 [hep-ph]} \BibitemShut
  {NoStop}%
\bibitem [{\citenamefont {Sjostrand}\ \emph {et~al.}(2008)\citenamefont
  {Sjostrand}, \citenamefont {Mrenna},\ and\ \citenamefont {Skands}}]{pythia8}%
  \BibitemOpen
  \bibfield  {author} {\bibinfo {author} {\bibfnamefont {Torbjorn}\
  \bibnamefont {Sjostrand}}, \bibinfo {author} {\bibfnamefont {Stephen}\
  \bibnamefont {Mrenna}}, \ and\ \bibinfo {author} {\bibfnamefont {Peter~Z.}\
  \bibnamefont {Skands}},\ }\bibfield  {title} {\enquote {\bibinfo {title} {{A
  Brief Introduction to PYTHIA 8.1}},}\ }\href {\doibase
  10.1016/j.cpc.2008.01.036} {\bibfield  {journal} {\bibinfo  {journal}
  {Comput.Phys.Commun.}\ }\textbf {\bibinfo {volume} {178}},\ \bibinfo {pages}
  {852--867} (\bibinfo {year} {2008})},\ \Eprint
  {http://arxiv.org/abs/0710.3820} {arXiv:0710.3820 [hep-ph]} \BibitemShut
  {NoStop}%
\bibitem [{\citenamefont {Cirelli}\ \emph {et~al.}(2011)\citenamefont
  {Cirelli}, \citenamefont {Corcella}, \citenamefont {Hektor}, \citenamefont
  {Hutsi}, \citenamefont {Kadastik} \emph {et~al.}}]{Cirelli}%
  \BibitemOpen
  \bibfield  {author} {\bibinfo {author} {\bibfnamefont {Marco}\ \bibnamefont
  {Cirelli}}, \bibinfo {author} {\bibfnamefont {Gennaro}\ \bibnamefont
  {Corcella}}, \bibinfo {author} {\bibfnamefont {Andi}\ \bibnamefont {Hektor}},
  \bibinfo {author} {\bibfnamefont {Gert}\ \bibnamefont {Hutsi}}, \bibinfo
  {author} {\bibfnamefont {Mario}\ \bibnamefont {Kadastik}},  \emph {et~al.},\
  }\bibfield  {title} {\enquote {\bibinfo {title} {{PPPC 4 DM ID: A Poor
  Particle Physicist Cookbook for Dark Matter Indirect Detection}},}\ }\href
  {\doibase 10.1088/1475-7516/2012/10/E01, 10.1088/1475-7516/2011/03/051}
  {\bibfield  {journal} {\bibinfo  {journal} {JCAP}\ }\textbf {\bibinfo
  {volume} {1103}},\ \bibinfo {pages} {051} (\bibinfo {year} {2011})},\ \Eprint
  {http://arxiv.org/abs/1012.4515} {arXiv:1012.4515 [hep-ph]} \BibitemShut
  {NoStop}%
\bibitem [{\citenamefont {Kachelriess}\ \emph {et~al.}(2009)\citenamefont
  {Kachelriess}, \citenamefont {Serpico},\ and\ \citenamefont
  {Solberg}}]{Kachelriess}%
  \BibitemOpen
  \bibfield  {author} {\bibinfo {author} {\bibfnamefont {M.}~\bibnamefont
  {Kachelriess}}, \bibinfo {author} {\bibfnamefont {P.D.}\ \bibnamefont
  {Serpico}}, \ and\ \bibinfo {author} {\bibfnamefont {M.~Aa.}\ \bibnamefont
  {Solberg}},\ }\bibfield  {title} {\enquote {\bibinfo {title} {{On the role of
  electroweak bremsstrahlung for indirect dark matter signatures}},}\ }\href
  {\doibase 10.1103/PhysRevD.80.123533} {\bibfield  {journal} {\bibinfo
  {journal} {Phys.Rev.}\ }\textbf {\bibinfo {volume} {D80}},\ \bibinfo {pages}
  {123533} (\bibinfo {year} {2009})},\ \Eprint {http://arxiv.org/abs/0911.0001}
  {arXiv:0911.0001 [hep-ph]} \BibitemShut {NoStop}%
\bibitem [{\citenamefont {Ciafaloni}\ \emph {et~al.}(2011)\citenamefont
  {Ciafaloni}, \citenamefont {Comelli}, \citenamefont {Riotto}, \citenamefont
  {Sala}, \citenamefont {Strumia} \emph {et~al.}}]{Ciafaloni}%
  \BibitemOpen
  \bibfield  {author} {\bibinfo {author} {\bibfnamefont {Paolo}\ \bibnamefont
  {Ciafaloni}}, \bibinfo {author} {\bibfnamefont {Denis}\ \bibnamefont
  {Comelli}}, \bibinfo {author} {\bibfnamefont {Antonio}\ \bibnamefont
  {Riotto}}, \bibinfo {author} {\bibfnamefont {Filippo}\ \bibnamefont {Sala}},
  \bibinfo {author} {\bibfnamefont {Alessandro}\ \bibnamefont {Strumia}},
  \emph {et~al.},\ }\bibfield  {title} {\enquote {\bibinfo {title} {{Weak
  Corrections are Relevant for Dark Matter Indirect Detection}},}\ }\href
  {\doibase 10.1088/1475-7516/2011/03/019} {\bibfield  {journal} {\bibinfo
  {journal} {JCAP}\ }\textbf {\bibinfo {volume} {1103}},\ \bibinfo {pages}
  {019} (\bibinfo {year} {2011})},\ \Eprint {http://arxiv.org/abs/1009.0224}
  {arXiv:1009.0224 [hep-ph]} \BibitemShut {NoStop}%
\bibitem [{\citenamefont {Ackermann}\ \emph
  {et~al.}(2012{\natexlab{b}})\citenamefont {Ackermann} \emph
  {et~al.}}]{ackermannajelloatwood2012}%
  \BibitemOpen
  \bibfield  {author} {\bibinfo {author} {\bibfnamefont {M.}~\bibnamefont
  {Ackermann}} \emph {et~al.},\ }\bibfield  {title} {\enquote {\bibinfo {title}
  {Fermi-lat observations of the diffuse gamma-ray emission: Implications for
  cosmic rays and the interstellar medium},}\ }\href
  {http://stacks.iop.org/0004-637X/750/i=1/a=3} {\bibfield  {journal} {\bibinfo
   {journal} {The Astrophysical Journal}\ }\textbf {\bibinfo {volume} {750}},\
  \bibinfo {pages} {3} (\bibinfo {year} {2012}{\natexlab{b}})}\BibitemShut
  {NoStop}%
\bibitem [{\citenamefont {Abdo}\ \emph {et~al.}(2010)\citenamefont {Abdo},
  \citenamefont {Ackermann}, \citenamefont {Ajello}, \citenamefont {Baldini},
  \citenamefont {Ballet} \emph {et~al.}}]{supernovaw49b}%
  \BibitemOpen
  \bibfield  {author} {\bibinfo {author} {\bibfnamefont {A.A.}\ \bibnamefont
  {Abdo}}, \bibinfo {author} {\bibfnamefont {M.}~\bibnamefont {Ackermann}},
  \bibinfo {author} {\bibfnamefont {M.}~\bibnamefont {Ajello}}, \bibinfo
  {author} {\bibfnamefont {L.}~\bibnamefont {Baldini}}, \bibinfo {author}
  {\bibfnamefont {J.}~\bibnamefont {Ballet}},  \emph {et~al.},\ }\bibfield
  {title} {\enquote {\bibinfo {title} {{Fermi-LAT Study of Gamma-ray Emission
  in the Direction of Supernova Remnant W49B}},}\ }\href {\doibase
  10.1088/0004-637X/722/2/1303} {\bibfield  {journal} {\bibinfo  {journal}
  {Astrophys.J.}\ }\textbf {\bibinfo {volume} {722}},\ \bibinfo {pages}
  {1303--1311} (\bibinfo {year} {2010})},\ \Eprint
  {http://arxiv.org/abs/1008.4190} {arXiv:1008.4190 [astro-ph.HE]} \BibitemShut
  {NoStop}%
\bibitem [{\citenamefont {Hooper}\ \emph {et~al.}(2013)\citenamefont {Hooper},
  \citenamefont {Cholis}, \citenamefont {Linden}, \citenamefont
  {Siegal-Gaskins},\ and\ \citenamefont {Slatyer}}]{HooperP}%
  \BibitemOpen
  \bibfield  {author} {\bibinfo {author} {\bibfnamefont {Dan}\ \bibnamefont
  {Hooper}}, \bibinfo {author} {\bibfnamefont {Ilias}\ \bibnamefont {Cholis}},
  \bibinfo {author} {\bibfnamefont {Tim}\ \bibnamefont {Linden}}, \bibinfo
  {author} {\bibfnamefont {Jennifer}\ \bibnamefont {Siegal-Gaskins}}, \ and\
  \bibinfo {author} {\bibfnamefont {Tracy}\ \bibnamefont {Slatyer}},\
  }\bibfield  {title} {\enquote {\bibinfo {title} {{Pulsars Cannot Account for
  the Inner Galaxy's GeV Excess}},}\ }\href@noop {} {\  (\bibinfo {year}
  {2013})},\ \Eprint {http://arxiv.org/abs/1305.0830} {arXiv:1305.0830
  [astro-ph.HE]} \BibitemShut {NoStop}%
\bibitem [{\citenamefont {James}\ and\ \citenamefont {Roos}(1975)}]{minuit}%
  \BibitemOpen
  \bibfield  {author} {\bibinfo {author} {\bibfnamefont {F.}~\bibnamefont
  {James}}\ and\ \bibinfo {author} {\bibfnamefont {M.}~\bibnamefont {Roos}},\
  }\bibfield  {title} {\enquote {\bibinfo {title} {{Minuit: A System for
  Function Minimization and Analysis of the Parameter Errors and
  Correlations}},}\ }\href {\doibase 10.1016/0010-4655(75)90039-9} {\bibfield
  {journal} {\bibinfo  {journal} {Comput.Phys.Commun.}\ }\textbf {\bibinfo
  {volume} {10}},\ \bibinfo {pages} {343--367} (\bibinfo {year}
  {1975})}\BibitemShut {NoStop}%
\bibitem [{\citenamefont {Beringer}\ \emph {et~al.}(2012)\citenamefont
  {Beringer} \emph {et~al.}}]{particledatagroup}%
  \BibitemOpen
  \bibfield  {author} {\bibinfo {author} {\bibfnamefont {J.}~\bibnamefont
  {Beringer}} \emph {et~al.} (\bibinfo {collaboration} {Particle Data Group}),\
  }\bibfield  {title} {\enquote {\bibinfo {title} {Review of particle
  physics},}\ }\href {\doibase 10.1103/PhysRevD.86.010001} {\bibfield
  {journal} {\bibinfo  {journal} {Phys. Rev. D}\ }\textbf {\bibinfo {volume}
  {86}},\ \bibinfo {pages} {010001} (\bibinfo {year} {2012})}\BibitemShut
  {NoStop}%
\bibitem [{\citenamefont {Ackermann}\ \emph
  {et~al.}(2011{\natexlab{b}})\citenamefont {Ackermann} \emph
  {et~al.}}]{dwarfs}%
  \BibitemOpen
  \bibfield  {author} {\bibinfo {author} {\bibfnamefont {M.}~\bibnamefont
  {Ackermann}} \emph {et~al.} (\bibinfo {collaboration} {Fermi-LAT
  collaboration}),\ }\bibfield  {title} {\enquote {\bibinfo {title}
  {{Constraining Dark Matter Models from a Combined Analysis of Milky Way
  Satellites with the Fermi Large Area Telescope}},}\ }\href {\doibase
  10.1103/PhysRevLett.107.241302} {\bibfield  {journal} {\bibinfo  {journal}
  {Phys.Rev.Lett.}\ }\textbf {\bibinfo {volume} {107}},\ \bibinfo {pages}
  {241302} (\bibinfo {year} {2011}{\natexlab{b}})},\ \Eprint
  {http://arxiv.org/abs/1108.3546} {arXiv:1108.3546 [astro-ph.HE]} \BibitemShut
  {NoStop}%
\bibitem [{\citenamefont {Profumo}(2013)}]{ProfumoTasi}%
  \BibitemOpen
  \bibfield  {author} {\bibinfo {author} {\bibfnamefont {Stefano}\ \bibnamefont
  {Profumo}},\ }\bibfield  {title} {\enquote {\bibinfo {title} {{TASI 2012
  Lectures on Astrophysical Probes of Dark Matter}},}\ }\href@noop {} {\
  (\bibinfo {year} {2013})},\ \Eprint {http://arxiv.org/abs/1301.0952}
  {arXiv:1301.0952 [hep-ph]} \BibitemShut {NoStop}%
\bibitem [{\citenamefont {Steigman}\ \emph {et~al.}(2012)\citenamefont
  {Steigman}, \citenamefont {Dasgupta},\ and\ \citenamefont
  {Beacom}}]{steigmandasguptabeacom2012}%
  \BibitemOpen
  \bibfield  {author} {\bibinfo {author} {\bibfnamefont {Gary}\ \bibnamefont
  {Steigman}}, \bibinfo {author} {\bibfnamefont {Basudeb}\ \bibnamefont
  {Dasgupta}}, \ and\ \bibinfo {author} {\bibfnamefont {John~F.}\ \bibnamefont
  {Beacom}},\ }\bibfield  {title} {\enquote {\bibinfo {title} {{Precise Relic
  WIMP Abundance and its Impact on Searches for Dark Matter Annihilation}},}\
  }\href {\doibase 10.1103/PhysRevD.86.023506} {\bibfield  {journal} {\bibinfo
  {journal} {Phys.Rev.}\ }\textbf {\bibinfo {volume} {D86}},\ \bibinfo {pages}
  {023506} (\bibinfo {year} {2012})},\ \Eprint {http://arxiv.org/abs/1204.3622}
  {arXiv:1204.3622 [hep-ph]} \BibitemShut {NoStop}%
\bibitem [{\citenamefont {Hooper}\ \emph {et~al.}(2012)\citenamefont {Hooper},
  \citenamefont {Kelso},\ and\ \citenamefont
  {Queiroz}}]{hooperkelsoqueiroz2012}%
  \BibitemOpen
  \bibfield  {author} {\bibinfo {author} {\bibfnamefont {Dan}\ \bibnamefont
  {Hooper}}, \bibinfo {author} {\bibfnamefont {Chris}\ \bibnamefont {Kelso}}, \
  and\ \bibinfo {author} {\bibfnamefont {Farinaldo~S.}\ \bibnamefont
  {Queiroz}},\ }\bibfield  {title} {\enquote {\bibinfo {title} {{Stringent and
  Robust Constraints on the Dark Matter Annihilation Cross Section From the
  Region of the Galactic Center}},}\ }\href@noop {} {\  (\bibinfo {year}
  {2012})},\ \Eprint {http://arxiv.org/abs/1209.3015} {arXiv:1209.3015
  [astro-ph.HE]} \BibitemShut {NoStop}%
\bibitem [{\citenamefont {{Aharonian}}\ \emph {et~al.}(2006)\citenamefont
  {{Aharonian}} \emph {et~al.}}]{Aharonian:2006au}%
  \BibitemOpen
  \bibfield  {author} {\bibinfo {author} {\bibfnamefont {F.}~\bibnamefont
  {{Aharonian}}} \emph {et~al.},\ }\bibfield  {title} {\enquote {\bibinfo
  {title} {{Discovery of very-high-energy {$\gamma$}-rays from the Galactic
  Centre ridge}},}\ }\href {\doibase 10.1038/nature04467} {\bibfield  {journal}
  {\bibinfo  {journal} {\nat}\ }\textbf {\bibinfo {volume} {439}},\ \bibinfo
  {pages} {695--698} (\bibinfo {year} {2006})},\ \Eprint
  {http://arxiv.org/abs/arXiv:astro-ph/0603021} {arXiv:astro-ph/0603021}
  \BibitemShut {NoStop}%
\bibitem [{\citenamefont {Hooper}\ and\ \citenamefont
  {Slatyer}(2013)}]{hooperslatyer2013}%
  \BibitemOpen
  \bibfield  {author} {\bibinfo {author} {\bibfnamefont {Dan}\ \bibnamefont
  {Hooper}}\ and\ \bibinfo {author} {\bibfnamefont {Tracy~R.}\ \bibnamefont
  {Slatyer}},\ }\bibfield  {title} {\enquote {\bibinfo {title} {{Two Emission
  Mechanisms in the Fermi Bubbles: A Possible Signal of Annihilating Dark
  Matter}},}\ }\href@noop {} {\  (\bibinfo {year} {2013})},\ \Eprint
  {http://arxiv.org/abs/1302.6589} {arXiv:1302.6589 [astro-ph.HE]} \BibitemShut
  {NoStop}%
\bibitem [{\citenamefont {{Abdo}}\ \emph {et~al.}(2009)\citenamefont {{Abdo}}
  \emph {et~al.}}]{2009Sci...325..845A}%
  \BibitemOpen
  \bibfield  {author} {\bibinfo {author} {\bibfnamefont {A.~A.}\ \bibnamefont
  {{Abdo}}} \emph {et~al.},\ }\bibfield  {title} {\enquote {\bibinfo {title}
  {{Detection of High-Energy Gamma-Ray Emission from the Globular Cluster 47
  Tucanae with Fermi}},}\ }\href {\doibase 10.1126/science.1177023} {\bibfield
  {journal} {\bibinfo  {journal} {Science}\ }\textbf {\bibinfo {volume}
  {325}},\ \bibinfo {pages} {845--} (\bibinfo {year} {2009})}\BibitemShut
  {NoStop}%
\end{thebibliography}%

\appendix

\begin{table*}[h!]
\section{Gamma Ray Excess Data}
\label{sec:data}

\begin{ruledtabular}
\begin{tabular}{|c|c|c|c|c|}
\centering 
$E_{\rm min}$ [GeV] &	$E_{\rm max}$ [GeV] &	$dN/dE$ [GeV$^{-1}$ cm$^{-2}$ s$^{-1}$]  &	Statistical Error [GeV$^{-1}$ cm$^{-2}$ s$^{-1}$] &	Systematic Error [GeV$^{-1}$ cm$^{-2}$ s$^{-1}$]\\ \hline
0.30&	0.40 &	$2.20\times 10^{-6}$ &	$-$ &	$-$ \\
0.40&	0.54 &	$9.69\times 10^{-7}$ &	$5.52\times 10^{-8}$ &	$2.45\times 10^{-7}$ \\
0.54&	0.72 &	$5.43\times 10^{-7}$ &	$9.89\times 10^{-8}$ &	$1.14\times 10^{-7}$ \\
0.72&	0.97 &	$3.20\times 10^{-7}$ &	$3.16\times 10^{-8}$ &	$7.94\times 10^{-8}$ \\
0.97&	1.29 &	$2.01\times 10^{-7}$ &	$2.44\times 10^{-8}$ &	$4.08\times 10^{-8}$ \\
1.29&	1.73 &	$1.39\times 10^{-7}$ &	$1.51\times 10^{-8}$ &	$2.07\times 10^{-8}$ \\
1.73&	2.32 &	$6.27\times 10^{-8}$ &	$1.48\times 10^{-9}$ &	$1.09\times 10^{-8}$ \\
2.32&	3.11 &	$3.67\times 10^{-8}$ &	$3.52\times 10^{-9}$ &	$5.11\times 10^{-9}$ \\
3.11&	4.16 &	$1.58\times 10^{-8}$ &	$1.69\times 10^{-9}$ &	$2.32\times 10^{-9}$ \\
4.16&	5.57 &	$4.67\times 10^{-9}$ &	$8.28\times 10^{-10}$ &	$1.01\times 10^{-9}$ \\
5.57&	7.47 &	$3.51\times 10^{-9}$ &	$4.82\times 10^{-10}$ &	$5.03\times 10^{-10}$ \\
7.47&	10.00 &	$1.23\times 10^{-9}$ &	$2.46\times 10^{-10}$ &	$1.53\times 10^{-10}$

\end{tabular}
\end{ruledtabular}

\caption{\label{tab:5} Summary of spectral data, statistical errors and systematic errors used in our analysis. A graphical representation of this data can be found in Fig.~\eqref{fig:SED}. The spectral point $dN/dE$ evaluated at the logarithmic midpoint of $E_{\rm min,i}$ and $E_{\rm max,i}$ has been evaluated from $F_i$ using $dN/dE\approx {F_i \over (E_{\rm max,i}-E_{\rm min,i})}$. Since the first band (0.30$-$0.40 GeV) had a $TS<10$ we report a $2\sigma$ upper limit instead.  }
\end{table*}

\end{document}